\documentclass[aps, prd, twocolumn, amsmath, floats,floatfix, superscriptaddress, nofootinbib]{revtex4-1}

\usepackage{amssymb}
\usepackage{amsmath}
\usepackage{verbatim}
\usepackage{mathrsfs}
\usepackage{amsfonts}
\usepackage{latexsym}
\usepackage{epsfig}
\usepackage{color}
\usepackage{graphicx,subfigure}
\usepackage{units}
\usepackage{envmath}
\usepackage{natbib}
\usepackage[dvipsnames]{xcolor}
\usepackage[normalem]{ulem}
\usepackage[linktocpage]{hyperref}

\begin{document}


\definecolor{orange}{rgb}{0.9,0.45,0}

\newcommand{\re}{\mbox{Re}}
\newcommand{\im}{\mbox{Im}}

\newcommand{\tf}[1]{\textcolor{red}{TF: #1}}
\newcommand{\nsg}[1]{\textcolor{cyan}{NSG: #1}}
\newcommand{\ch}[1]{\textcolor{blue}{#1}}
\newcommand{\fdg}[1]{\textcolor{magenta}{FDG: #1}}
\newcommand{\mz}[1]{\textcolor{purple}{Miguel: #1}}
\newcommand{\pcd}[1]{\textcolor{Green}{#1}}

\def\CovDev{D}
\def\Res{{\mathcal R}}
\def\Gammaflat{\hat \Gamma}
\def\metricflat{\hat \gamma}
\def\Dflat{\hat {\mathcal D}}
\def\part_n{\partial_\perp}

\def\Lie{\mathcal{L}}
\def\A{\mathcal{X}}
\def\Aphi{\A_{\phi}}
\def\hAphi{\hat{\A}_{\phi}}
\def\E{\mathcal{E}}
\def\Ham{\mathcal{H}}
\def\M{\mathcal{M}}
\def\R{\mathcal{R}}
\def\p{\partial}

\def\hg{\hat{\gamma}}
\def\hA{\hat{A}}
\def\hD{\hat{D}}
\def\hE{\hat{E}}
\def\hR{\hat{R}}
\def\hcA{\hat{\mathcal{A}}}
\def\hDelt{\hat{\triangle}}

\def\na{\nabla}
\def\dif{{\rm{d}}}
\def\non{\nonumber}
\newcommand{\erf}{\textrm{erf}}

\renewcommand{\t}{\times}



\title{On the dynamical bar-mode instability in spinning bosonic stars}
 
\author{Fabrizio Di Giovanni}
\affiliation{Departamento de
  Astronom\'{\i}a y Astrof\'{\i}sica, Universitat de Val\`encia,
  Dr. Moliner 50, 46100, Burjassot (Val\`encia), Spain}

\author{Nicolas Sanchis-Gual}
\affiliation{Centro de Astrof\'\i sica e Gravita\c c\~ao - CENTRA, Departamento de F\'\i sica,
Instituto Superior T\'ecnico - IST, Universidade de Lisboa - UL, Avenida
Rovisco Pais 1, 1049-001, Portugal}

 \author{Pablo Cerd\'a-Dur\'an
}
\affiliation{Departamento de
  Astronom\'{\i}a y Astrof\'{\i}sica, Universitat de Val\`encia,
  Dr. Moliner 50, 46100, Burjassot (Val\`encia), Spain}

 \author{Miguel Zilh\~ao}
\affiliation{Centro de Astrof\'\i sica e Gravita\c c\~ao - CENTRA, Departamento de F\'\i sica,
Instituto Superior T\'ecnico - IST, Universidade de Lisboa - UL, Avenida
Rovisco Pais 1, 1049-001, Portugal}

    \author{Carlos Herdeiro}
\affiliation{Departamento de Matem\'atica da Universidade de Aveiro and 
Centre for Research and Development in Mathematics and Applications (CIDMA), 
Campus de Santiago, 
3810-183 Aveiro, Portugal}

\author{Jos\'e A. Font}
\affiliation{Departamento de
  Astronom\'{\i}a y Astrof\'{\i}sica, Universitat de Val\`encia,
  Dr. Moliner 50, 46100, Burjassot (Val\`encia), Spain}
\affiliation{Observatori Astron\`omic, Universitat de Val\`encia, C/ Catedr\'atico 
  Jos\'e Beltr\'an 2, 46980, Paterna (Val\`encia), Spain}

 \author{Eugen Radu}
\affiliation{Departamento de Matem\'atica da Universidade de Aveiro and 
Centre for Research and Development in Mathematics and Applications (CIDMA), 
Campus de Santiago, 
3810-183 Aveiro, Portugal}


\date{October 2020}


\begin{abstract} 
Spinning bosonic stars  (SBSs) can form from the gravitational collapse  of a dilute cloud of scalar/Proca particles with non-zero angular momentum, via gravitational  cooling.  The scalar stars are,  however, transient due to a non-axisymmetric instability which triggers the loss of angular momentum. By contrast, no such instability was observed for the fundamental ($m=1$) Proca stars. In~\cite{sanchis:2019dynamics} we tentatively related the different stability properties to the different toroidal/spheroidal morphology of the scalar/Proca models. 
Here, we continue this investigation, using  three-dimensional numerical-relativity simulations of the Einstein-(massive, complex)Klein-Gordon system and of the  Einstein-(complex)Proca system. Firstly, we incorporate a quartic self-interaction potential in the scalar case to gauge its effect on the instability. Secondly, we investigate toroidal ($m=2$) Proca stars to assess their stability. Thirdly, we attempt to relate the instability of SBSs to the growth rate of azimuthal density modes and the existence of a corotation point in the unstable models.  Our results show that: (a) the self-interaction potential can only delay the instability in scalar SBSs but cannot quench it completely; (b)  $m=2$ Proca stars always migrate to the stable $m=1$ spheroidal family; (c) unstable $m=2$ Proca stars and $m=1$ scalar boson stars exhibit a pattern of frequencies for the azimuthal density modes which crosses the angular velocity profile of the stars in the corotation point. This establishes a parallelism with rotating neutron stars affected by dynamical bar-mode instabilities. Finally, we compute the gravitational waves emitted by SBSs due to the non-axisymmetric instability. We investigate the detectability of the waveforms comparing the characteristic strain of the signal with the sensitivity curves of a variety of detectors, computing the signal-to-noise ratio for different ranges of masses and for different source distances.  Moreover, by assuming that the characteristic damping timescale of the bar-like deformation in SBSs is only set by gravitational-wave emission and not by viscosity (unlike in neutron stars), we find that the post-collapse emission could be orders of magnitude more energetic than that of the bar-mode instability itself. Our results indicate that gravitational-wave observations of SBSs might be within  the reach of future experiments, offering a potential means to establish the existence of such stars and to place tight constraints on the mass of the bosonic particle.

\end{abstract}




\maketitle

\vspace{0.8cm}

\section{Introduction}

The brand new field of 
gravitational-wave (GW) astronomy~\cite{Abbott:2016blz,Abbott:2016nmj,Abbott:2017vtc,Abbott:2017gyy,TheLIGOScientific:2017qsa,Abbott:2017oio} is allowing for new explorations of the Universe from an astrophysical scale to a cosmological scale. The recent LIGO-Virgo detections of GW signals from coalescing compact binaries along with the Event Horizon Telescope observations of the center of the galaxy M87~\cite{Akiyama:2019cqa} provide firm evidence to the black hole (BH) hypothesis. A picture is emerging that BHs seem to populate the cosmos in large numbers and they are widely regarded as the main type of dark compact object, i.e.~an object that barely interacts with baryonic matter except through gravity. Notwithstanding the prominent place that BHs currently occupy in our standard model, a good many varieties of {\it exotic} dark compact objects have been proposed in the past (see e.g.~\cite{Cardoso2019} and references therein). The study of these so-called BH ``mimickers'' is interesting from a number of perspectives, chiefly to test General Relativity in the strong-field regime, possibly through the detection of GWs, but also to assess their potential  relevance as alternative candidates to explain the nature of dark matter (DM). 

In particular, the introduction of new fields not included in the Standard Model of fundamental interactions is necessary in cosmology to explain the mounting evidence supporting the existence of DM. The simplest possible theory which minimally couples a massive bosonic field, either scalar~\cite{Kaup:1968zz,Ruffini:1969qy} or vector~\cite{Brito:2015pxa}, to Einstein's gravity, gives rise to self-gravitating compact objects. These are known as bosonic stars (BSs) or oscillatons~\cite{Page:2003rd}, depending on whether the field is complex or real, respectively.\footnote{See also~\cite{Hawley:2002zn} for multiscalar stars that may interpolate between oscillatons  and boson stars.} They are dark as far as their interaction with the Standard Model particles is considered to be weak. The dynamical features of BSs have been deeply studied (see e.g.~\cite{Schunck:2003kk,Liebling:2012fv} and references therein) in the static, spherically symmetric case. For some range of the model parameters, the fundamental family (FF) can form dynamically through the so-called gravitational cooling mechanism~\cite{seidel1994formation,di2018dynamical} and are stable under perturbations~\cite{Gleiser:1988ih,Lee:1988av,Brito:2015pxa,sanchis2017numerical,Guzman:2019mat}. Spherical BS models have moreover been considered to build orbiting binaries, from which GWs have been extracted and compared to BHs signals~\cite{palenzuela2007head,palenzuela2008orbital,sanchis2019head}. All existing studies within spherical symmetry have shown a remarkable parallelism between the dynamics of scalar and vector BS models.

Models of axisymmetric, spinning bosonic stars (SBSs) have also been constructed for a scalar field~\cite{Schunck:1996he,Yoshida:1997qf,Kleihaus:2007vk}, a vector field~\cite{Brito:2015pxa,Herdeiro:2016tmi,Santos:2020pmh} (see also~\cite{Herdeiro:2019mbz}), and some of their phenomenology has been studied, including geodesic motion~\cite{Meliani:2015zta,Herdeiro:2015gia,Herdeiro:2016gxs,Franchini:2016yvq,Delgado:2020udb}, lensing~\cite{Vincent:2015xta,Cunha:2015yba} and  properties of the X-ray spectrum due to an accretion disk~\cite{Cao:2016zbh,Shen:2016acv}. Recently~\cite{sanchis:2019dynamics} we studied the dynamical properties of SBSs by performing fully non-linear numerical-relativity simulations. The goal of that study was to answer two fundamental questions: (i) are SBSs stable? and (ii) may they form dynamically from the gravitational collapse of a bosonic cloud? Our study revealed that the parallelism between scalar and vector fields in the spherically-symmetric case breaks down when we consider spinning models. We found that scalar SBSs in the FF always develop a non-axisymmetric instability. Moreover, in the formation scenario, the collapse of the cloud leads only to a transient SBS, which then splits into an orbiting binary which eventually re-collapses into a non-spinning scalar boson star, ejecting all the angular momentum. The evolution of an already formed stationary SBS triggers the same type of non-axisymmetric instability and the collapse to a BH, even considering models which were thought to be stable to linear axisymmetric perturbations - see the discussion in Sec.~6.2 in~\cite{Herdeiro:2015gia}. By contrast, the vector SBS models we considered, also known as spinning Proca stars, did not show any instability. As a result, in~\cite{sanchis:2019dynamics} we put forward the hypothesis that the different dynamical properties of these two families of SBSs were related to their different morphology (the energy density profile of scalar SBSs has a toroidal shape while vector SBSs exhibit a spheroidal one) and possibly to the existence of a corotational instability in the scalar case.

In this work we further investigate this issue, extending our previous investigation along different directions. Firstly, we carry out a  deeper exploration of the two families of SBSs by taking into account the dynamics of a larger set of new models; secondly,  we provide a qualitative description of the growth of the non-axisymmetric instability of scalar (and vector) SBSs and compare our findings with well-known results for differentially rotating neutron stars~\cite{Watts:2005,CERDADURAN2007288,Paschalidis:2016vmz}. Our dynamical study is  focused only on the formation scenario. Comparing with our previous work, we construct here new models of scalar bosonic clouds with a quartic self-interaction potential and study if the instability found in the scalar case is affected by increasing the contribution from the self-interaction term. For the vector field case, we consider Proca clouds belonging to the $m=2$ family of solutions which, unlike the $m=1$ case discussed in~\cite{sanchis:2019dynamics}, show a toroidal morphology and may be subject of the same type of instability that affects the scalar case, which would support our conjecture. 

We furthermore study the GWs emitted by unstable SBSs, computing the mode decomposition of the Newman-Penrose scalar $\Psi_4$. We evaluate the characteristic GW strain $h_{\rm char}$ for some of our models and we compare it with the sensitivity curves of current and future ground-based and space interferometric detectors, as well as from future observational projects based on Pulsar Timing Arrays. For each detector and for different ranges of masses of the SBS, we compute the horizon distance, defined as the distance between the observer and the source at which the signal-to-noise ratio (SNR) is equal to a certain threshold value. Our results show that the GW signals produced by the bar-mode instability in SBSs are within reach of future detectors which offers the intriguing possibility of an eventual detection of such exotic objects and might help place constraints on the mass of the bosonic particle. In this context it is worth pointing out our recent proposal to estimate such a mass by considering collisions of Proca stars to explain GW signal GW190521~\cite{GW190521D, GW190521I, bustillo2020ultra}.

This paper is organized as follows: in Section~\ref{sec2} we briefly present the basic set of equations we solve, for both the scalar field and the vector field. In Section~\ref{sec3} we construct the initial data for the bosonic cloud. Section~\ref{numerics} presents our numerical framework and in Section~\ref{results} we discuss the main results of our work. Finally,  our findings are summarized in Section~\ref{conclusions}. We use geometrized units, $G=c=\hbar=1$,  $G$ being Newton's constant and $c$ the speed of light. This choice makes the Planck mass equal to one, effectively disappearing from all equations. Latin (Greek) indices run from 1 (0) to 3.

\section{Formalism}
\label{sec2}

In this paper we study the dynamics of a scalar/Proca field minimally coupled to gravity by solving numerically the Einstein-Klein-Gordon and Einstein-Proca systems respectively. In both cases, the bosonic field is assumed to be complex and massive.
The systems are described by the action 
\begin{eqnarray}
\mathcal{S}=\int d^4x \sqrt{-g}\left(\frac{R}{16G\pi} + \mathcal{L}_{(S)}\right)\,, 
\end{eqnarray}
where the subscript $(S)$ for the Lagrangian densities refers to the spin of the particles, i.e.~$0$ for the scalar field and $1$ for the Proca field. The spacetime line element reads 
\begin{eqnarray}
ds^2&=&g_{\mu\nu}dx^{\mu}dx^{\nu} \\
    &=&-(\alpha^2-\beta_{i}\beta^{i})dt^2 + 2\gamma_{ij}\beta^idtdx^j + \gamma_{ij}dx^idx^j, \nonumber 
\end{eqnarray}
where $\alpha$ is the lapse function, $\beta^i$ is the shift vector, and $\gamma_{ij}$ is the spatial metric. We cast the field equations into a 3+1 form, introducing the extrinsic curvature (conjugated momentum of the 3-metric) $K_{ij}$, defined as
\begin{equation}
K_{ij}=-\frac{1}{2\alpha}(\partial_t - \mathcal{L}_{\beta})\gamma_{ij},
\end{equation}
where $\mathcal{L}_{\beta}$ is the Lie derivative along $\beta^i$. We use the Baumgarte-Shapiro-Shibata-Nakamura (BSSN) formulation of Einstein's equations~\cite{nakamura1987general,Shibata95,baumgarte1998numerical}. The matter source terms appearing in the BSSN equations depend on the energy density $\rho_e$, the momentum density $j_{i}$ measured by an observer $n^{\mu}$ normal to the spatial hypersurfaces defining the spacetime foliation, and the spatial projection of the energy-momentum tensor $S_{ij}$, namely
\begin{align}
\rho_e&= n^{\mu}n^{\nu}T_{\mu \nu}, \\
j_i&=-\gamma_{i}^{\mu}n^{\nu}T_{\mu \nu}, \\
S_{ij}&= \gamma_{i}^{\mu} \gamma_{j}^{\nu} T_{\mu \nu},
\end{align}
 where the unit normal vector is $n^{\mu} = \frac{1}{\alpha}(1,-\beta^{i})$ and $\gamma_{i}^{\mu}$ is the spatial projection operator.

\subsection{Einstein-Klein-Gordon system}
The Lagrangian density for a scalar field $\phi$ with a quartic self-interaction potential is given by
\begin{equation}
\mathcal{L}_{(0)}= -\frac{1}{2}\partial^\alpha \phi \,\partial_\alpha \bar{\phi}-\frac{1}{2}\mu_0^2 \phi\bar{\phi} - \frac{1}{4}\lambda(\phi\bar{\phi})^2\,, 
\label{scalar-Lagrangian} 
\end{equation}
where the bar denotes complex conjugation, $\mu_0$ is the mass parameter of the scalar field, and $\lambda$ is the coupling constant of the self-interaction term. The energy-momentum tensor associated with this field is
\begin{eqnarray} 
T_{\mu\nu}&=& \frac{1}{2}g_{\mu\nu}(\partial_{\lambda}\bar{\phi}\partial^{\lambda}\phi + \mu_0^2\bar{\phi}\phi+\frac{1}{2}\lambda(\bar{\phi}\phi)^2) \nonumber \\
 &+& \frac{1}{2}(\partial_{\mu}\bar{\phi}\partial_{\nu}\phi+\partial_{\mu}\phi\partial_{\nu}\bar{\phi}) .
\end{eqnarray}
After introducing the conjugated momentum of the scalar field, $\Pi$, defined as
\begin{equation}\label{scalarPi}
\Pi = -\frac{1}{\alpha}(\partial_t - \mathcal{L}_{\beta})\phi\,,
\end{equation}
it can be shown that in this case
\begin{align}
\rho_e&= \frac{1}{2} \left(\bar{\Pi}\Pi + \mu_{0}^{2} \bar{\phi}\phi + \frac{1}{2} \lambda (\bar{\phi}\phi)^{2} + D^{i}\bar{\phi}D_i\phi\right)\,, \\
j_i&=\frac{1}{2}(\bar{\Pi}D_{i}\phi + \Pi D_{i}\bar{\phi})\,, \\
S_{ij}&= \frac{1}{2} (D_{i}\bar{\phi}D_j\phi + D_{j}\bar{\phi}D_i\phi) + \frac{1}{2}\gamma_{ij}(\bar{\Pi}\Pi \nonumber \\
	& -\mu_{0}^{2} \bar{\phi}\phi - \frac{1}{2} \lambda (\bar{\phi}\phi)^{2} - D^k\bar{\phi}D_{k}\phi)\,,
\end{align}
where $D_{i}$ stands for the covariant derivative associated with $\gamma_{ij}$.

\subsection{Einstein-Proca system}
Correspondingly, the Lagrangian density for a Proca field $\mathcal{A}^{\alpha}$ reads as
\begin{equation}
\mathcal{L}_{(1)}=-\frac{1}{4}\mathcal{F}_{\alpha\beta}\bar{\mathcal 
{F}}^{\alpha\beta}-\frac{1 } {2}\mu_1^2\mathcal{A}_\alpha\bar{\mathcal{A}}^\alpha \,,
\label{Proca-Lagrangian}
\end{equation}
where the bar denotes complex conjugation, $\mathcal{F}=d\mathcal{A}$ is the Proca field strength, and $\mu_1$ is the Proca mass parameter. From the variation of this Lagrangian we can build the energy-momentum tensor of the Proca field,
\begin{eqnarray} 
T_{\mu\nu}&=& -\mathcal{F}_{\lambda(\mu}  \bar {\mathcal{F}}_{\nu)}^{\,\,\lambda}-\frac{1}{4}g_{\mu\nu}\mathcal{F}_{\lambda\alpha}\bar{\mathcal{F}}^{\lambda\alpha} 
\nonumber \\
&+& \mu^2_1 \left[
\mathcal{A}_{(\mu}\bar{\mathcal{A}}_{\nu)}-\frac{1}{2}g_{\mu\nu}\mathcal{A}_{\lambda}\bar{\mathcal{A}}^{\lambda}
\right]\, .
\end{eqnarray}
The index notation $(\mu,\nu)$ indicates, as usual, index symmetrization. The Proca 1-form $\mathcal{A}_{\mu}$ can be split into its scalar potential $\mathcal{X}_{\phi}$, its $3$-vector potential $\mathcal{X}_{i}$, and the 3-dimensional electric $E_{i}$ and magnetic $B_i$ field, defined by
\begin{eqnarray}
\mathcal{X}_{\phi}&=&-n^{\mu}\mathcal{A}_{\mu}\  , \label{propot}\\
\mathcal{X}_{i}&=&\gamma^{\mu}_{i}\mathcal{A}_{\mu}\ ,\\
E^{i}&=&-i\,\frac{\gamma^{ij}}{\alpha}\,\biggl(D_{j} (\alpha\mathcal{X}_{\phi})+\partial_{t}\mathcal{X}_{j}\biggl) ,\\ 
B^{i}&=& \epsilon^{ijk} D_{j}\mathcal{X}_k,
\end{eqnarray}
where $\epsilon^{ijk}$ is the Levi-Civita tensor.

Finally, in this case
\begin{align}
8\pi\rho_e&=\gamma_{ij}(\bar{E}^iE^j + \bar{B}^iB^j) + \mu_{1}^2(\bar{\mathcal{X}}_{\phi}\mathcal{X}_{\phi} + \gamma^{ij}\bar{\mathcal{X}}_{i}\mathcal{X}_{j}) , \\
4\pi j_i&=\frac{1}{2}\mu_{1}^2 (\bar{\mathcal{X}_{\phi}}\mathcal{X}_{i} + \mathcal{X}_{\phi}\bar{\mathcal{X}}_{i}), \\
4 \pi S_{ij}&= -\gamma_{ik}\gamma_{jl}(\bar{E}^k E^l + \bar{B}^k B^l) + \frac{1}{2}\gamma_{ij}(\bar{E}^k E_{k}  \nonumber \\
&+ \bar{B}^k B_{k} + \mu_{1}^2 \bar{\mathcal{X}}_{\phi}\mathcal{X}_{\phi} - \mu_{1}^2 \bar{\mathcal{X}}^k \mathcal{X}_{k}) + \mu_{1}^{2}\bar{\mathcal{X}}_{i} \mathcal{X}_{j}.
\end{align}

\section{Initial Data}
\label{sec3}

To study the dynamical formation of SBSs we must first construct the initial configurations of the fields, both for the spacetime and the matter. As in~\cite{sanchis:2019dynamics} our choice of initial data is a cloud of bosonic matter with non-zero angular momentum. The initial data must satisfy the constraint equations of the system, namely the Hamiltonian constraint, the momentum  constraint (see Eqs. (15)-(17) of~\citep{CorderoCarrion:2008nf}), and, for the Proca case, the Gauss constraint which reads as
\begin{equation}\label{Gauss-constraint}
D_{i} E^{i} = \mu_{1}^2 \mathcal{X}_{\phi}.
\end{equation}

 To build our initial configuration we assume an ansatz for the scalar/Proca field and we then build the spacetime fields by solving the Einstein constraint equations using the extended conformally flatness condition approximation~\citep{CorderoCarrion:2008nf}. We refer the interested reader to the supplementary material of~\cite{sanchis:2019dynamics} for the procedure we follow to construct the initial data. 

For completeness, we report here the ansatz for the scalar and the Proca fields. 
For the scalar field case we specify the ``shape" of the scalar cloud as in~\cite{sanchis:2019dynamics}
\begin{equation}\label{Phi1}
 \phi(t, r, \theta, \varphi) = R(r) Y_{11}(\theta,\varphi)\,e^{-i\omega t} \ ,
\end{equation}
where $Y_{11}(\theta,\varphi)=\sin{\theta}\,e^{i\varphi}$ is the $\ell=m=1$ spherical harmonic and $R(r) = A_{0}\,r\, e^{-\frac{r^2}{\sigma^2}}$. The width of the Gaussian cloud $\sigma$ is a free parameter of the initial data. At $t=0$
\begin{equation}\label{Pi2}
\Pi = -\frac{i}{\alpha} (\omega + \beta^{\varphi})\,\phi\ ,
\end{equation}
where we use that $\beta^{\varphi}$  is the only non-zero component of the shift vector,
a consequence of the axisymmetry invariance of the energy-momentum tensor.

For the components of the Proca field, we must also solve the Gauss constraint. In~\cite{sanchis:2019dynamics} we only considered the $m=1$ Proca field. Here, we make a new ansatz for the scalar potential, to describe the $m=2$ Proca field, namely
\begin{equation}  \label{Xphi}
\mathcal{X}_{\phi}(t, r, \theta, \varphi)  = R(r) Y_{22}(\theta,\varphi)\,e^{-i\omega t}  \ ,
\end{equation}
where $R(r) = A_{1}\,r e^{-\frac{r^2}{\sigma^2}}$ and $Y_{22}(\theta,\varphi)=\sin^2{\theta}\,e^{2 i\varphi}$ is the $\ell=m=2$ spherical harmonic. We assume the electric field $E^i$ is conservative, thus it can be written as the gradient of a potential. In this way the Gauss constraint can be solved analytically and yields 
\begin{align}
E^r(t,r,\theta,\varphi) &= \frac{A_{1} \sigma}{10 r^4} \biggl(-2 \sqrt{\pi}r^5 + 6 \sigma^{5} - 3 \sigma e^{-\frac{r^2}{\sigma^2}}(r^4 + \nonumber \\		
     & \qquad \qquad 2r^2\sigma^2 + 2\sigma^4 ) + 2\sqrt{\pi}r^5 \rm{Erf}(\frac{r}{\sigma})\biggr) \nonumber \\
     & \quad \times \sin{\theta}^2 e^{i(\omega t + 2\varphi)} \,, \\
  E^{\theta}(t,r,\theta,\varphi)
  &= \frac{A_{1} \sigma}{5 r^5} \biggl(- \sqrt{\pi}r^5  - 2 \sigma^{5} + \sigma e^{-\frac{r^2}{\sigma^2}}(r^4 + \nonumber \\		
     & \qquad \qquad 2r^2\sigma^2 + 2\sigma^4 ) + \sqrt{\pi}r^5 \rm{Erf}(\frac{r}{\sigma})\biggr)  \nonumber \\
     & \quad \times \sin{\theta}\cos{\theta} e^{i(\omega t + 2\varphi)}\,,  \\
  E^{\varphi}(t, r,\theta,\varphi)
  &= \frac{A_{1} \sigma}{5 r^5} \biggl(- \sqrt{\pi}r^5 - 2 \sigma^{5} + \sigma e^{-\frac{r^2}{\sigma^2}}(r^4 + \label{Ephi} \nonumber \\		
     & \qquad \qquad 2r^2\sigma^2 + 2\sigma^4 ) + \sqrt{\pi}r^5 \rm{Erf}(\frac{r}{\sigma})\biggr) \nonumber \\
     & \quad \times e^{i(\omega t + 2\varphi)}\,.  
\end{align}
The vector potential can be obtained following the same reasoning we used in the supplementary material of~\cite{sanchis:2019dynamics}, that gives us this relation
\begin{equation}
\mathcal{X}_{i}=\frac{i}{\alpha}(\omega + 2 \beta^{\phi})\gamma_{ij} E^{i}.
\end{equation}


\section{Numerical considerations} 
\label{numerics}

The numerical evolutions of the initial data are performed using the community-driven software platform \textsc{EinsteinToolkit}~\cite{EinsteinToolkit:2019_10,Loffler:2011ay,Zilhao:2013hia} which is based on the \textsc{Cactus} framework
with
\textsc{Carpet}~\cite{Schnetter:2003rb,CarpetCode:web} for mesh-refinement
capabilities. The spacetime variables in the BSSN formulation are solved using the \textsc{McLachlan} infrastructure~\cite{reisswig2011gravitational,brown2009turduckening}. The numerical code used in this work was originally assessed in~\cite{Zilhao:2015tya} and is currently publicly in available in~\cite{Canuda_2020_3565475} and distributed within each new release of the \textsc{EinsteinToolkit};  as in~\cite{sanchis2019head} we specifically employ a version of this code which was extended to take into account a complex Proca field.   

Our numerical grid uses four refinement levels, each spanning a different spatial domain and each discretized with a different resolution. From the outermost grid to the innermost one, the spatial domains of the grids are $\lbrace 512, 256, 128, 32\rbrace$ and the corresponding grid resolutions of each level are $\lbrace 6.4,3.2,1.6, 0.8\rbrace$. We consider a larger numerical grid for the computation of the GWs; the spatial domains are $\lbrace 1024, 512, 256, 32\rbrace$, and the corresponding grid resolutions of each level are $\lbrace 6.4,3.2,1.6, 0.8\rbrace$.
The time step is set as $\Delta_{t} = 0.125 \Delta_{x} = 0.1$, where $\Delta_x$ is the grid spacing of the innermost grid along the $x$ direction. All grids are equally spaced in all three spatial directions. Due to the geometry of the systems we investigate we assume reflection symmetry with respect to the equatorial plane. We use radiative (Sommerfeld) outer boundary condition implemented in the  \textsc{EinsteinToolkit} thorn \textsc{NewRad} for the evolutions.

We set the value of the mass of the particle to $\mu_{0}=\mu_{1}=1$ for all simulations. This sets the scale of the total mass of the systems under consideration at $M_0\sim 1$.
However, the simulations can be rescaled to obtain the results corresponding to different choices of $\mu_{0|1}$ by making the transformation
$r \to r \times \mu_{0|1}$, $M_0 \to M_0 \times \mu_{0|1}$, $t\to t\times  \mu_{0|1}$ and $\omega \to \omega / \mu_{0|1}$.

\begin{figure*}
\centering
\includegraphics[height=1.0in]{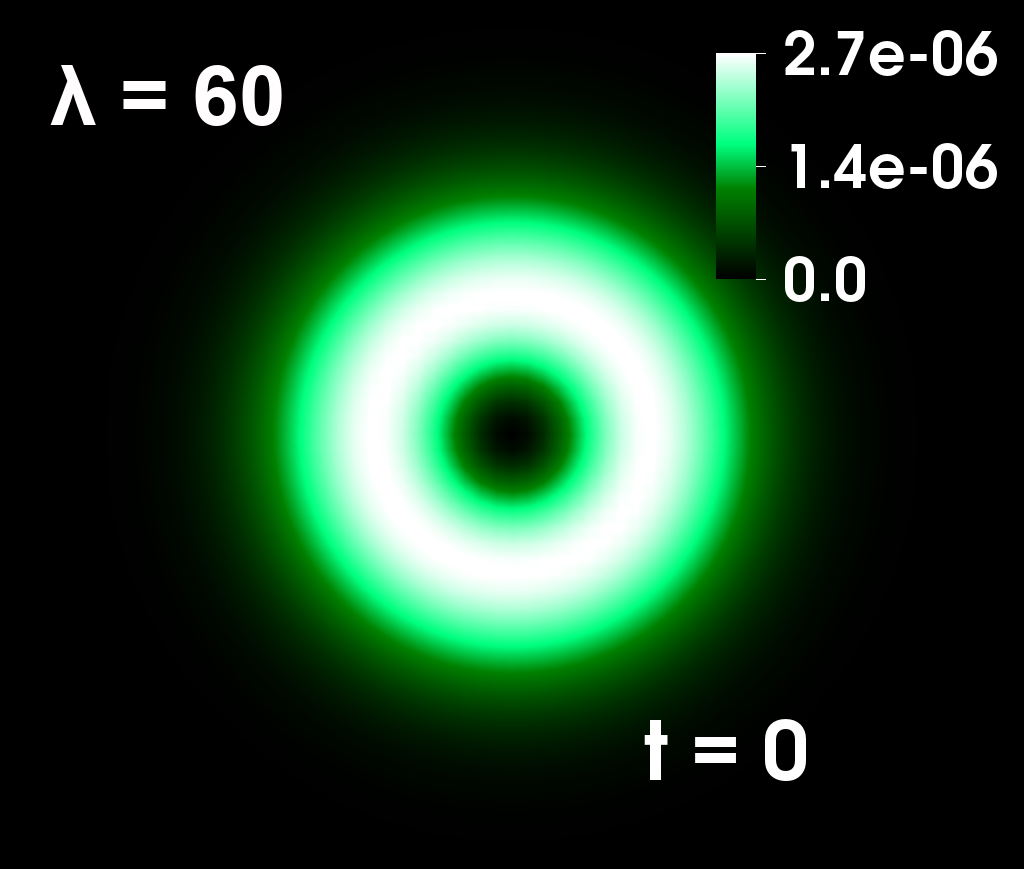} \hspace{0.15cm} \includegraphics[height=1.0in]{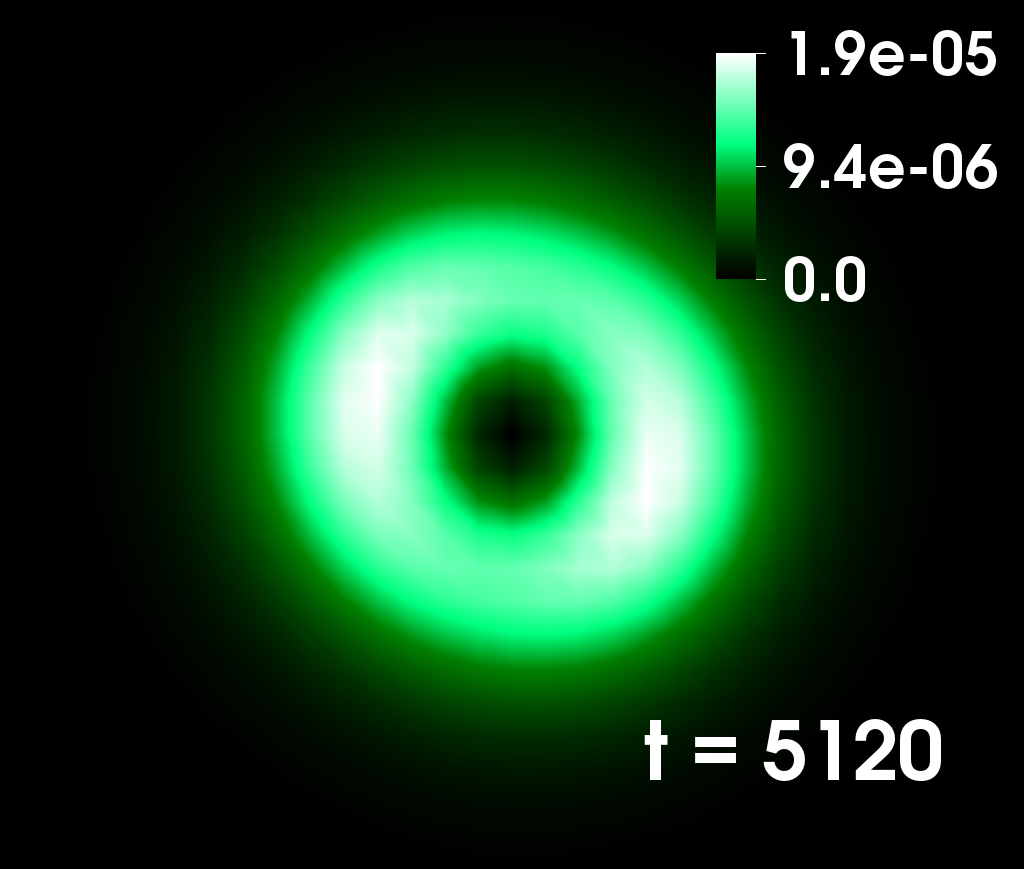}  \includegraphics[height=1.0in]{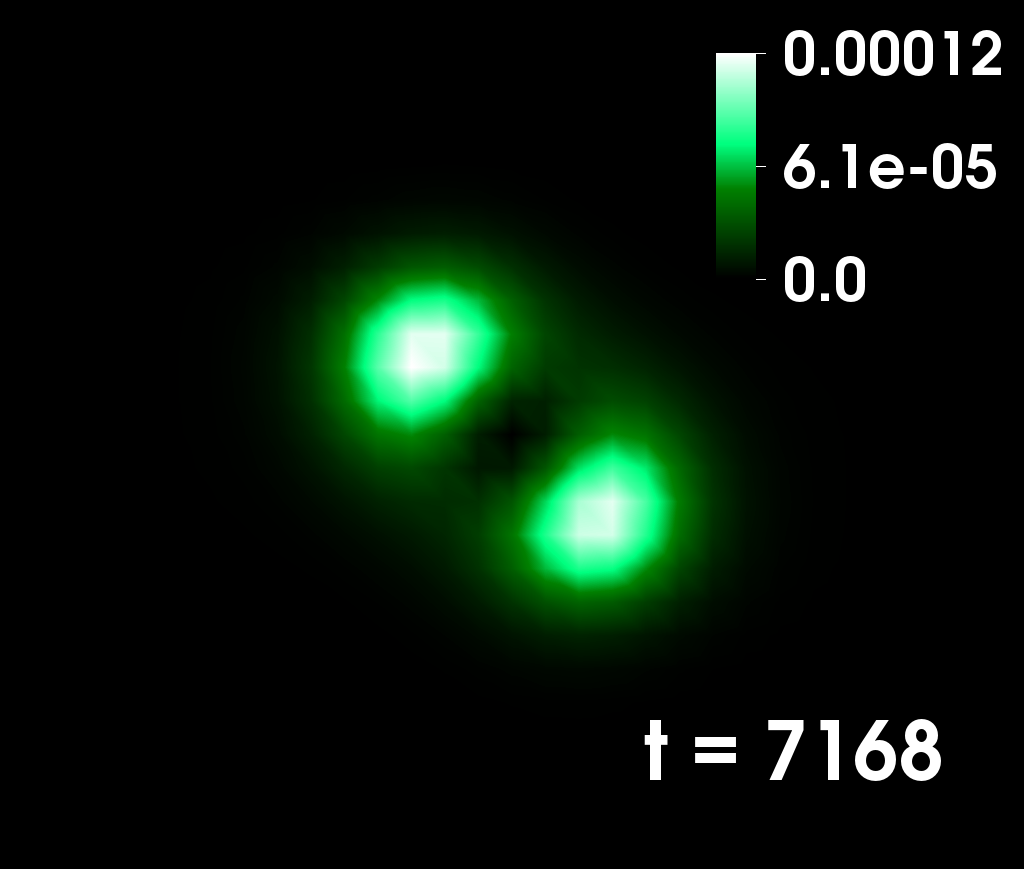} \includegraphics[height=1.0in]{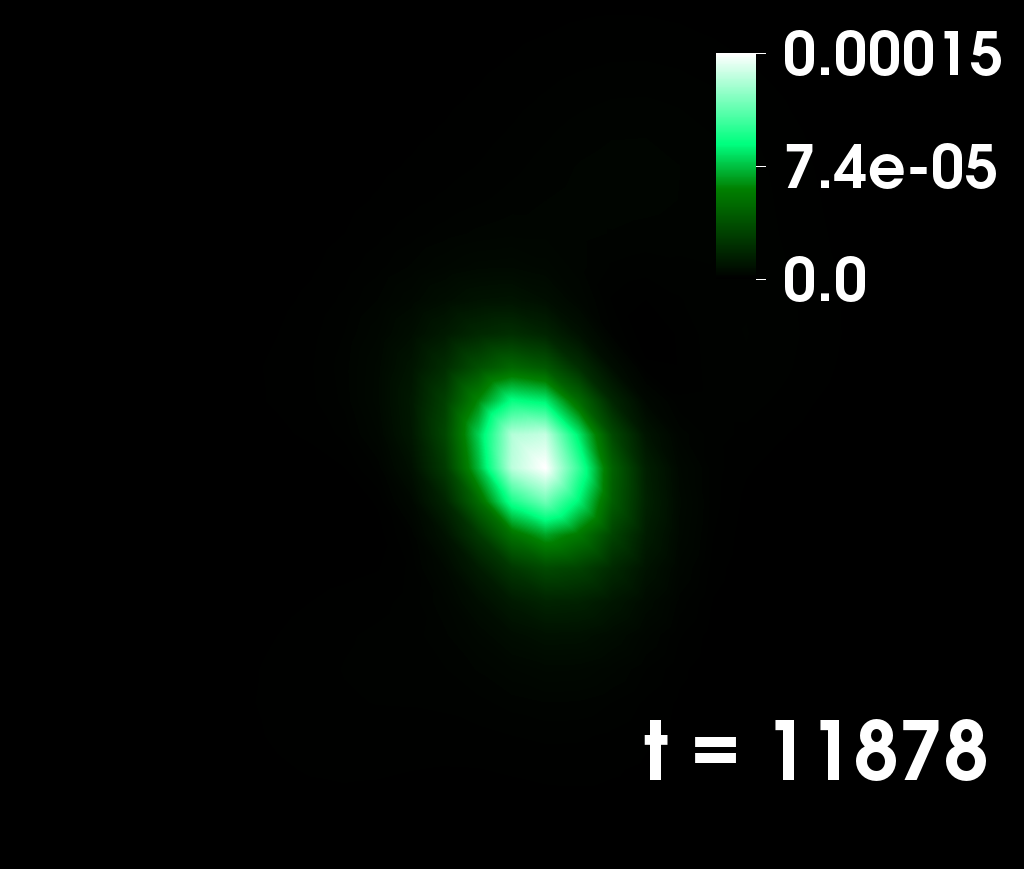} \includegraphics[height=1.0in]{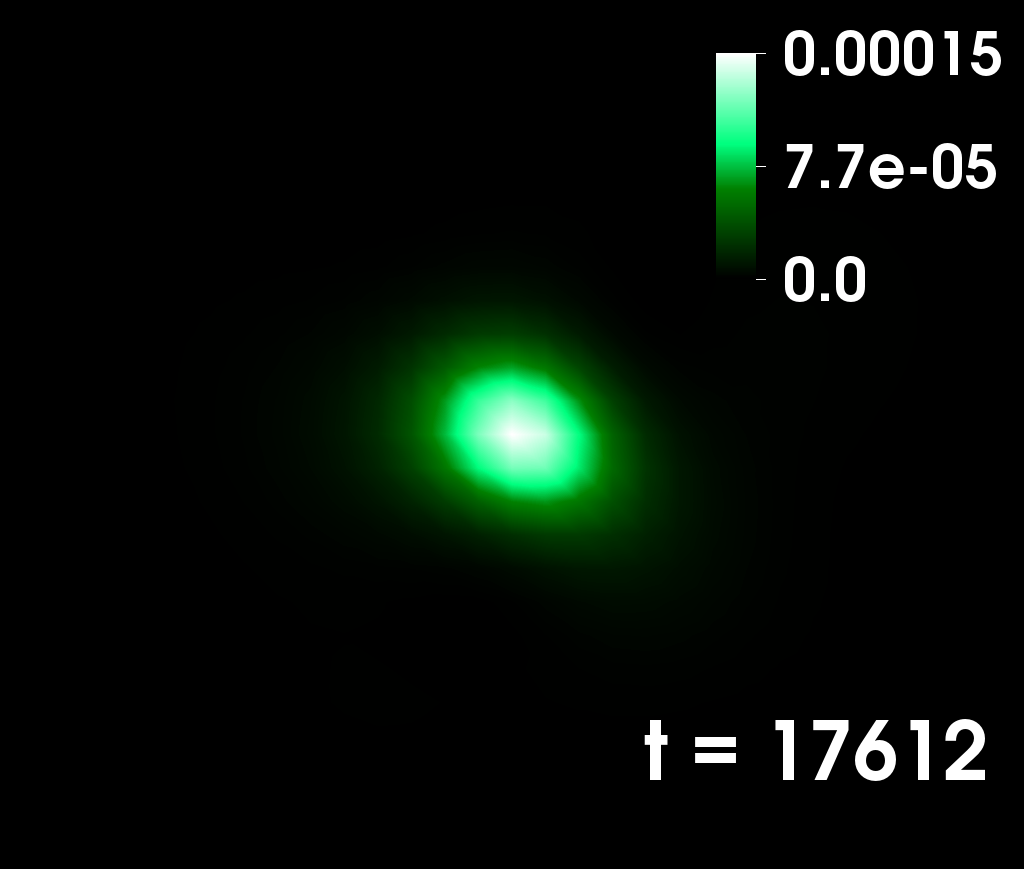} \\
\smallskip
\includegraphics[height=1.0in, width = 1.18in]{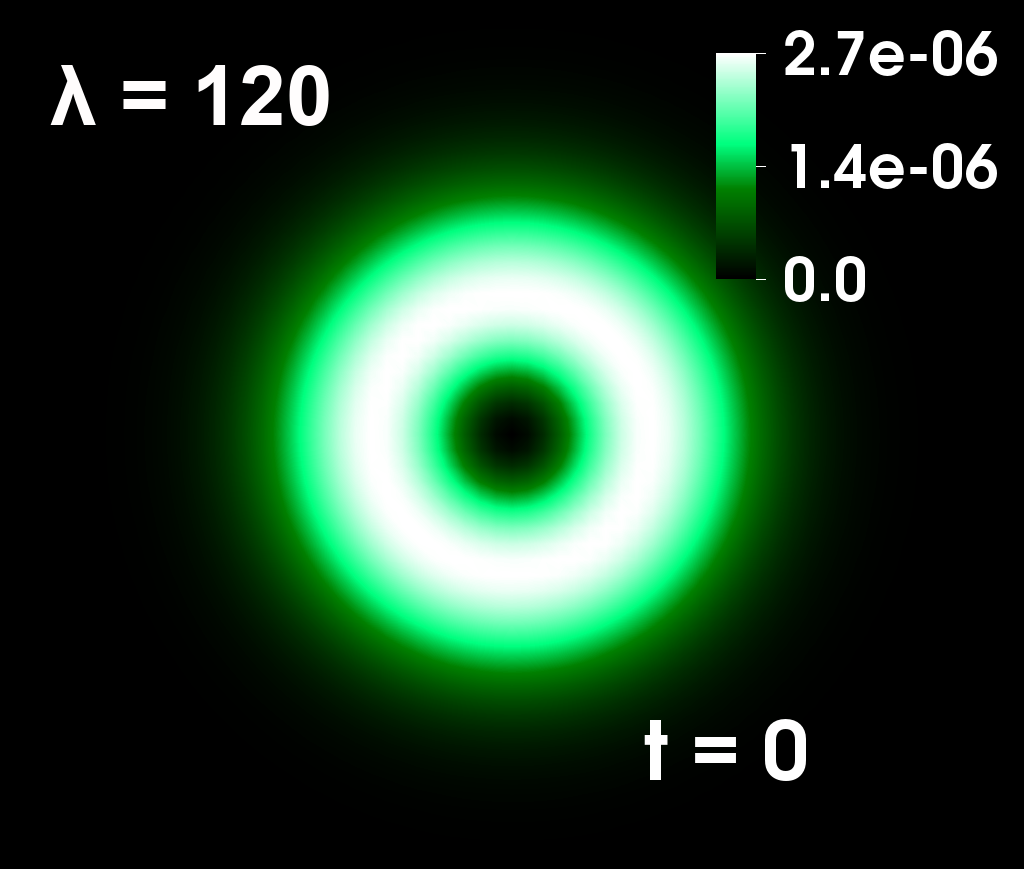} \hspace{0.15cm} \includegraphics[height=1.0in, width = 1.18in]{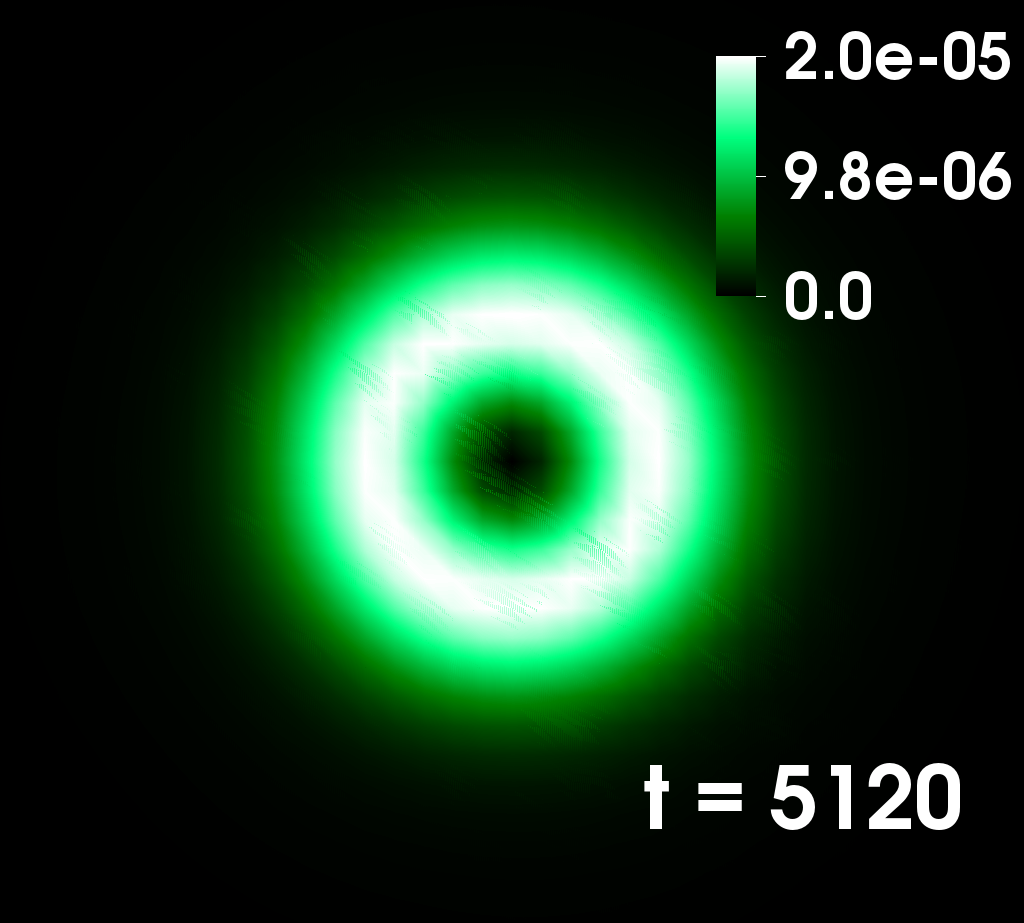}  \includegraphics[height=1.0in, width = 1.18in]{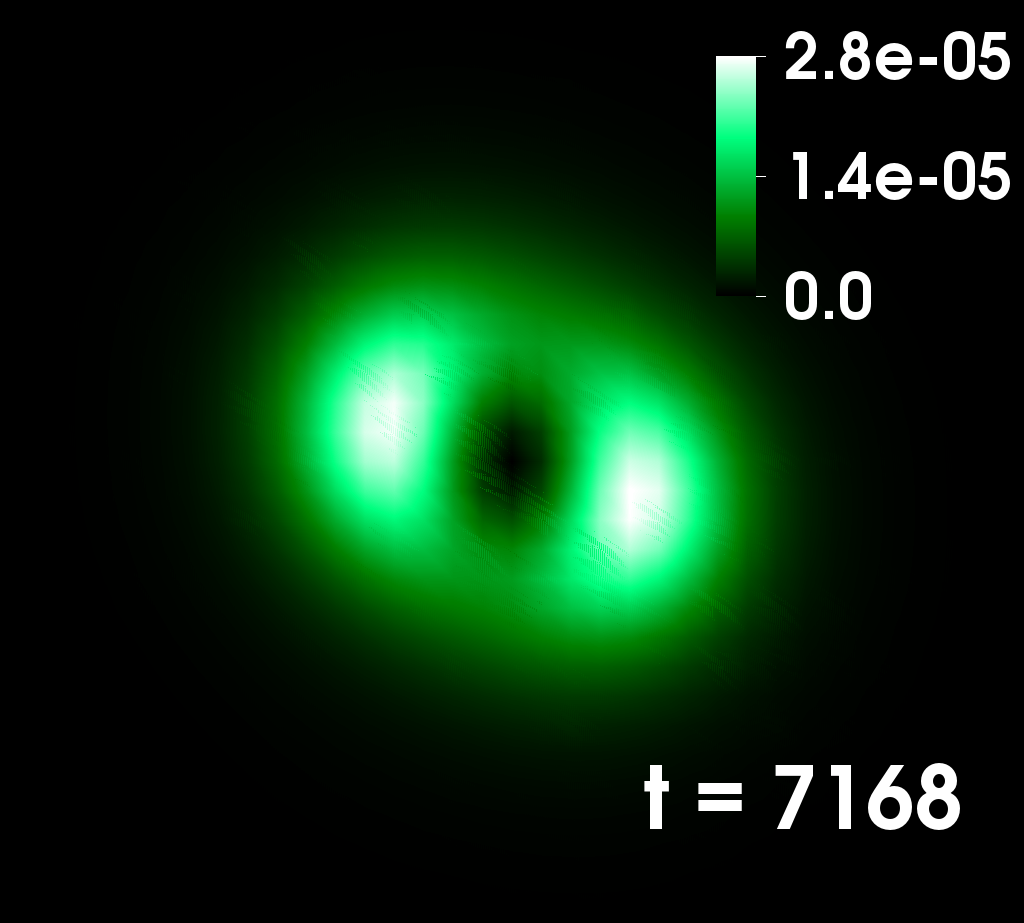} \includegraphics[height=1.0in, width = 1.18in]{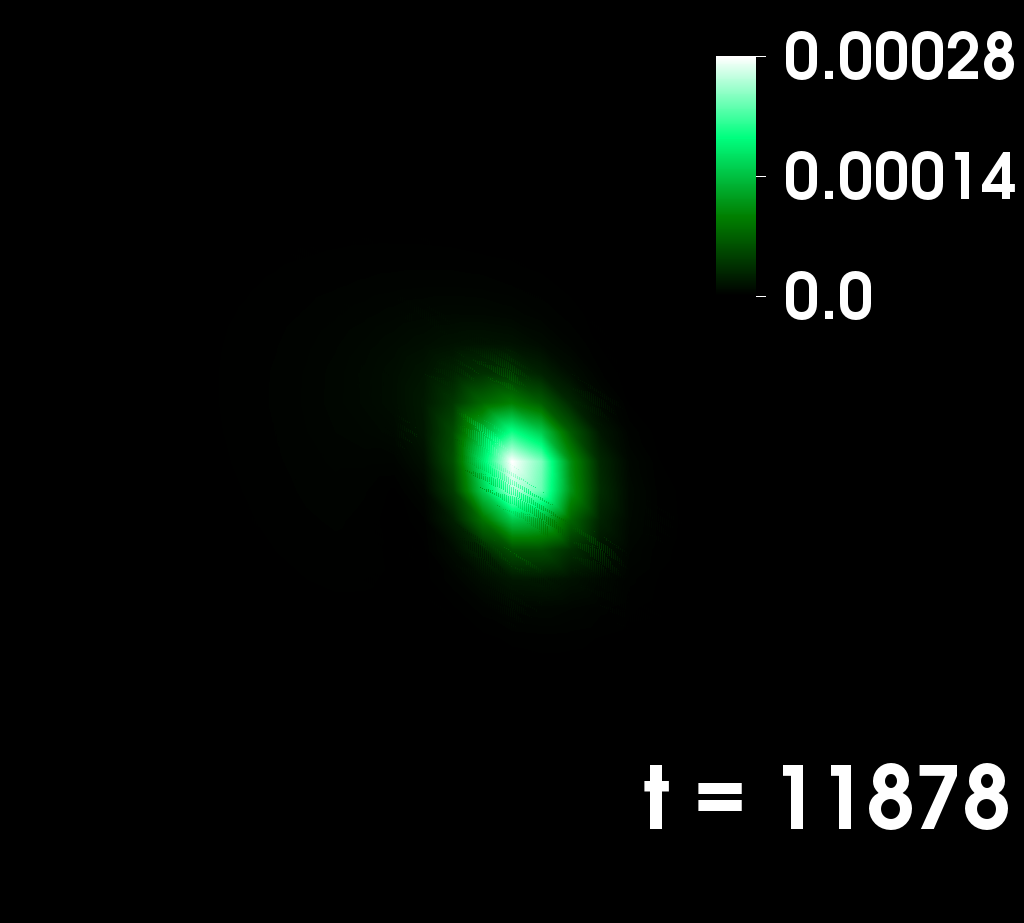} \includegraphics[height=1.0in, width = 1.18in]{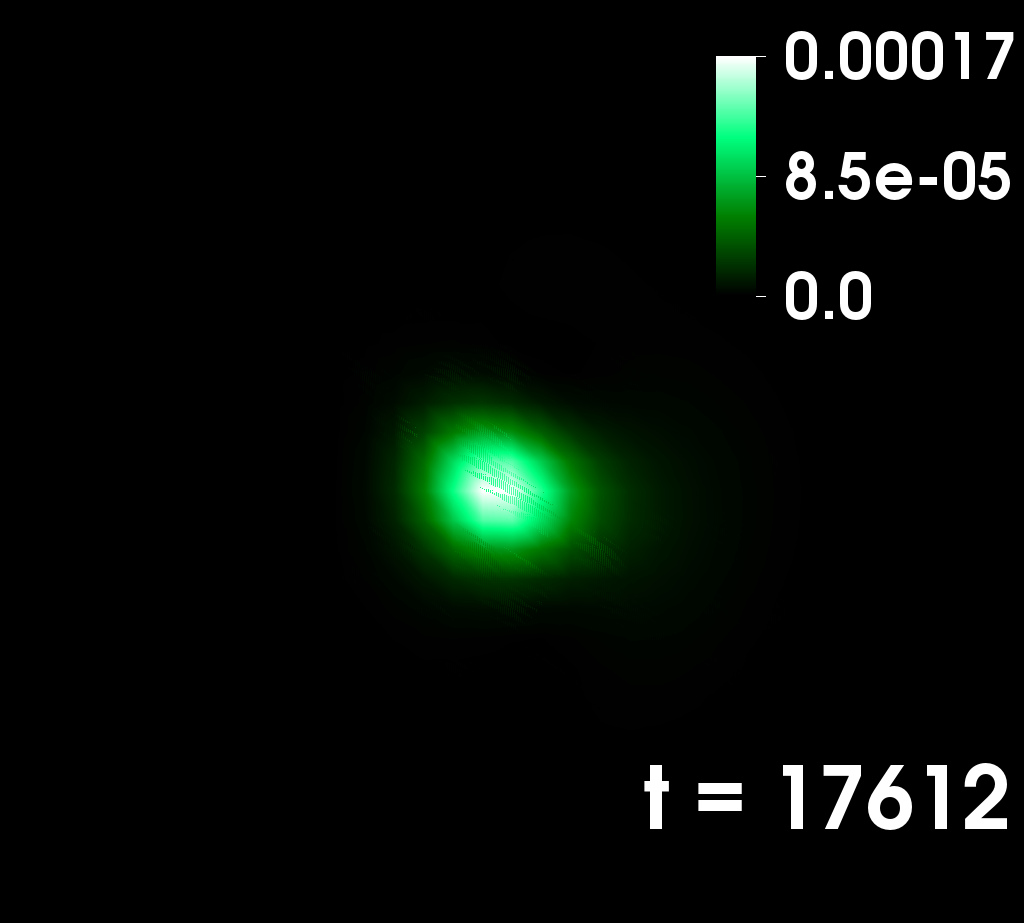} \\
\smallskip
\includegraphics[height=1.0in]{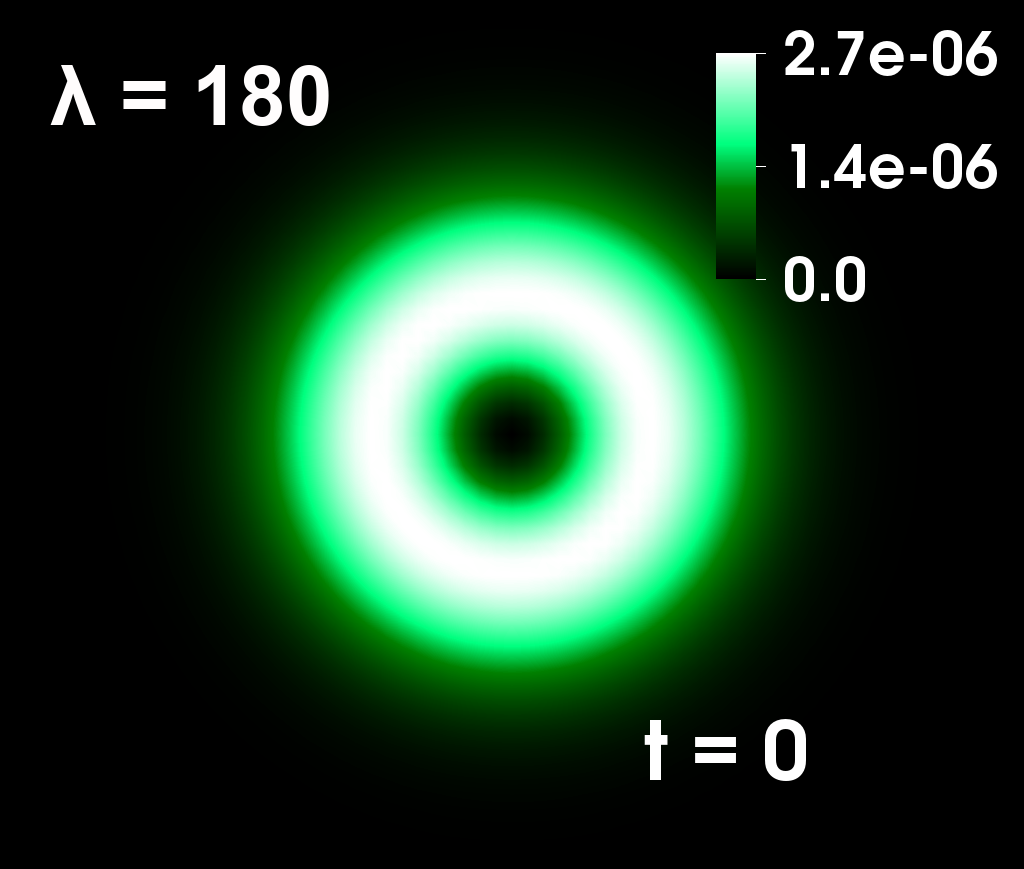} \hspace{0.15cm} \includegraphics[height=1.0in]{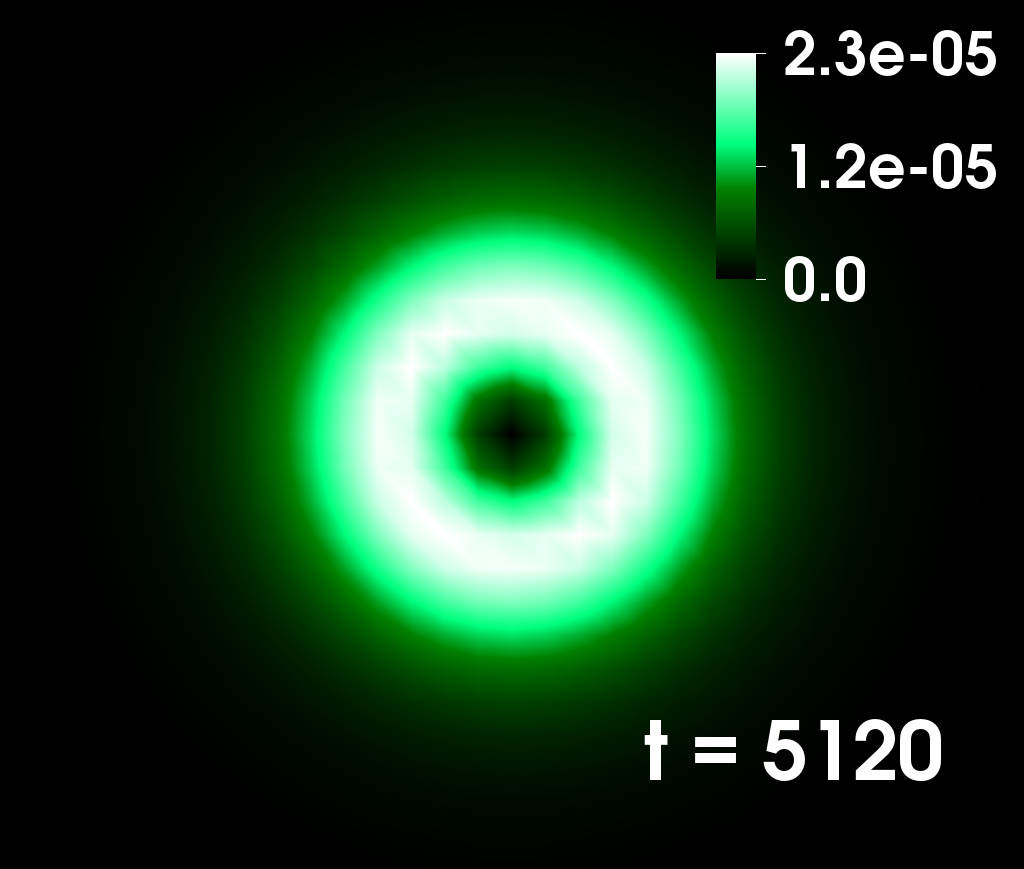}  \includegraphics[height=1.0in]{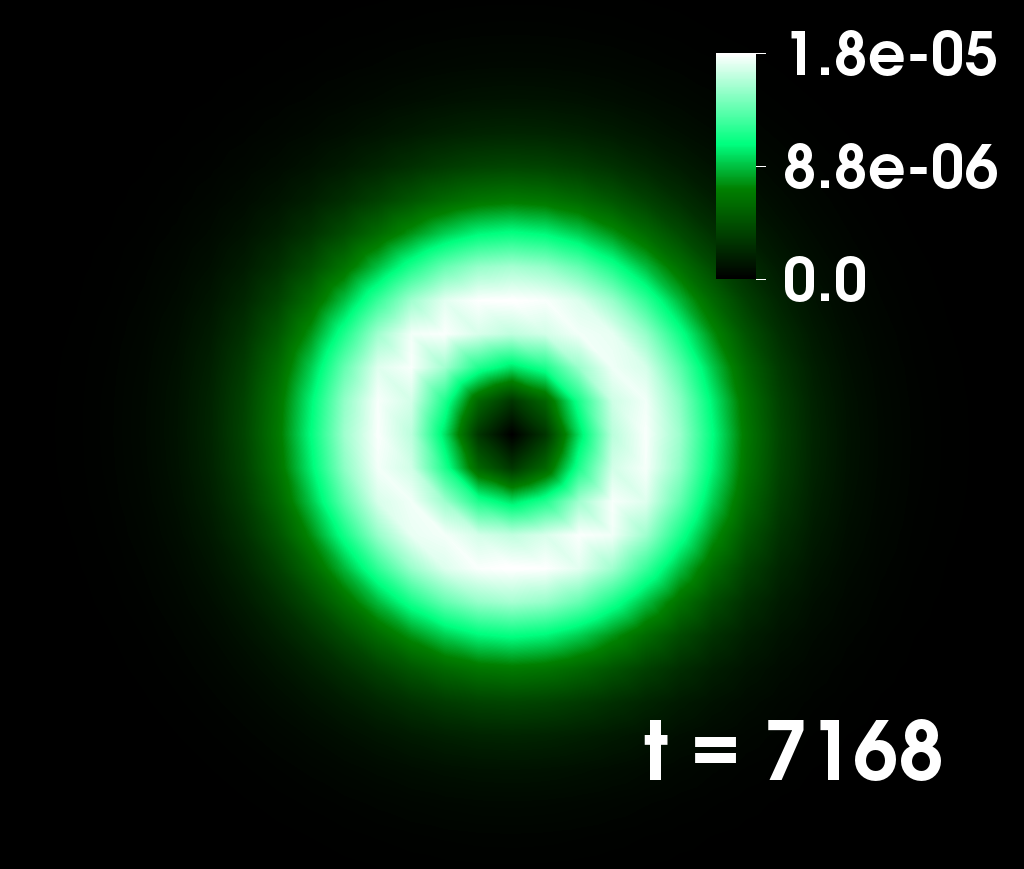} \includegraphics[height=1.0in]{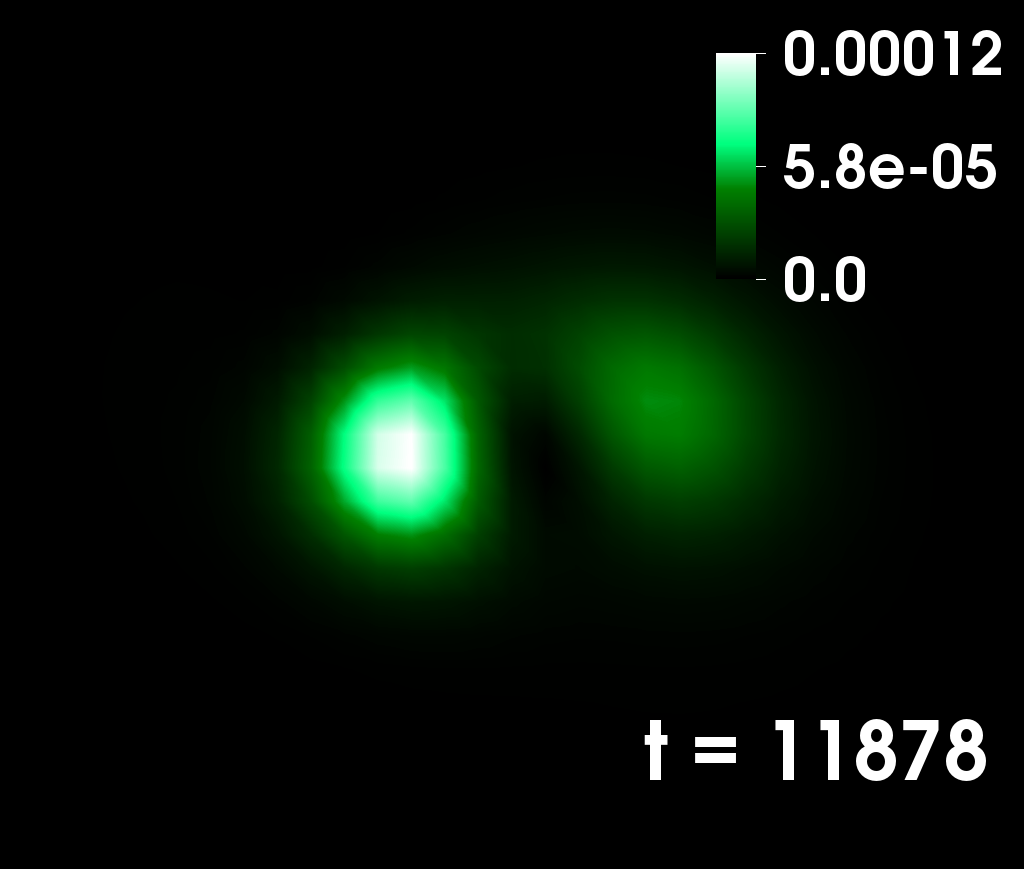} \includegraphics[height=1.0in]{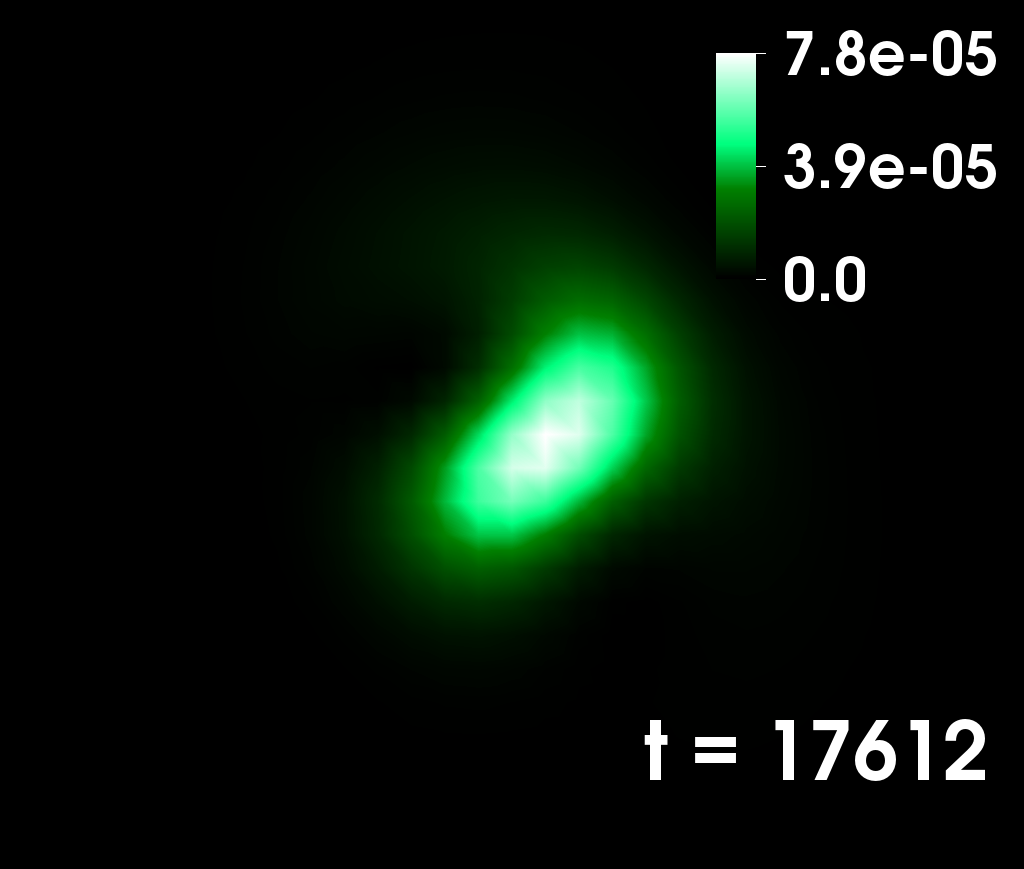} \\
\smallskip
\includegraphics[height=1.0in, width = 1.18in]{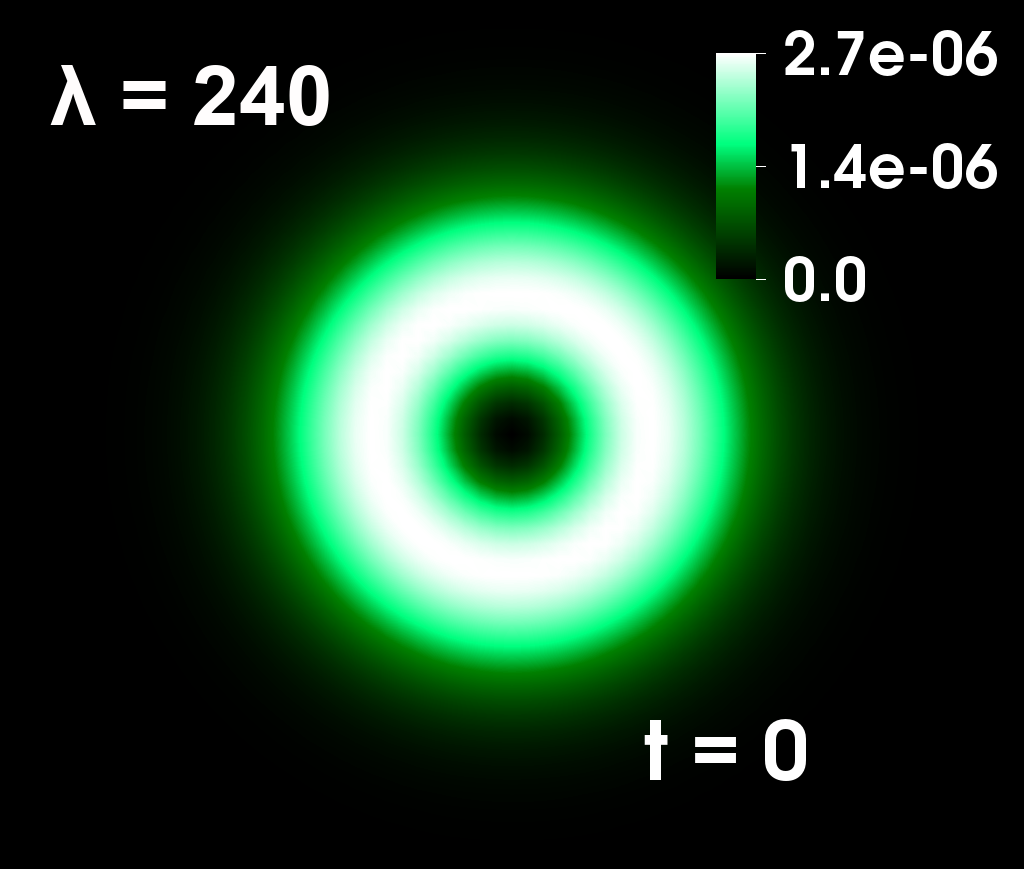}  \hspace{0.15cm} \includegraphics[height=1.0in, width = 1.18in]{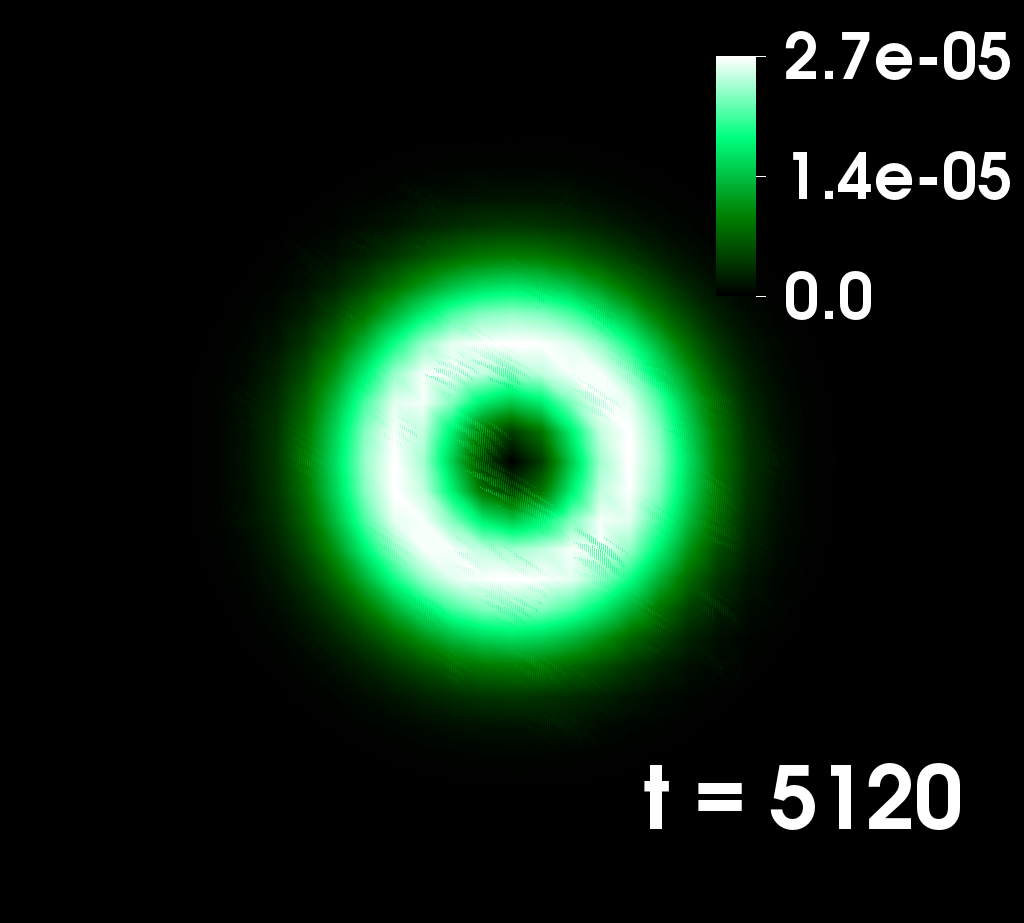}  \includegraphics[height=1.0in, width = 1.18in]{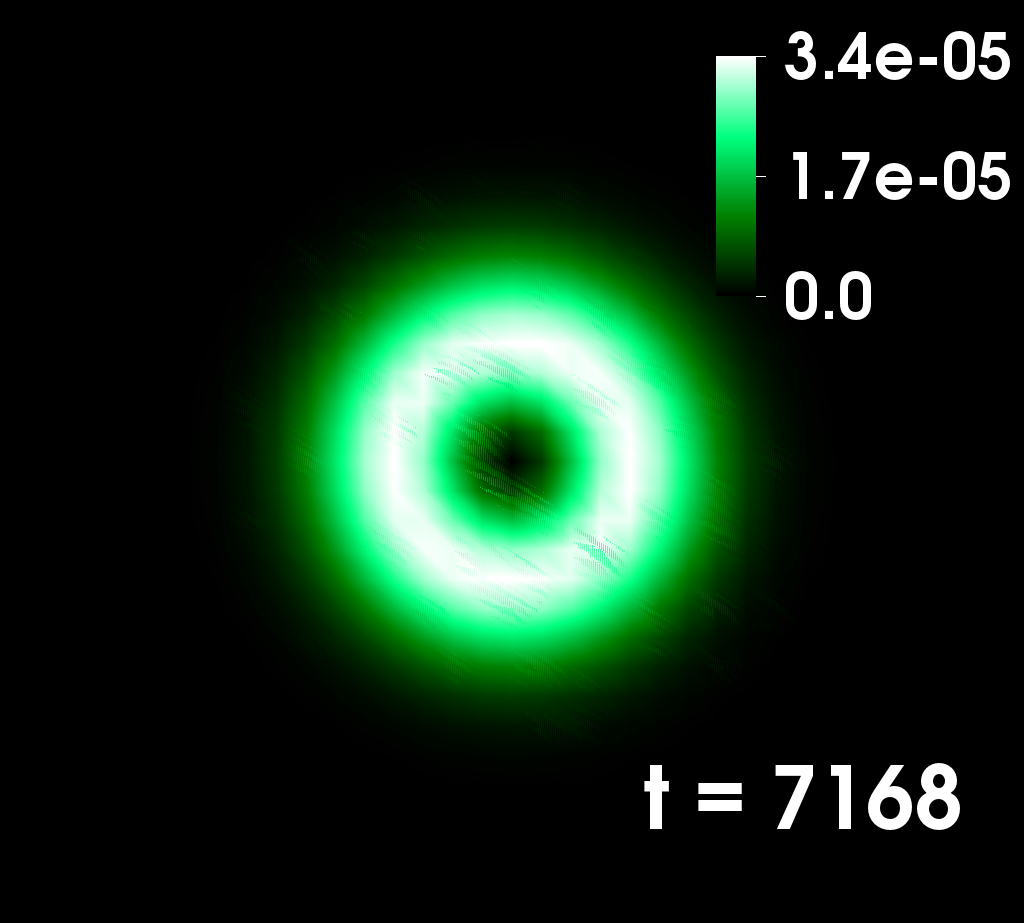} \includegraphics[height=1.0in, width = 1.18in]{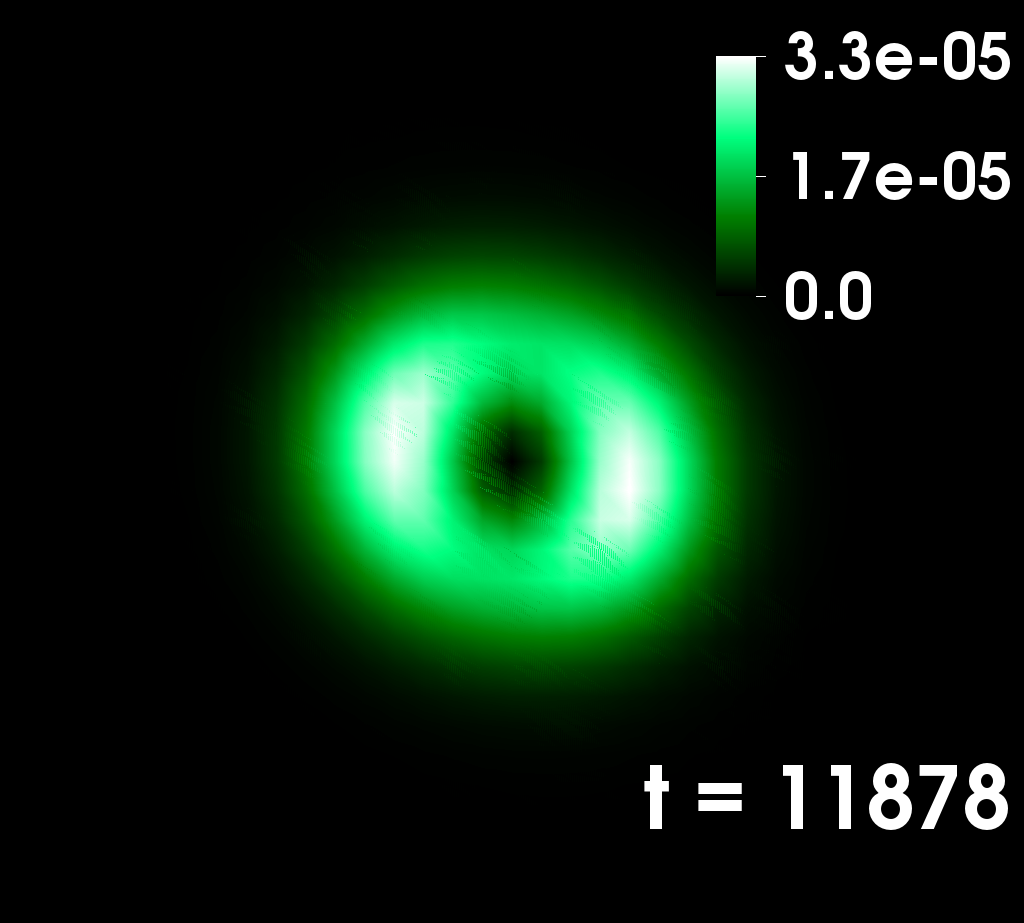} \includegraphics[height=1.0in, width = 1.18in]{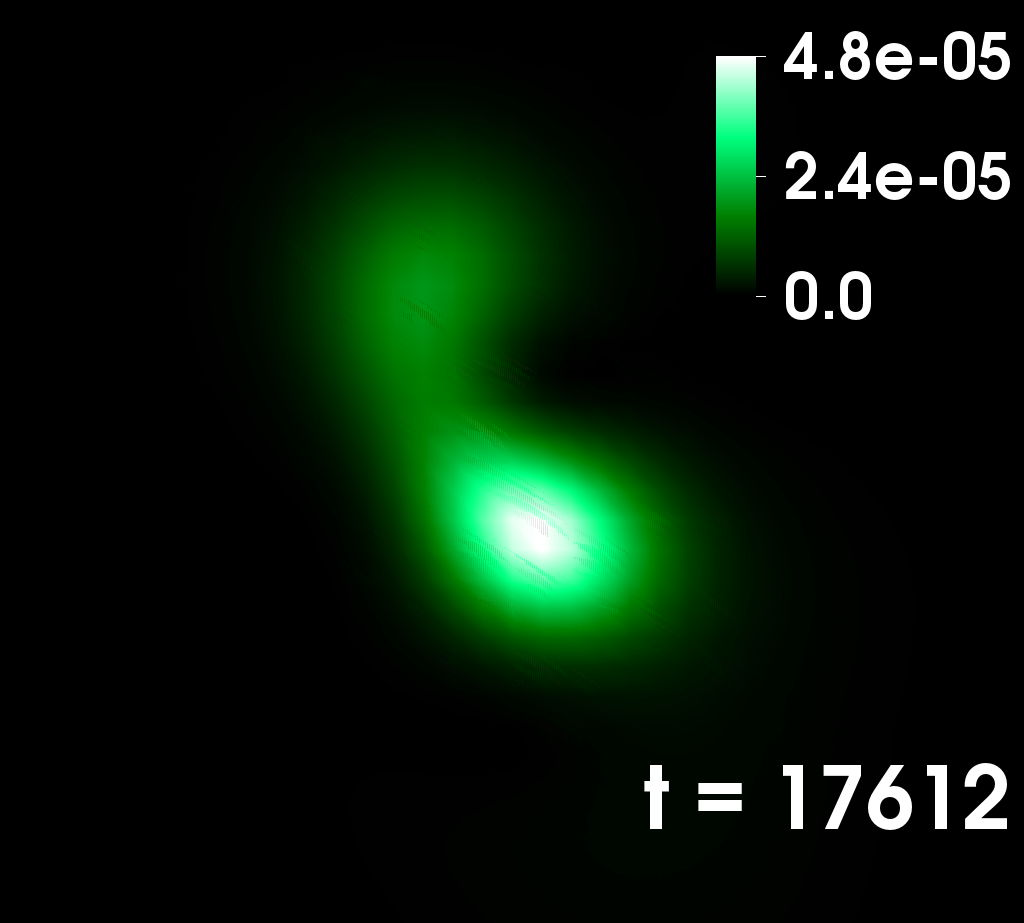} \\
\smallskip
\includegraphics[height=1.0in]{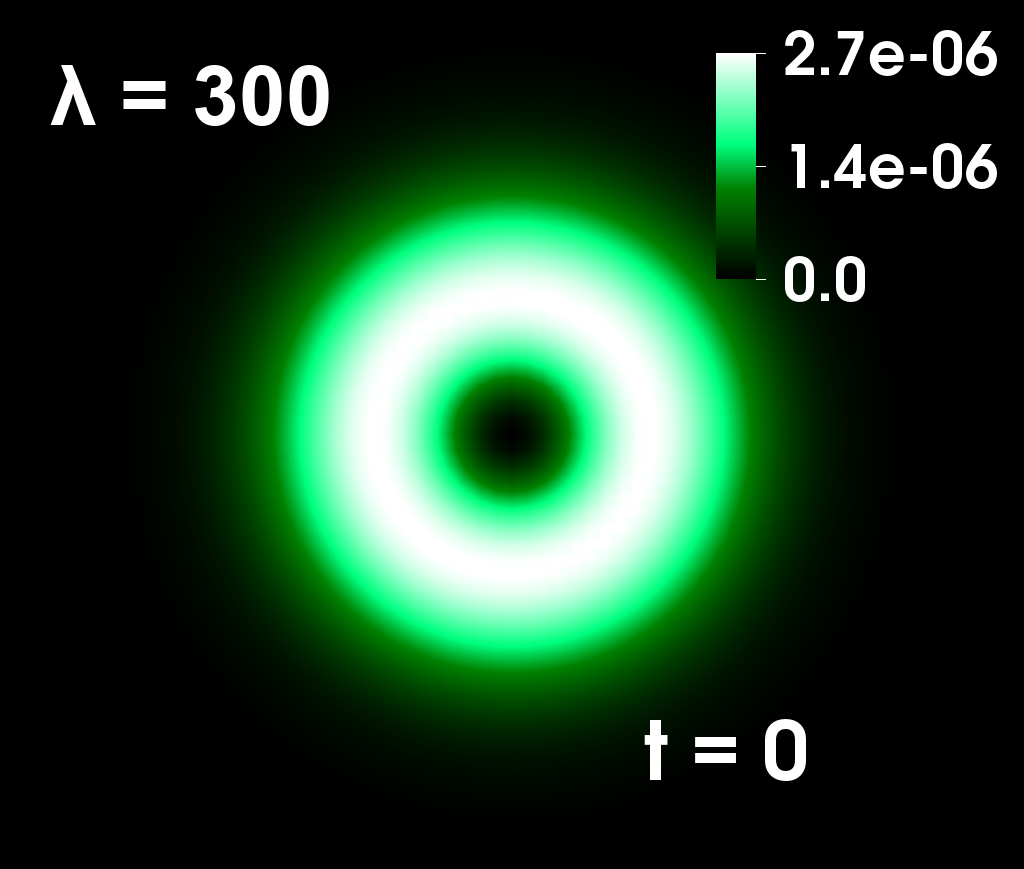}  \hspace{0.15cm} \includegraphics[height=1.0in]{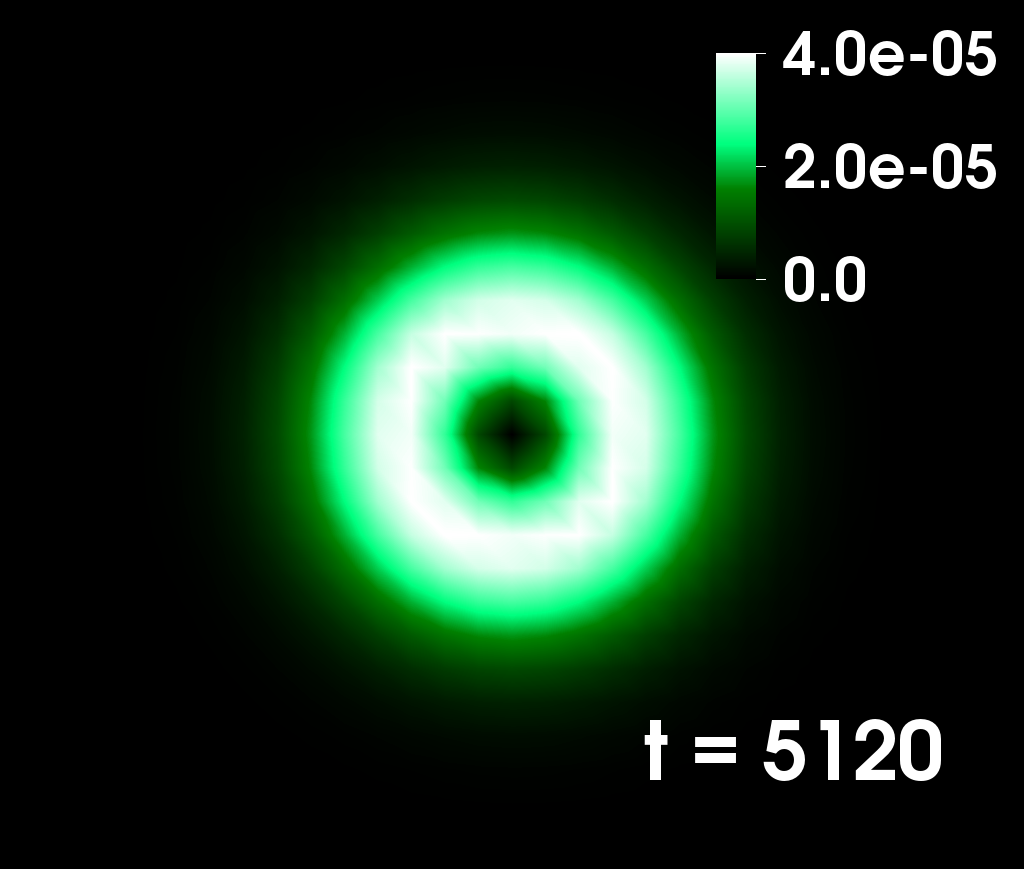}  \includegraphics[height=1.0in]{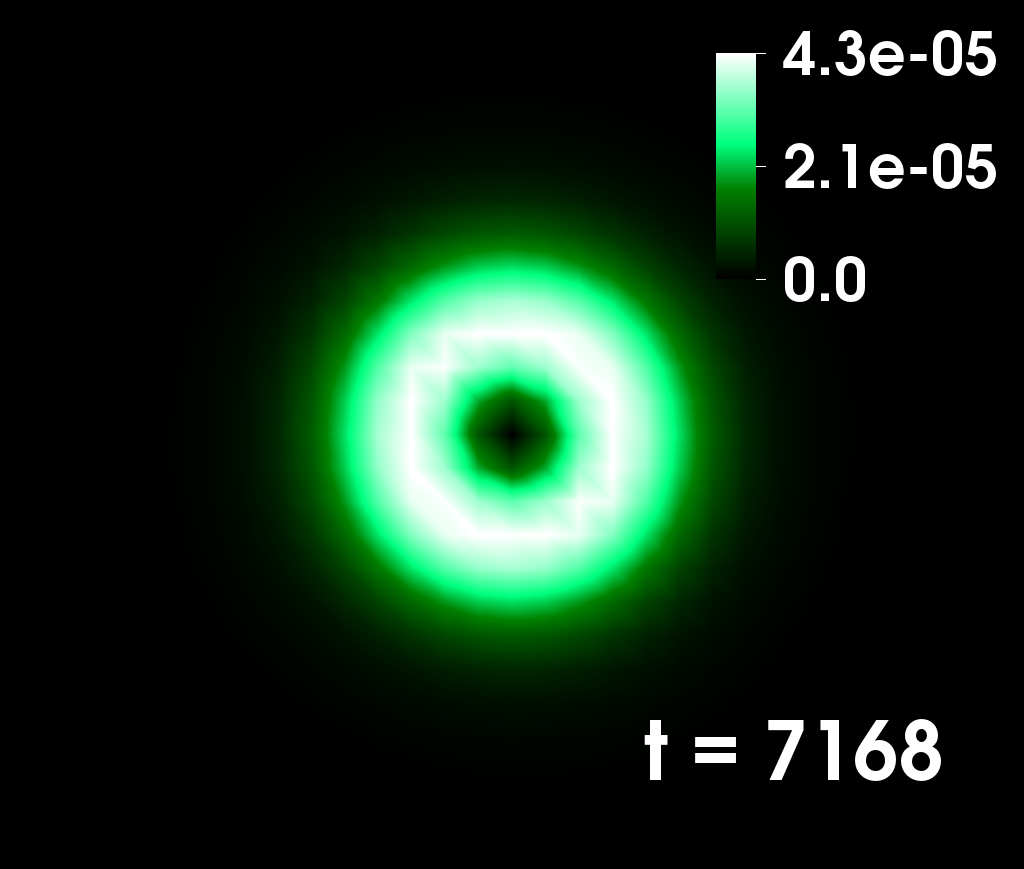}  \includegraphics[height=1.0in]{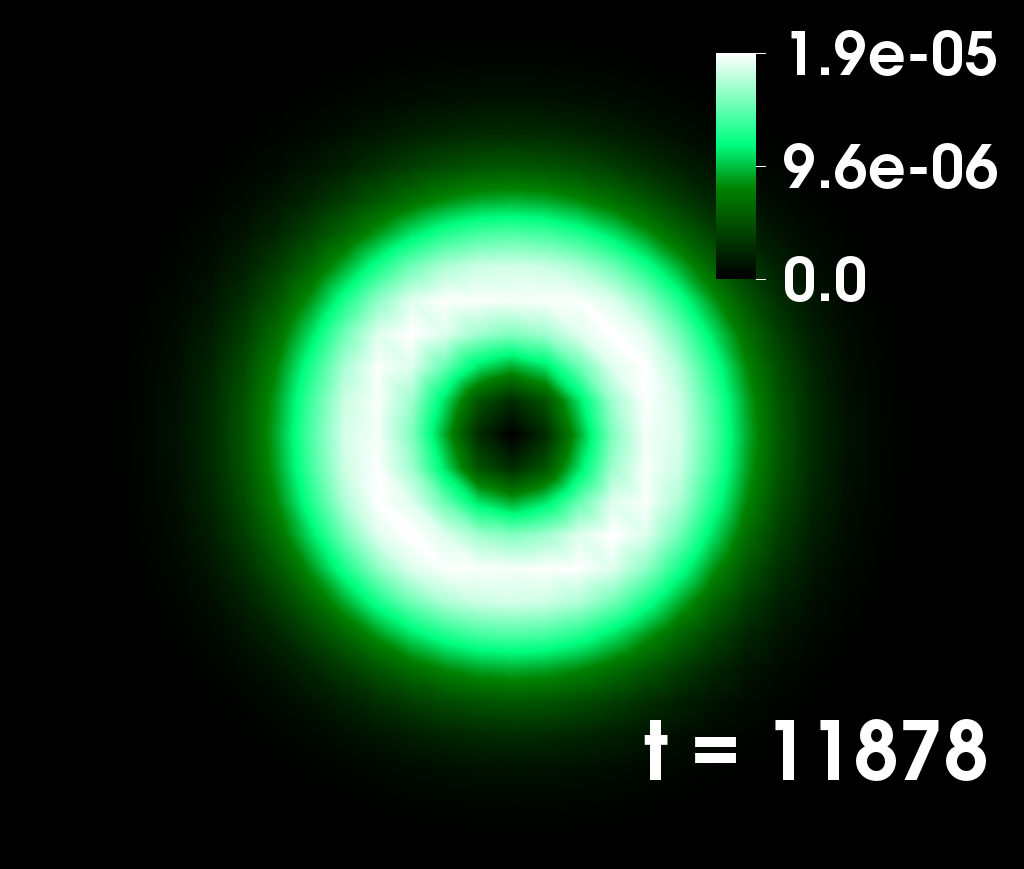} \includegraphics[height=1.0in]{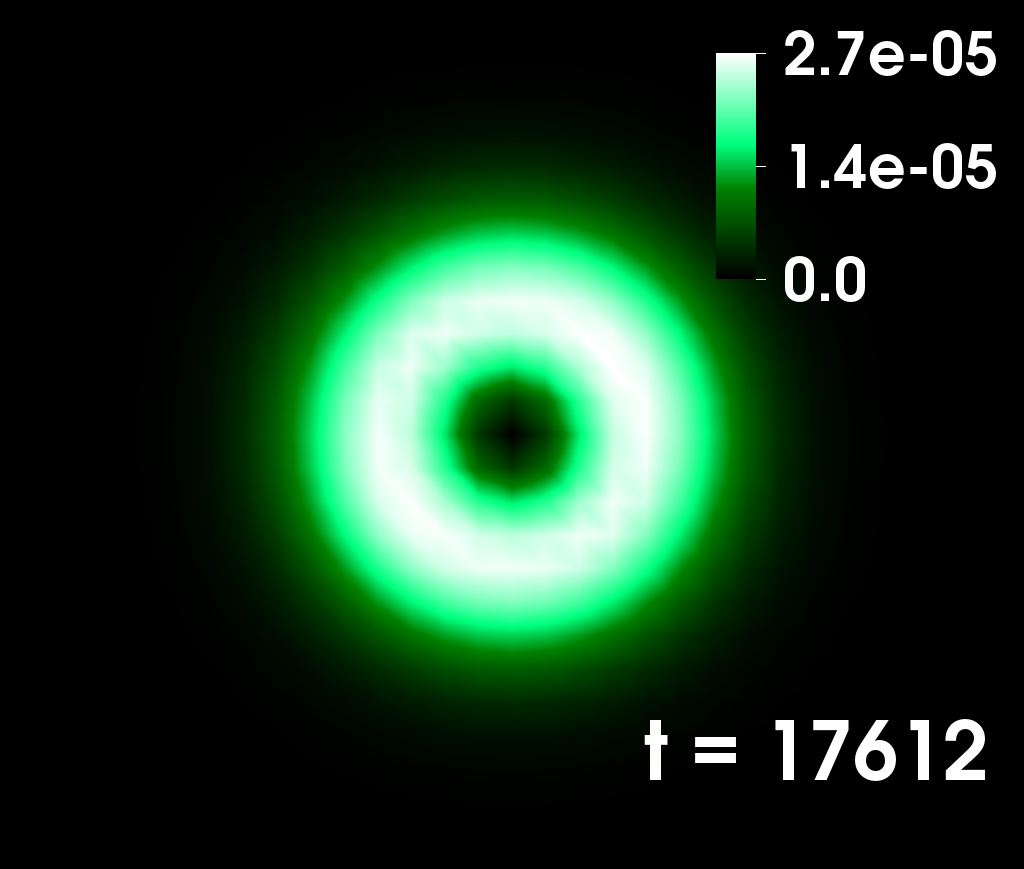} \\
\caption{Snapshots of the energy density at the equatorial plane for model BS2 with $\lambda = \lbrace 60, 120, 180, 240, 300 \rbrace $ (from top to bottom). The vertical axis corresponds to the $y-$direction and the horizontal to the $x-$direction. The spatial domain for the $t=0$ snapshots is $[-80,80]\times[-80,80]$. The subsequent time snapshots are zoomed in the domain $[-40,40]\times[-40,40]$. The time of each snapshot is indicated in the panels.
}
\label{fig1}
\end{figure*}

\section{Results}
\label{results}

\begin{table}[t!] 
\caption{Parameters of the initial models used in this study: S and P refer to scalar or Proca stars, respectively; $\sigma$ is the width of the cloud; $M_{0}$ and $J_{0}$ indicate the initial mass and angular momentum of the cloud. All cases are for $\mu_0=\mu_1=1$.
}
\label{table1}
\begin{tabular}{l|l|l|l|l|l|l}
\hline

Model & Type  & mode & $A_{0/1}$ & $\sigma$ & $M_{0}$ & $J_{0}$ \\
\hline 
BS2 & S & 1 & $16 \times 10^{-5}$  & 40 & 0.88 & 0.89 \\
PS5 & P & 2 & $42 \times 10^{-7}$  & 70 & 1.48 & 2.93 \\
PS6 & P  & 2 & $51 \times 10^{-7}$ & 70 & 2.27 & 4.48 \\

\hline
\end{tabular}
\end{table}

\begin{figure}[]
\centering
\includegraphics[scale = 0.30]{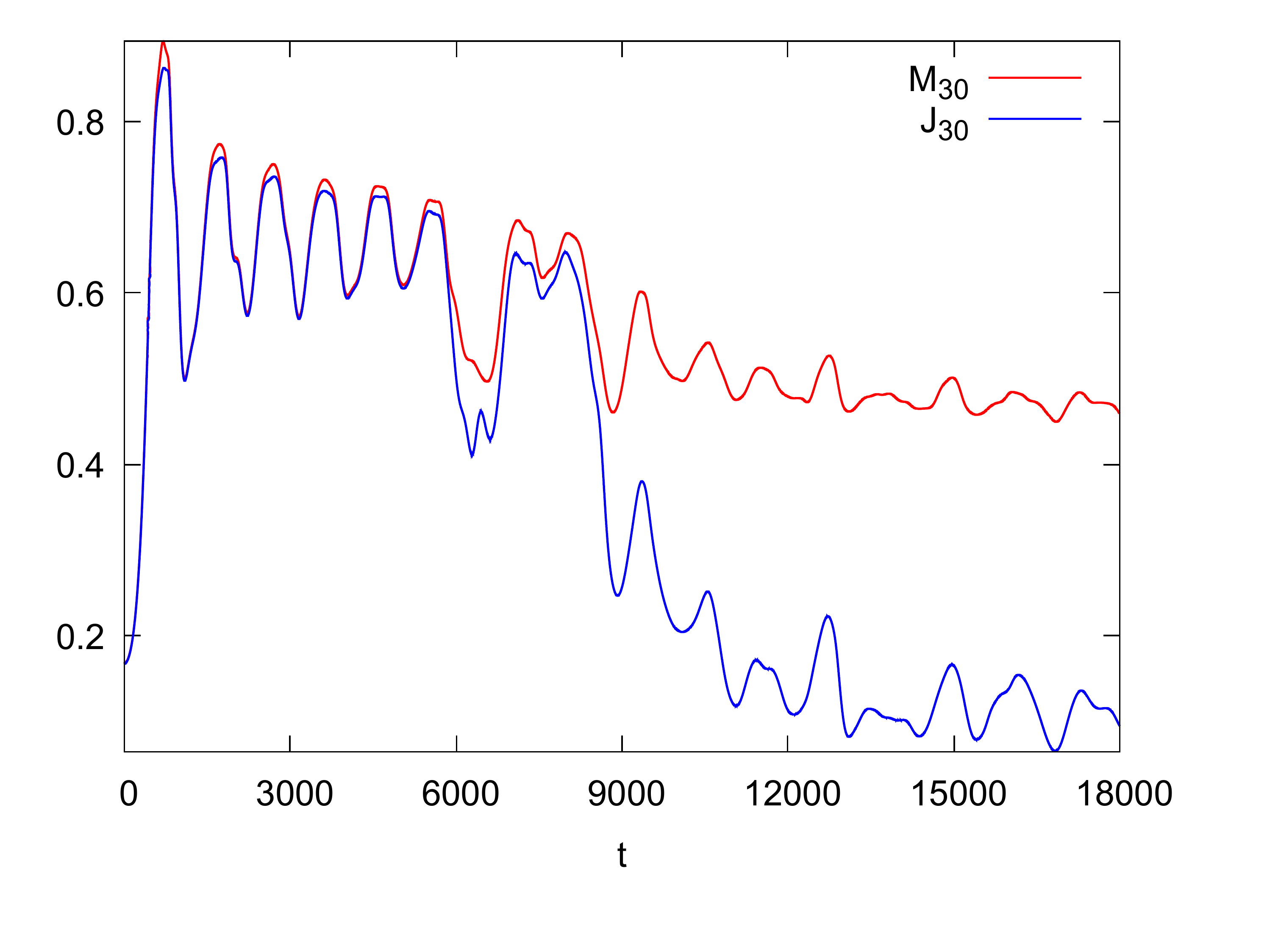}
\caption{Evolution of the mass and angular momentum contained in a sphere of radius 30 for the model BS2 with $\lambda=60$.}
\label{fig2}
\end{figure}

The numerical simulations start with a Gaussian cloud of bosonic matter, built as described in Section~\ref{sec3}, which then collapses  due to its own gravity. If enough energy is radiated away during this highly dynamical process through the mechanism of gravitational cooling~\cite{seidel1994formation}, a compact bosonic star will form. As we already showed in~\cite{sanchis:2019dynamics} the presence of rotation may trigger the appearance of instabilities in the newly formed spinning compact object. More precisely, we found that scalar boson stars are affected by a non-axisymmetric instability which triggers the loss of angular momentum and the reshaping of the energy density profile from a toroidal shape into a spheroidal one. This behaviour was not observed for $m=1$ spinning Proca stars, which have a spheroidal shape. As a consequence, we conjectured that this morphological difference was related to the dissimilar  stability properties of these objects. 

Here, we investigate if a self-interaction potential in the Klein-Gordon equation can quench the instability found in the scalar case and if our hypothesis that relates the instability to the toroidal shape of the energy density still holds when considering $m=2$ spinning Proca stars, which have a toroidal shape. Table~\ref{table1} summarizes the different models we consider to describe the initial cloud of bosonic matter. We use the same initial model BS2 in the scalar case as in our previous paper,  building initial data for five  different values of the self-interaction parameter, namely $\lambda=\lbrace 60, 120, 180, 240, 300 \rbrace $.

\subsection{Boson stars with self-interaction}

\begin{figure*}[]
\centering
\includegraphics[height=1.1in]{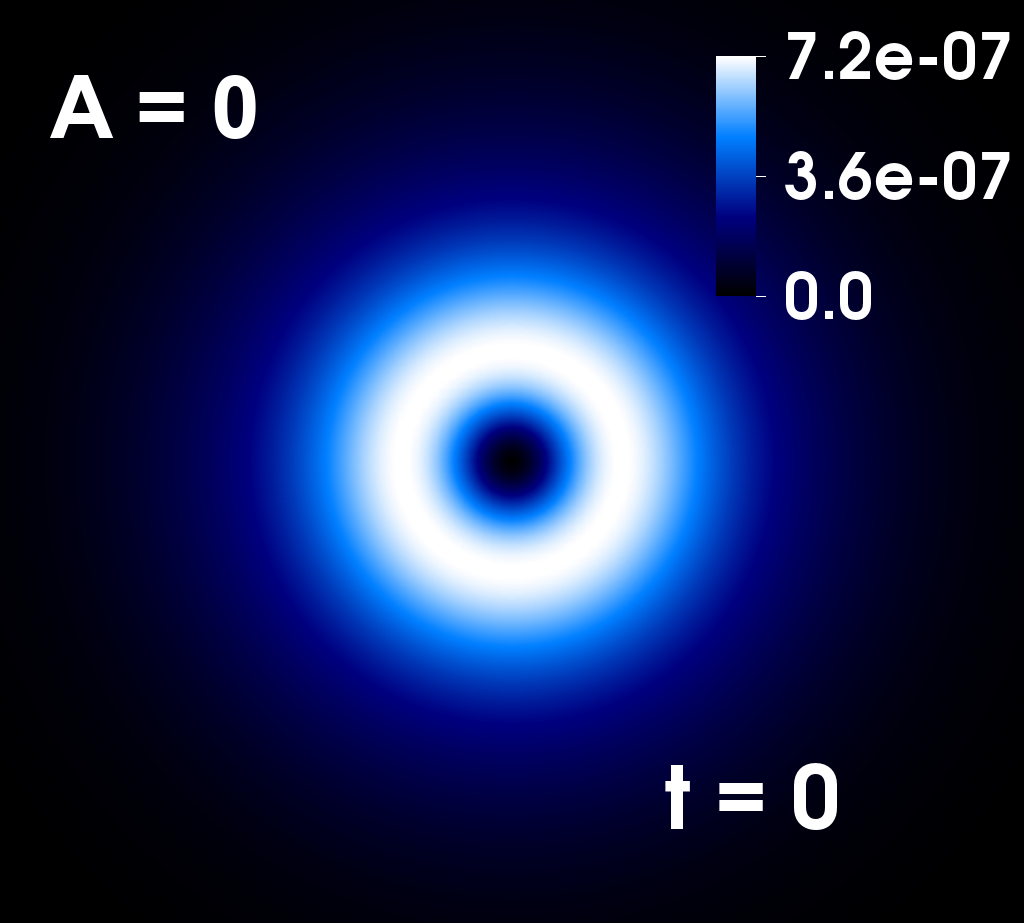} \hspace{0.15cm} \includegraphics[height=1.1in]{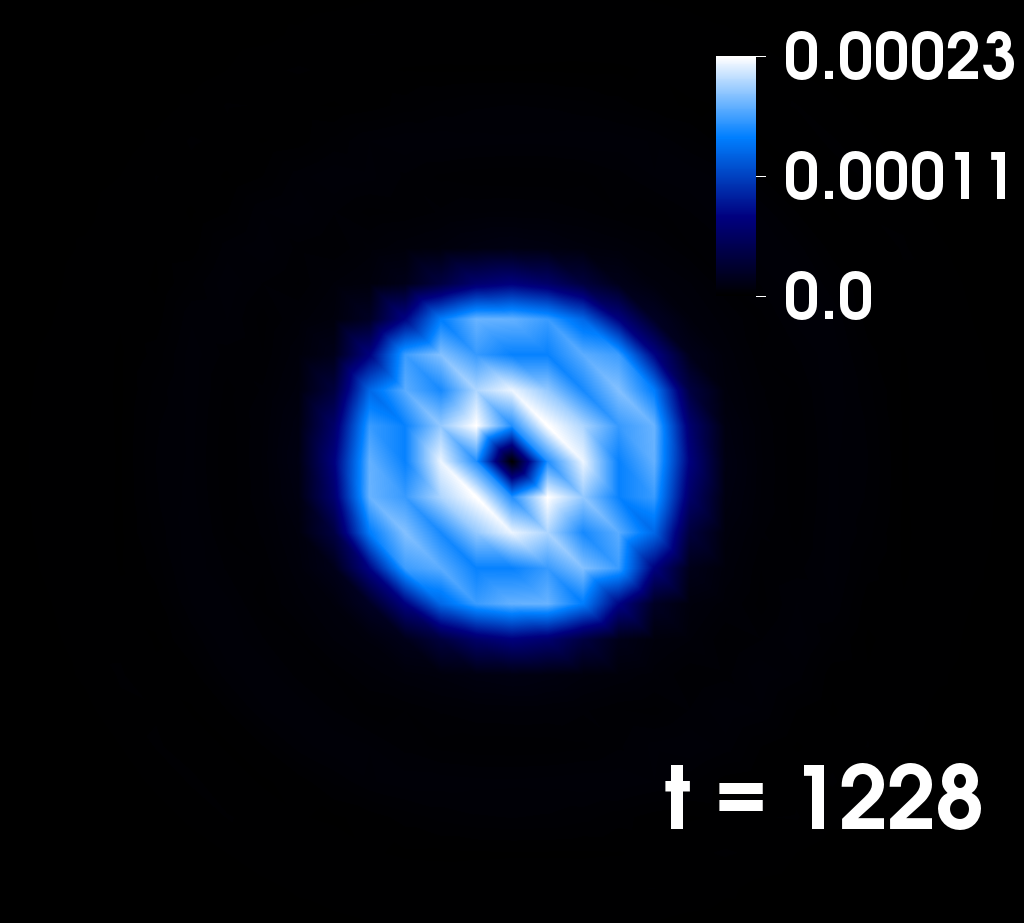} \includegraphics[height=1.1in]{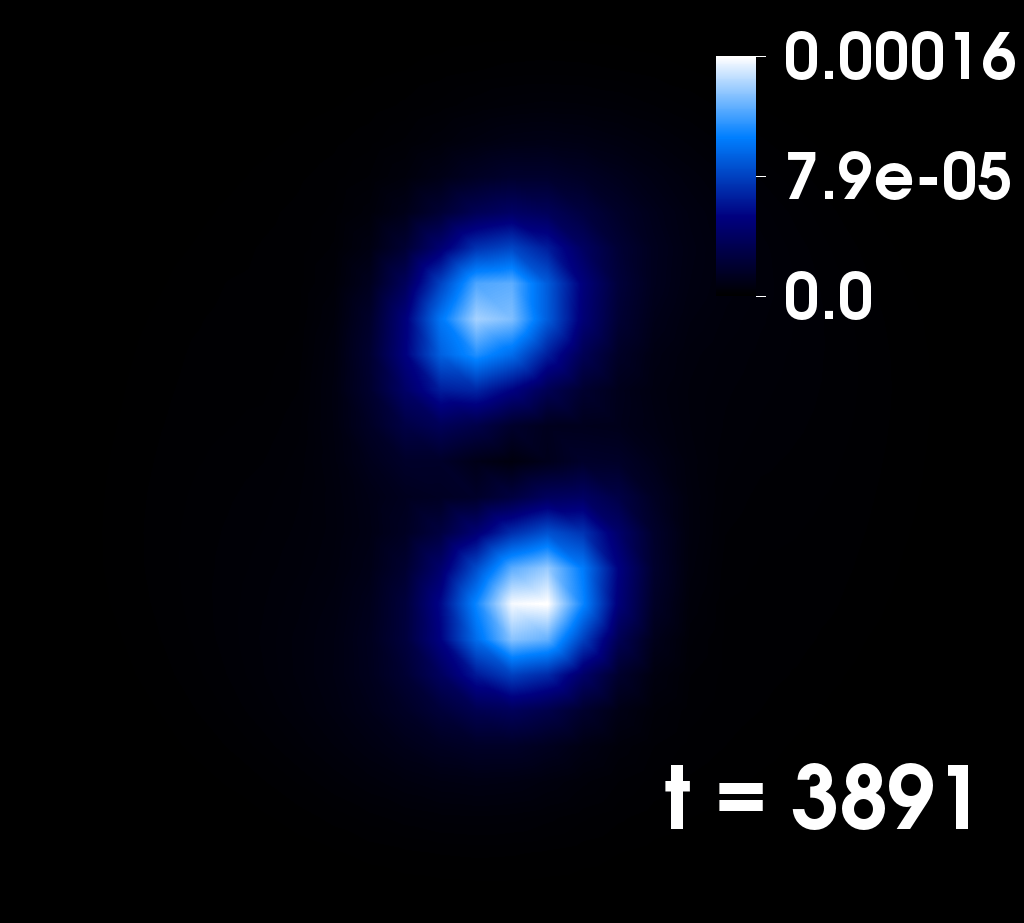} \includegraphics[height=1.1in]{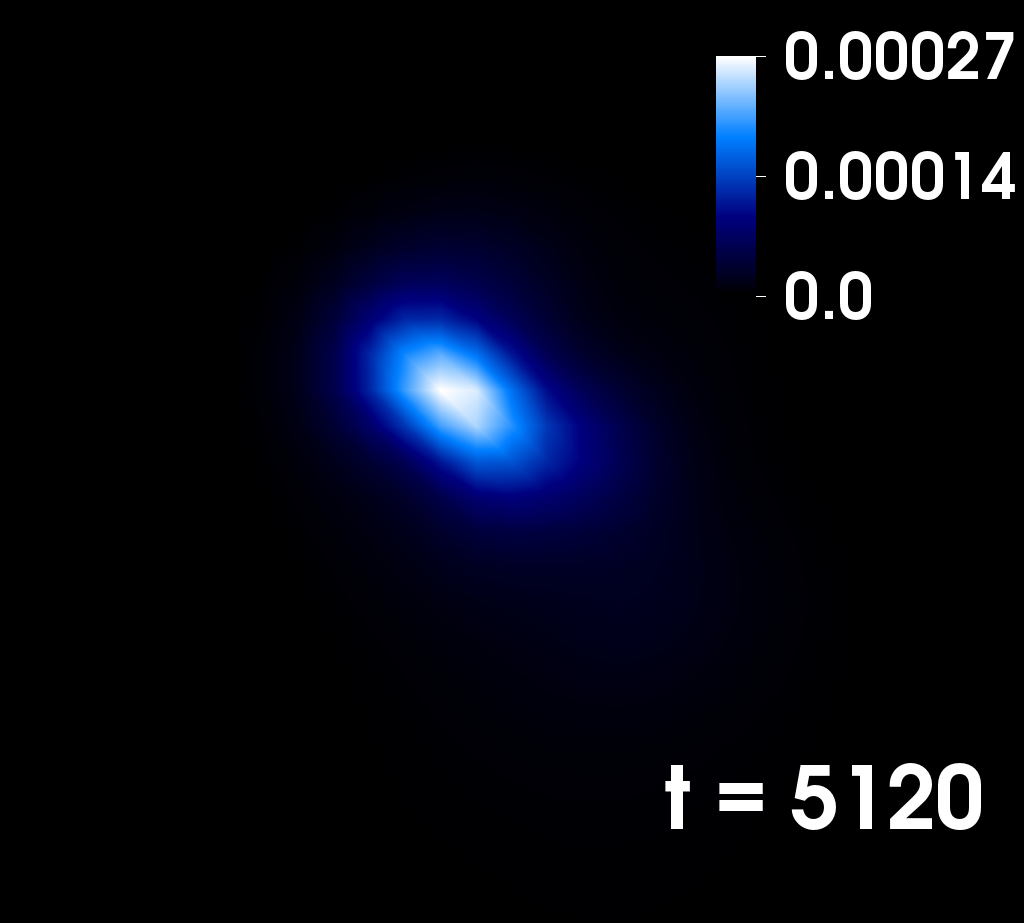} \includegraphics[height=1.1in]{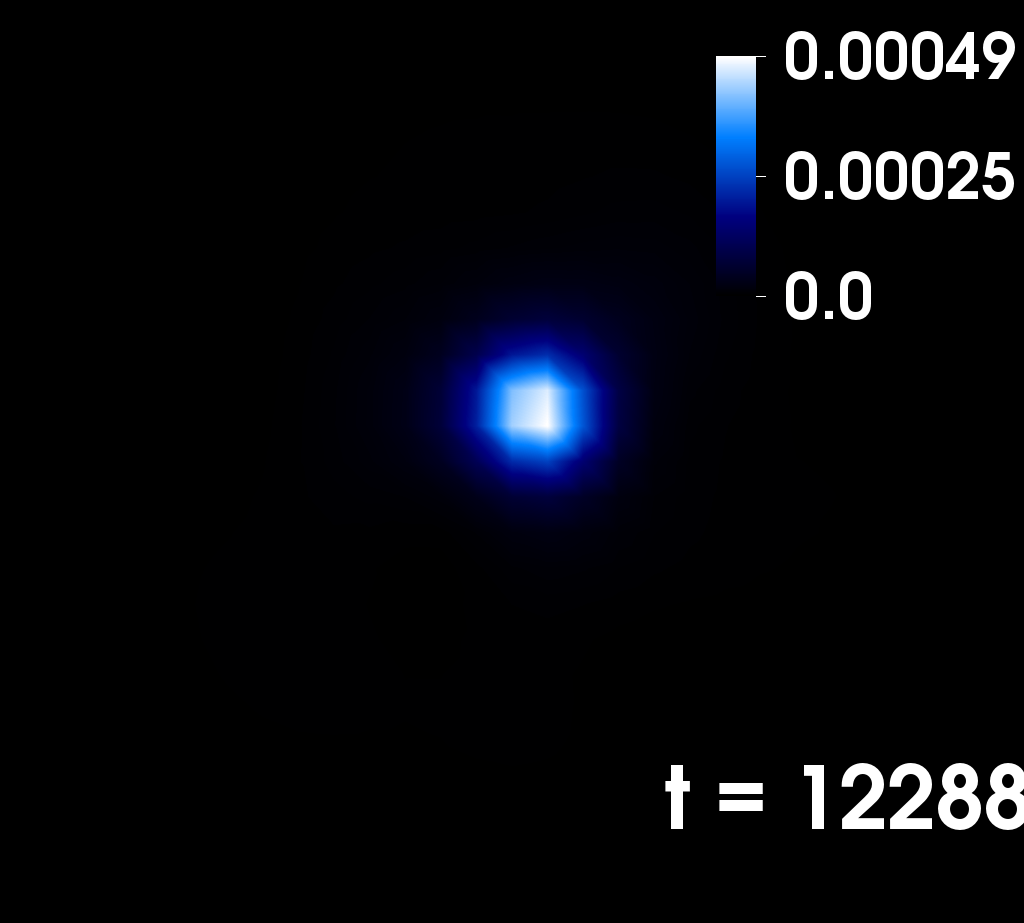}\\
\smallskip
\includegraphics[height=1.1in]{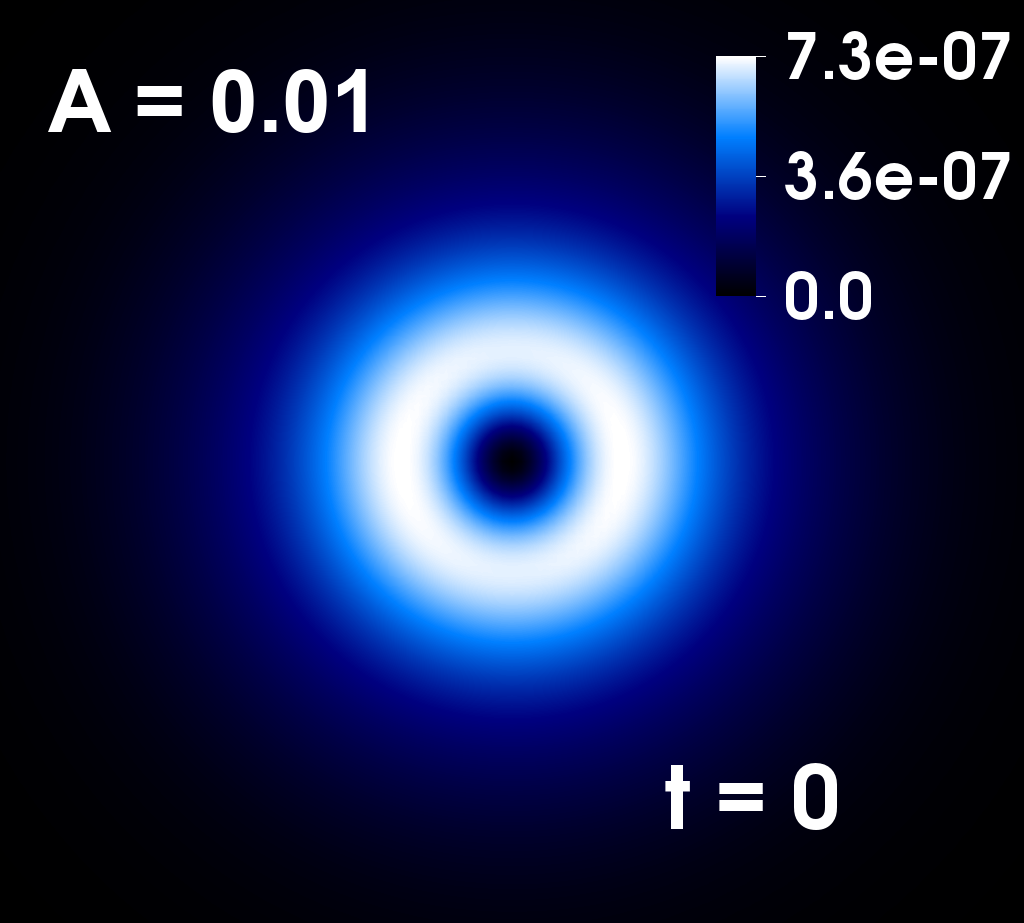} \hspace{0.15cm} \includegraphics[height=1.1in]{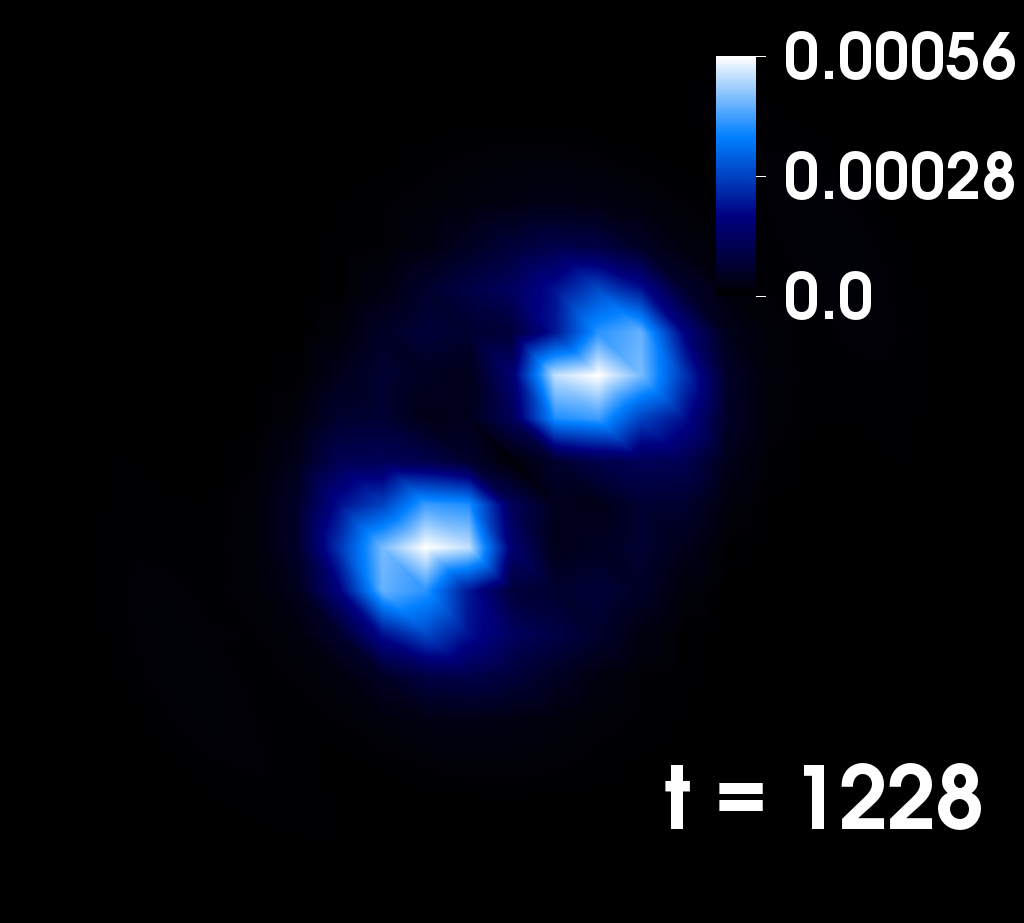} \includegraphics[height=1.1in]{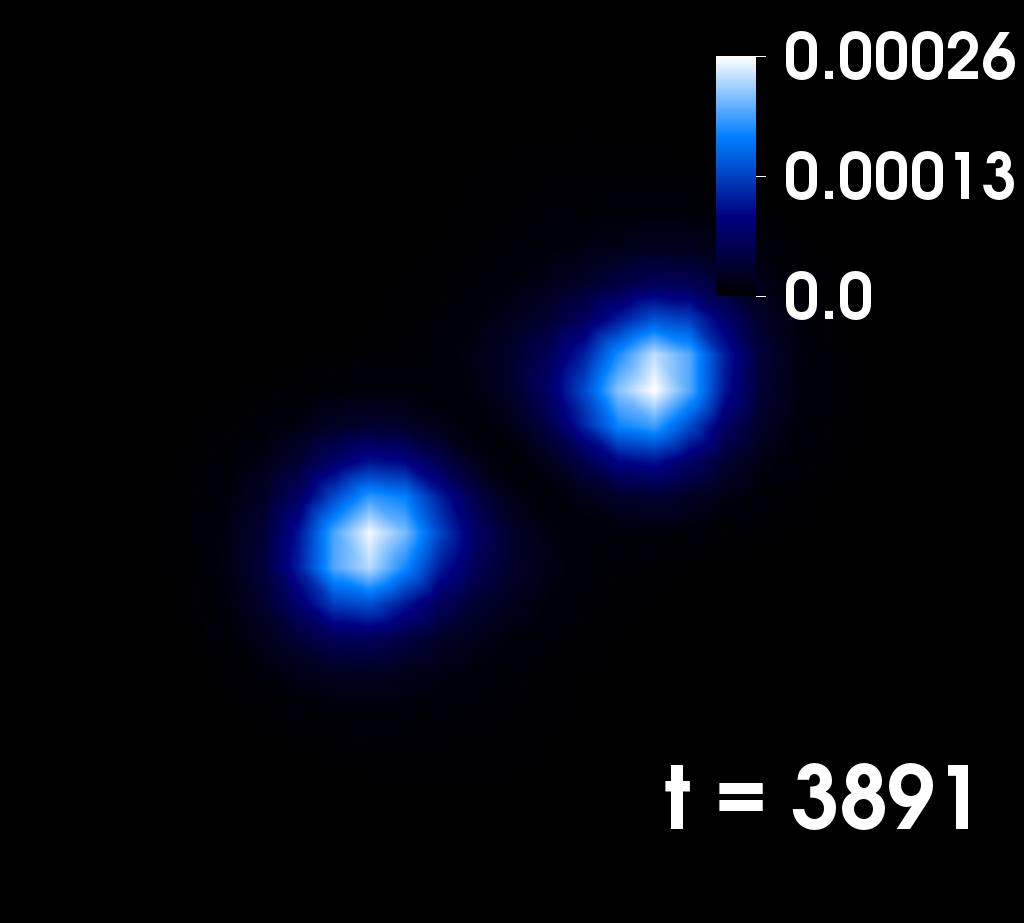} \includegraphics[height=1.1in]{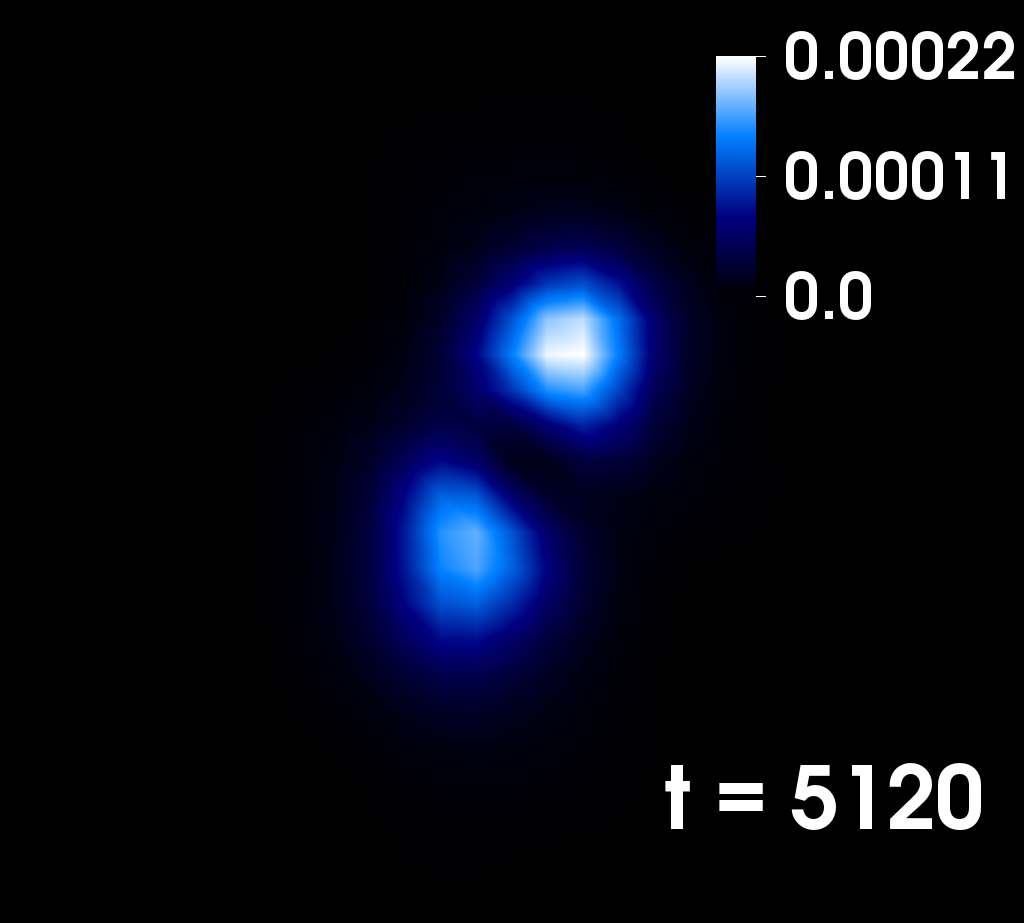} \includegraphics[height=1.1in]{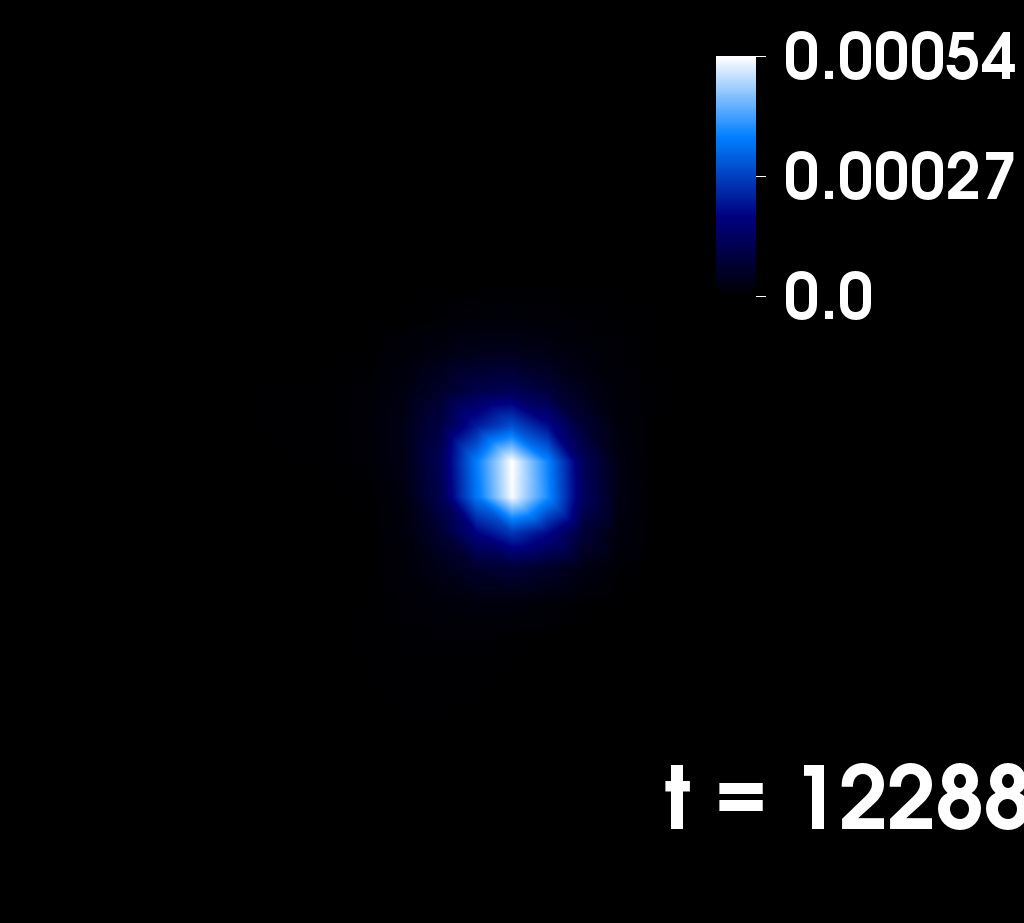} \\
\smallskip
\includegraphics[height=1.1in]{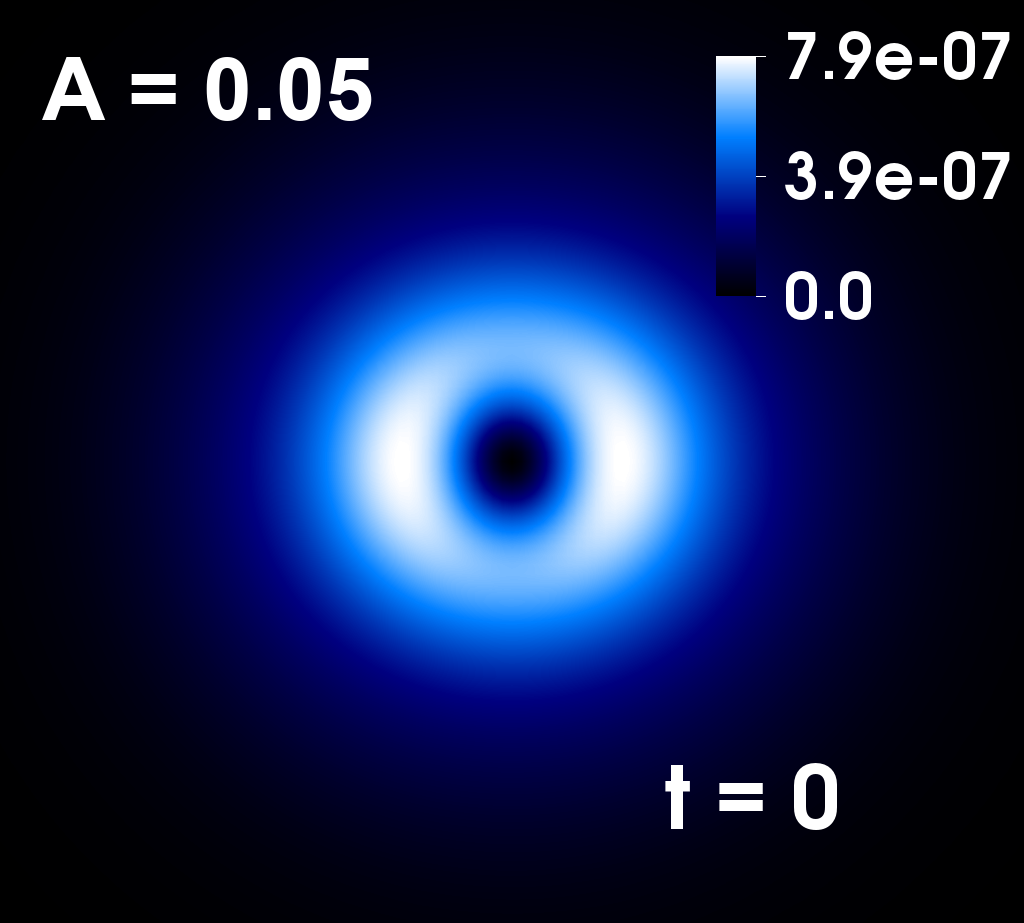} \hspace{0.15cm} \includegraphics[height=1.1in]{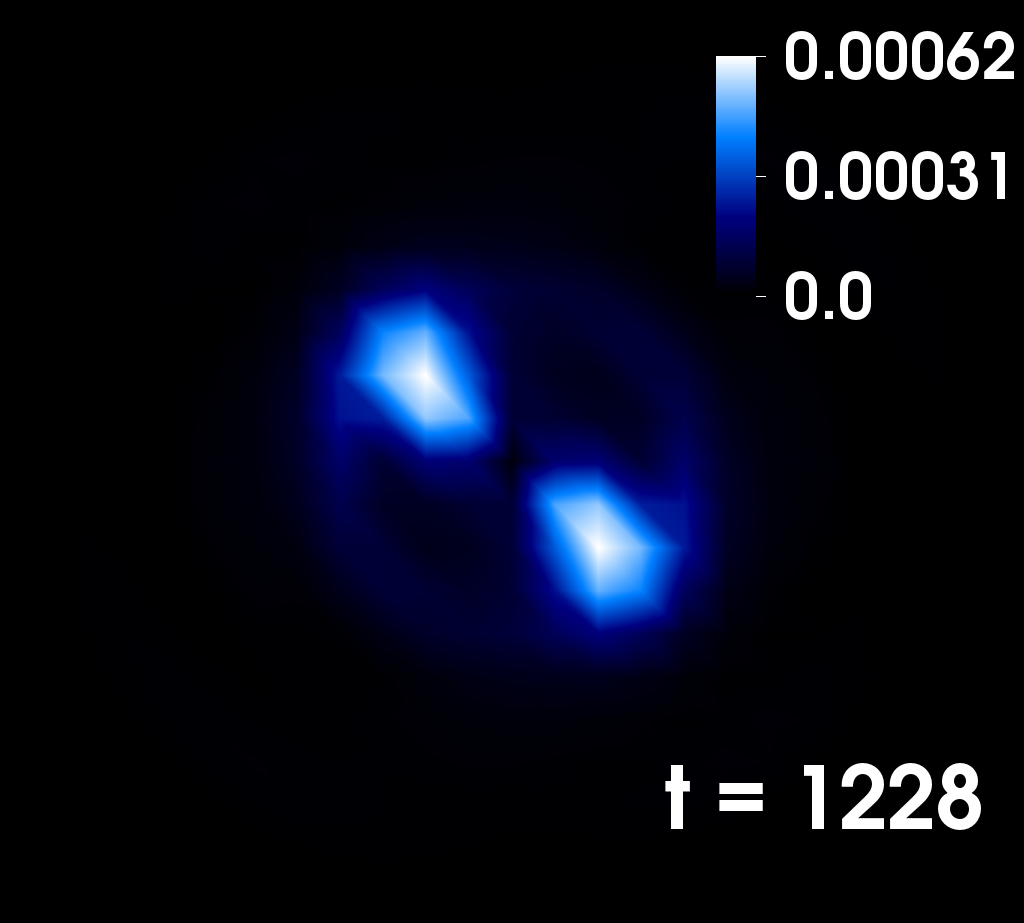} \includegraphics[height=1.1in]{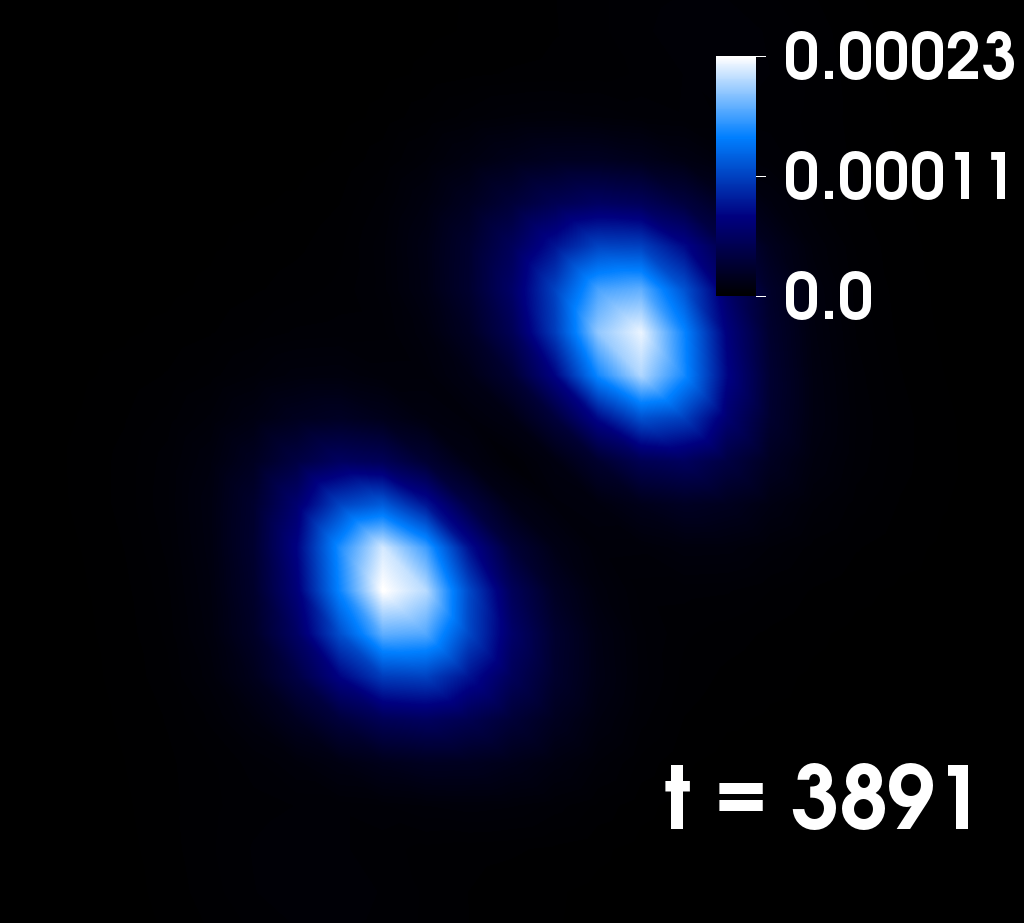} \includegraphics[height=1.1in]{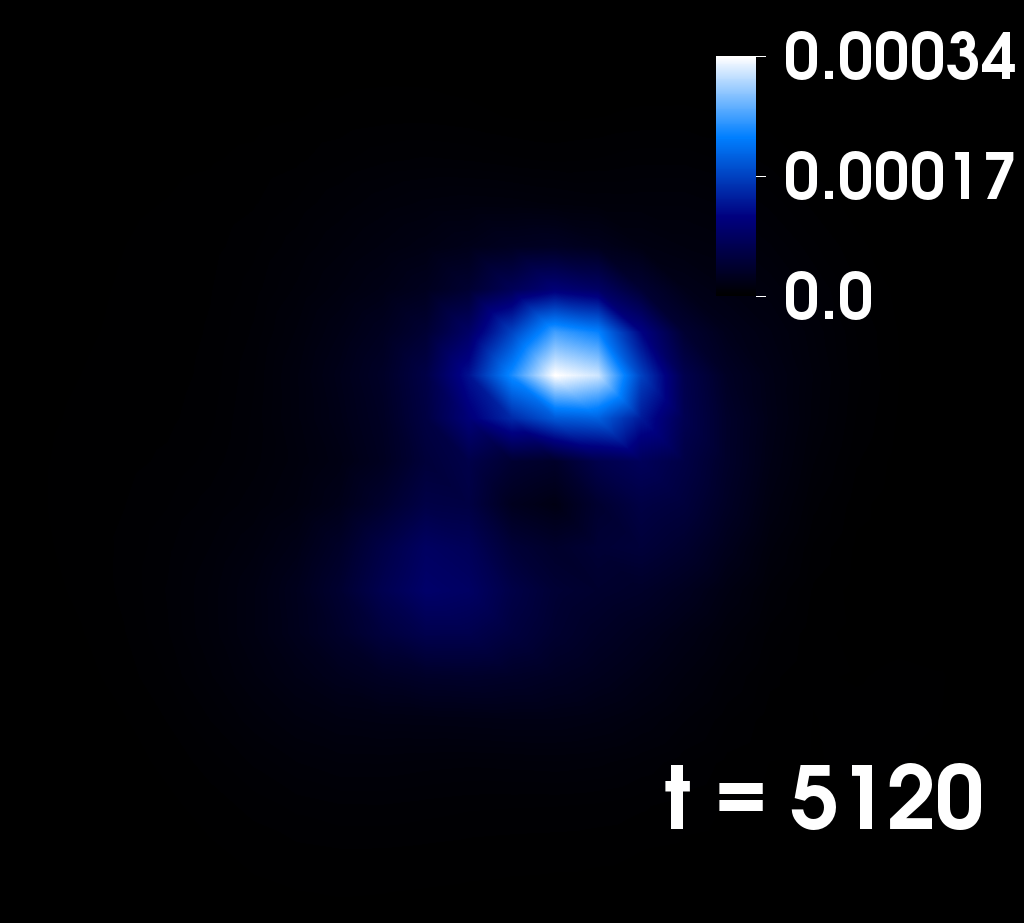} \includegraphics[height=1.1in]{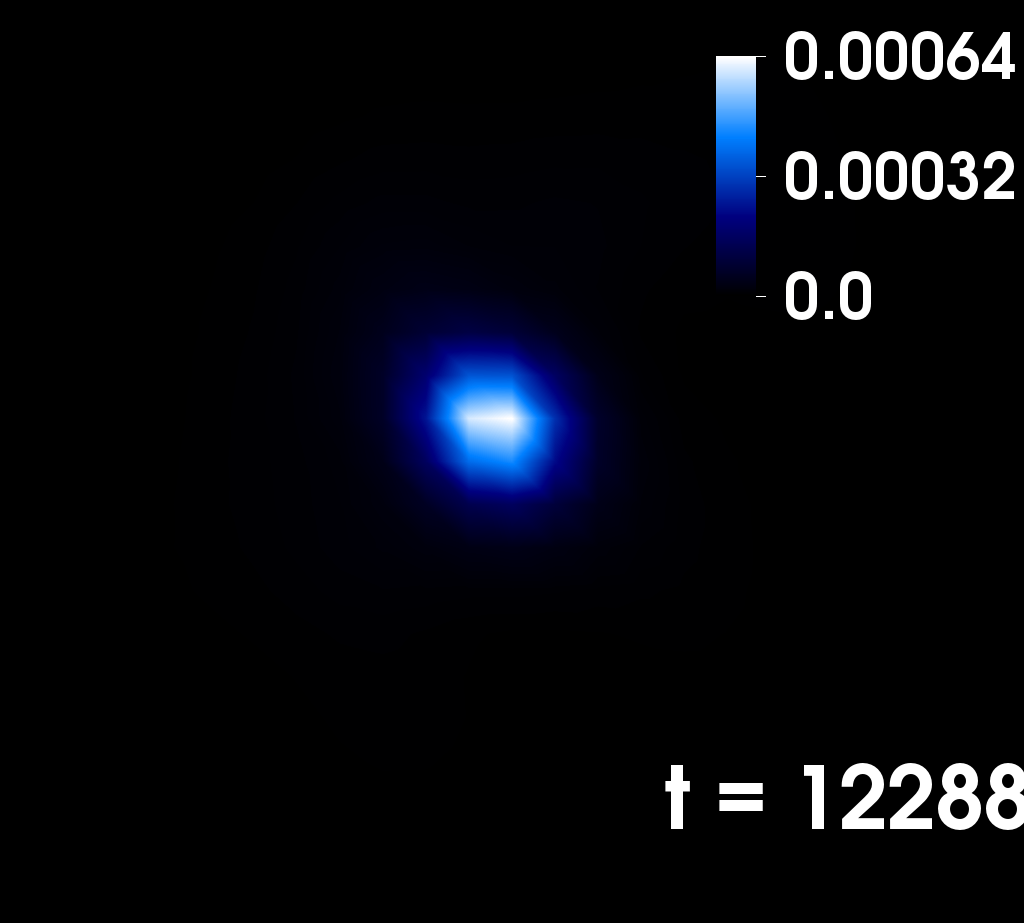} \\
\caption{Snapshots of the energy density in the equatorial plane for model PS6  and for three different perturbation parameters, $A=0, 0.01$ and $0.05$ (from top to bottom). The vertical axis correspond to the $y$ direction and the horizontal to the $x$ direction. Times are indicated in the legends. The spatial domain for the $t=0$ snapshots is $[-160,160]\times[-160,160]$. The subsequent time snapshots are zoomed in the domain $[-30,30]\times[-30,30]$. The time of each snapshot is indicated in the panels.
}
\label{fig3}
\end{figure*}

Figure~\ref{fig1} shows snapshots of the energy density distribution at the equatorial plane for model BS2 for evolutions with the five different values of $\lambda$. In all cases we observe that the final compact object is always affected by a non-axisymmetric instability whose time of appearance depends on the value of the self-interaction parameter. The instability makes the energy density distribution to break into two pieces which subsequently recombine into a spheroidal, smaller piece, while the angular momentum is ejected from the region where the compact object forms. By increasing the contribution of the self-interaction term it is possible to delay this occurrence. For the case $\lambda=300$ (final row) the instability is not visible simply because the evolution of the model is not sufficiently long. 

Figure~\ref{fig2} depicts the evolution of the mass $M_{30}$ and angular momentum $J_{30}$ of the bosonic matter enclosed inside a sphere of radius $r=30$ for the model with $\lambda=60$.  These two quantities are evaluated by means of the following integrals
\begin{align}
M_{r^*} &= - 2 \int_{0}^{r^*} dr \int_{0}^{\frac{\pi}{2}} d\theta \int_{0}^{2\pi} d\varphi \, (2 T_{t}^{t} - T_{\alpha}^{\alpha}) \, \sqrt{-g}\,, \\
J_{r^*} &= 2 \int_{0}^{r^*} dr \int_{0}^{\frac{\pi}{2}} d\theta \int_{0}^{2\pi} d\varphi \, T_{t}^{\varphi} \, \sqrt{-g}\,,
\end{align}
where we take into account the reflection symmetry with respect to the equatorial plane we enforce in our numerical simulations. 

We will use the notation $M$ and $J$ to refer to the total mass and angular momentum, evaluated up to the outer boundary of our numerical grid. Fig.~\ref{fig2} shows that when the instability is triggered and the morphology of the object is reshaped into a spheroidal form, there is an abrupt loss of angular momentum which subsequently approaches zero. These two features -- spheroidal shape and angular momentum loss -- suggest that the final object approaches a non-spinning $l=m=0$ boson star. While the model depicted in Fig.~\ref{fig2} corresponds to the case $\lambda=60$, the results obtained for the other values of $\lambda$ are remarkably similar. We note  that it is possible to construct a countable number of families of boson stars and Proca stars labelled by the azimuthal number $m$. The ADM mass and angular momentum for these stars obey the simple rule $J = m Q$, where $Q$ is their Noether  charge, which means that there are no solutions with a single bosonic field with intermediate values of the angular momentum between 0 and $Q$.

\subsection{$m=2$ spinning Proca stars}

The initial data for the Proca field are described by Eqs.~\eqref{Xphi}-\eqref{Ephi}. Besides evolving the unperturbed case we also consider perturbed initial data. The latter are obtained by replacing in the field equations $e^{2 i \varphi} \rightarrow e^{2 i \varphi} + A e^{2 i \varphi}$, which explicitly breaks  the axisymmetry of the energy density distribution. In Fig.~\ref{fig3} we display time snapshots of the energy density at the equatorial plane for model PS6 in the unperturbed case ($A=0$; top row) and for two different values of the perturbation factor, namely $A=0.01$ (middle row) and $A=0.05$ (bottom row). We can observe that these stars undergo the same fragmentation process that happens for scalar boson stars. The larger the initial perturbation the sooner the instability that breaks the energy distribution  into two pieces occurs. For our most extreme case ($A=0.05$) this phenomenon happens during the gravitational collapse of the initial cloud and before the final compact object is formed. Nonetheless, in all three cases the two pieces remain bounded for a while during the evolution and the timescale at which they rejoin into a spheroidal Proca star is almost the same ($t\approx 4900$ for the unperturbed case, $t\approx 5500$ for both the perturbed cases) regardless of the initial perturbation. We speculate that the initial perturbation amplitude plays a role in the timescale on which the instability grows, but is not relevant for the timescale of the recombination (energy radiation timescale) because the latter is much longer than the first one.

\begin{figure}[]
\centering
\includegraphics[scale = 0.29]{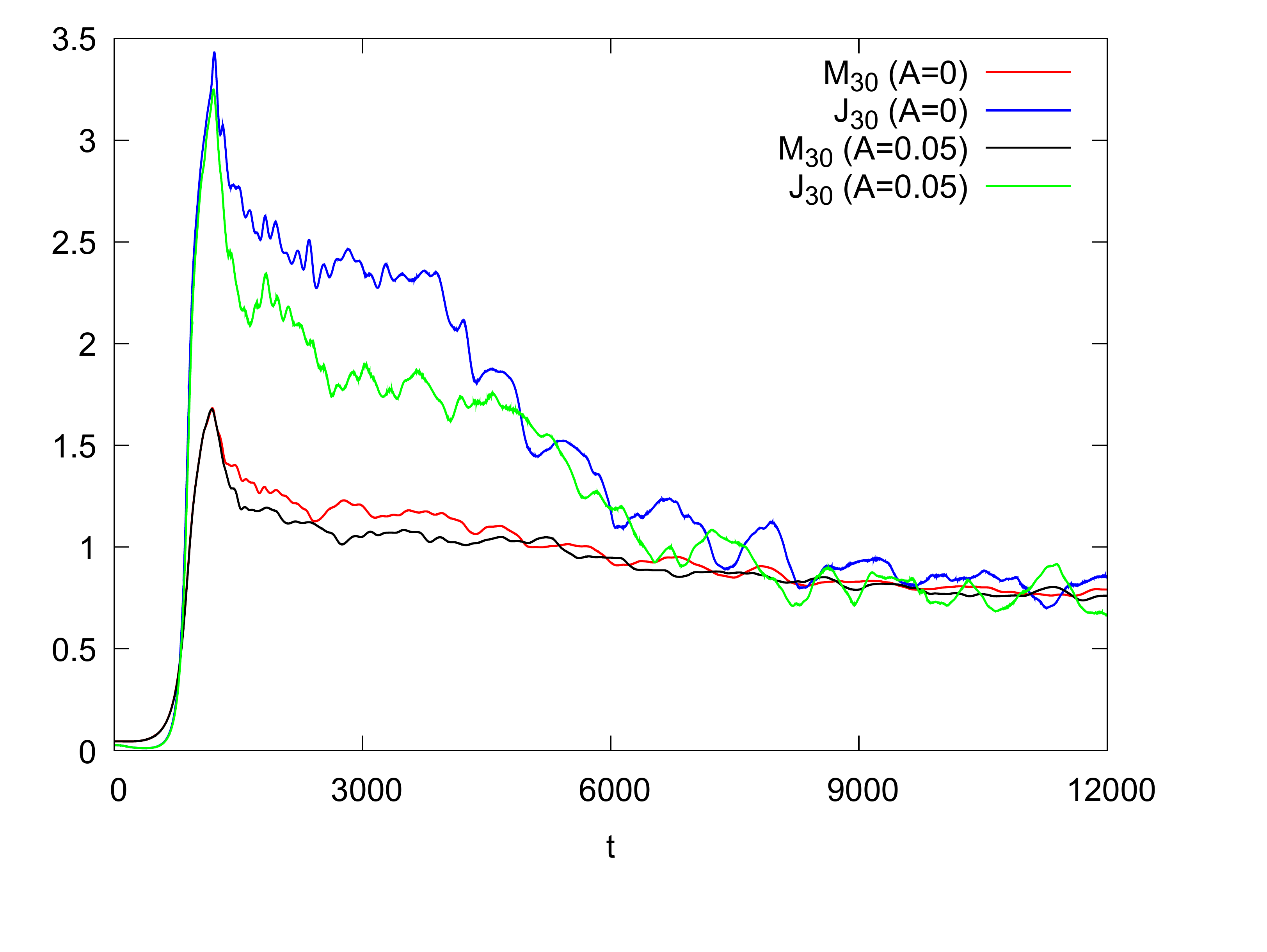} 
\caption{
Evolution of the mass and angular momentum enclosed in a sphere of radius 30 for model PS6, for the cases $A=0$ (unperturbed) and $A=0.05$. The different curves are indicated in the legend.}
\label{fig4}
\end{figure}

Figure~\ref{fig4} depicts the mass $M_{30}$ and angular momentum $J_{30}$ enclosed inside a volume of radius $r=30$ for model PS6 with $A=0$ and $A=0.05$.  We observe that if we perturb the object, it loses smoothly angular momentum from $t\approx 1800$. In the unperturbed case we have a meta-stable phase  from $t\approx 2000$ to $t\approx 3900$ during which the relation $J=mM$ is fulfilled and no loss of angular momentum is found. At $t\approx 3900$, this phase is lost and the angular momentum drops and rapidly reaches the same values as the perturbed case. Comparing this figure with Fig.~\ref{fig2} the different outcomes of spinning scalar and vector clouds becomes manifest. In the case of $m=2$ Proca stars we observe that, for the two models, at the end of the evolution the angular momentum $J_{30}$  is converging to a value similar to that of the mass $M_{30}$. This observation, together with the final spheroidal shape typical of $l=m=1$ Proca solutions, suggest that the models dynamically approach a spinning $m=1$ Proca star. Those stars are stable, as shown in~\cite{sanchis:2019dynamics}.

\subsection{Growth of non-axisymmetric modes}

We turn now to assess the nature of the non-axisymmetry instability found for $m=1$ scalar boson stars and for $m=2$ Proca stars. In particular, we further elaborate on the analogy we first put forward in~\cite{sanchis:2019dynamics} between this dynamical phenomenon in rotating boson stars and differentially rotating neutron stars. It is well-known that differentially rotating  neutron stars  can be subject to various non-axisymmetric instabilities depending on the amount and degree of differential rotation (for a review see~\cite{Paschalidis:2016vmz} and references therein).  For highly differentially rotating stars, an $m= 2$ {\it dynamical} bar-mode instability sets in, driven by hydrodynamics and gravity, $m$ being the order of the azimuthal non-axisymmetric  fluid  mode $e^{\pm im\varphi}$. While we follow the standard notation of using the letter $m$ to indicate the azimuthal number of the perturbation, we warn the reader not to confuse it with the notation we also follow to denote the different families of bosonic stars in the manuscript. The distinction should be clear from the context. Moreover, highly differentially rotating neutron stars can also become unstable to a dynamical one-arm ($m=1$) ``spiral" instability. At lower rotation rates gravitational radiation and viscosity can drive a neutron star {\it secularly} unstable against bar-mode deformation. The occurrence of either kind of bar-mode instability depends on the particular value of the ratio $\beta=T/|W|$ of rotational kinetic energy $T$ to  gravitational  potential  energy $W$ (see~\cite{Paschalidis:2016vmz} for details).

\begin{figure}[]
\centering
\includegraphics[height = 2.37in, width = 3.4in]{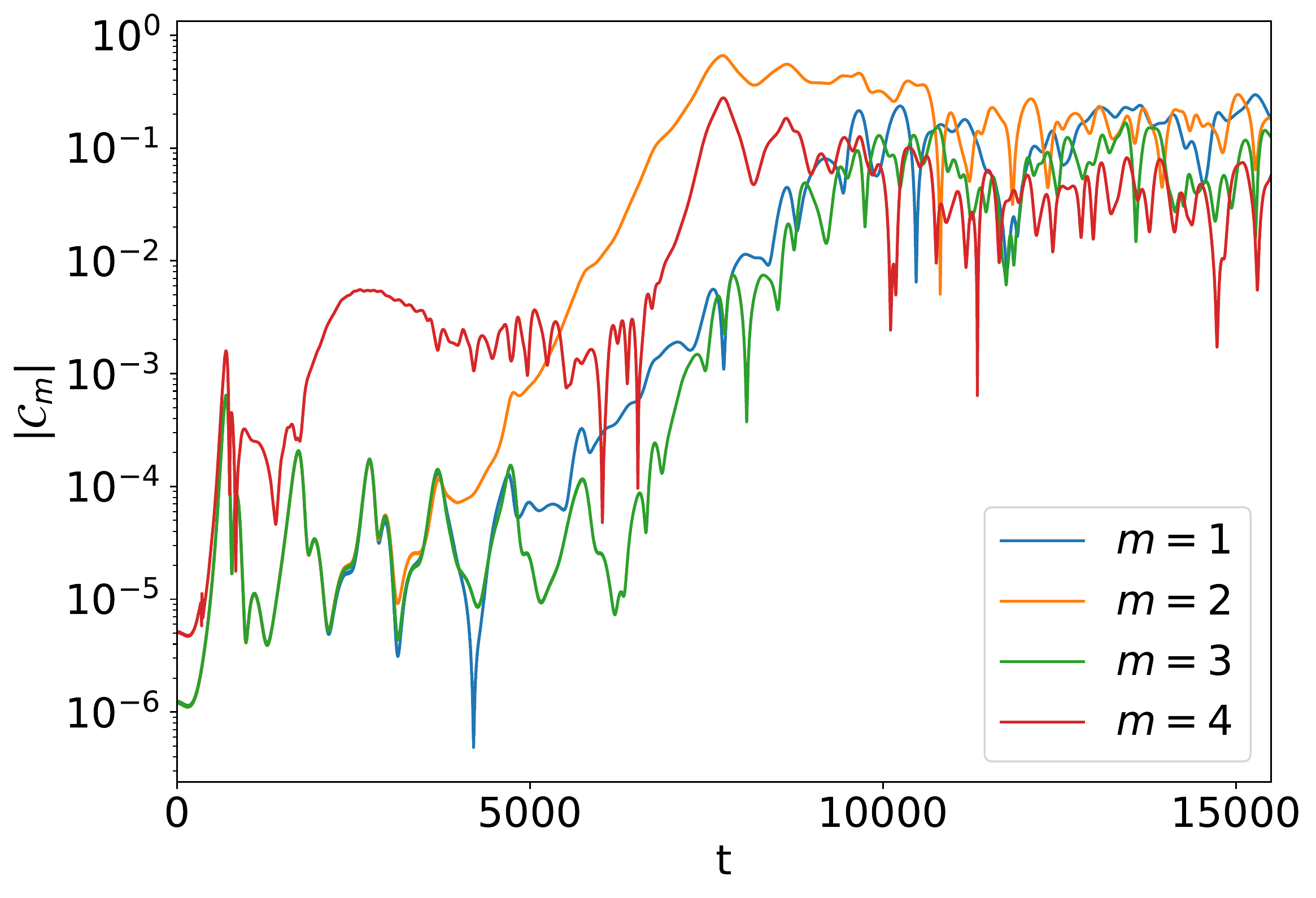} 
\caption{Time evolution of the first four azimuthal modes $|\mathcal{C}_{m}|$  $(m=\lbrace1,2,3,4\rbrace)$ for model BS2 with $\lambda=120$.}
\label{fig5}
\end{figure}

As customary when studying the appearance of non-axisymmetric instabilities in differentially rotating fluids~\cite{CERDADURAN2007288, Baiotti2007, Camarda:2009, Kiuchi:2011, Espino:2019xcl} we monitor the growth of the amplitude of the first few non-axisymmetric modes. To this aim we define the volume-integrated azimuthal density (Fourier) mode decomposition as
\begin{equation}\label{Am}
C_{m} = \int d\textbf{x}^3 \rho_{e}(\textbf{x}) e^{i m \varphi}\,,
\end{equation}
and the corresponding normalized quantity $\mathcal{C}_{m}=\frac{C_{m}}{C_{0}}$. Note that $C_{0}$ is  a measure of the total energy of the system. For our study we consider the first four modes $m=\lbrace1,2,3,4\rbrace$. While we focus our discussion on model BS2 with $\lambda=120$ as a an illustrative case, the results are qualitatively similar for all values of $\lambda$ considered. 

\begin{figure}[t]
\centering
\includegraphics[width=3.2in]{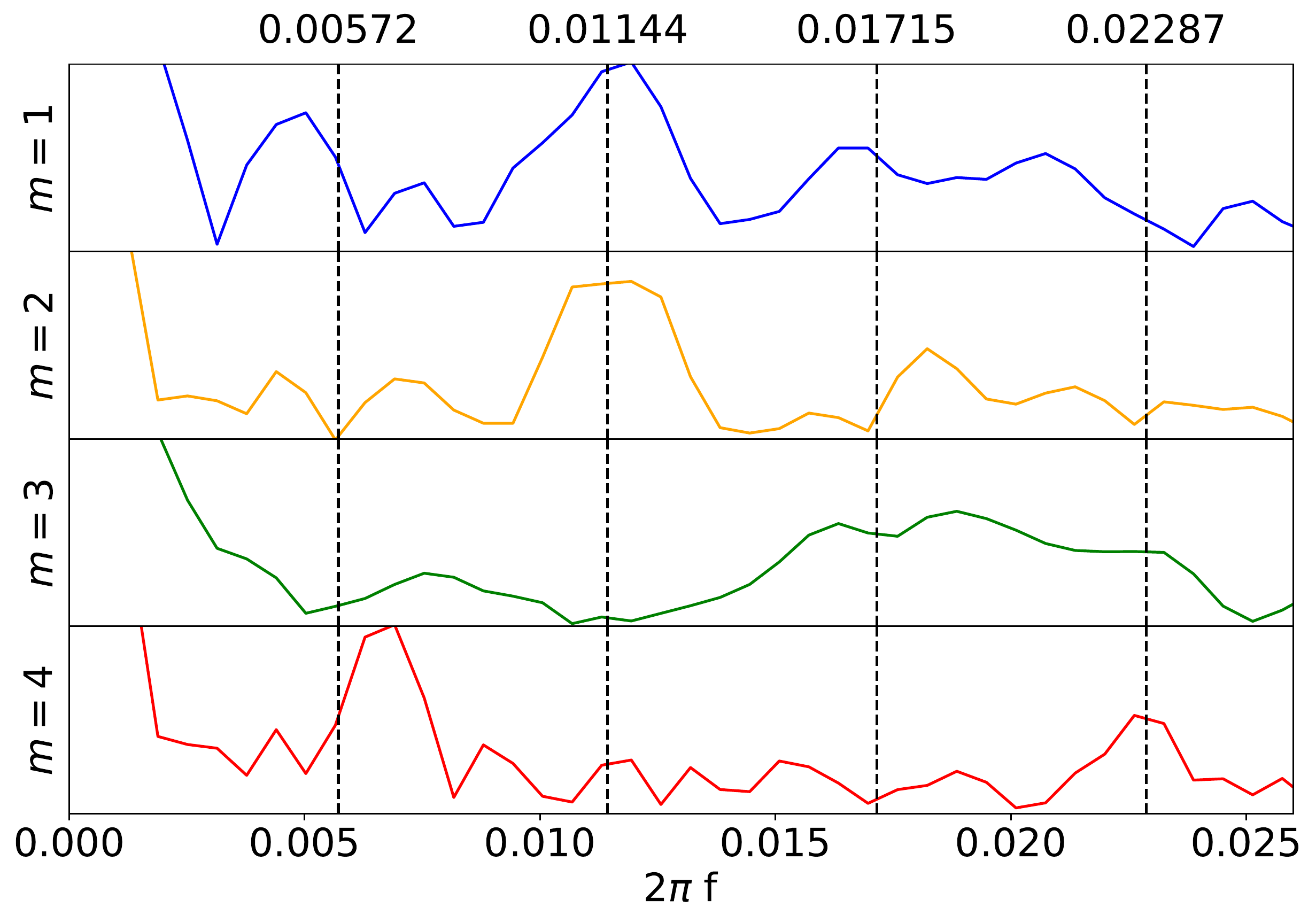} 
\caption{Absolute value of the Fourier transform of $\mathcal{C}_{m}$ for model BS2 with $\lambda=120$. 
The dashed vertical lines (and the values on top of them) are integer multiples of the pattern frequency estimated using the $m=2$
mode frequency ($\omega_{\rm p} = 0.0057$). The amplitude of the Fourier transform is normalised to fit all curves within the plot.}
\label{fig7}
\end{figure}

In Fig.~\ref{fig5} we depict the time evolution of $|\mathcal{C}_{m}|$ for the modes considered, in a logarithmic scale. As the initial data we construct have an axisymmetric energy-momentum tensor, the values of the mode amplitudes are initially  close to zero. We first observe the growth of the $m=4$ mode  whose rapid excitation we attribute to the perturbation triggered by the Cartesian numerical grid employed in our code. Around time $t\approx 3000$ the other three modes start to be excited too, and around $t\approx 4000$ the $m=2$ mode starts growing exponentially to soon become the dominant mode. At mode growth saturation the amplitude of the $m=4$ mode is about one order of magnitude smaller than that of the other modes.

As the $m=1$ mode starts to increase the boson star acquires a non-zero linear momentum. As a result, it undergoes a kick which displaces its center of mass from the origin of the numerical grid. This displacement has to be taken into account in order to properly estimate $C_{m}$ with respect to the center of mass of the star \citep{CERDADURAN2007288}. To this end we redefine the azimuthal coordinate 
\begin{eqnarray}
\varphi = \arctan{\left(\frac{y}{x}\right)} \rightarrow \varphi = \arctan{\left(\frac{y - y_{\rm CM}}{x - x_{\rm CM}}\right)} \,, 
\end{eqnarray}
where $(x_{\rm CM}, y_{\rm CM})$ are the coordinates of the center of mass evaluated as
\begin{align}
x_{\rm CM} &= \frac{1}{M} \int d\textbf{x}^3 \rho_{e}(\textbf{x})x\,,\\
y_{\rm CM} &= \frac{1}{M} \int d\textbf{x}^3 \rho_{e}(\textbf{x})y\,.
\end{align}
The coefficients shown in Fig.~\ref{fig5}, including this correction, show that the $m=2$ mode dominates over all the other modes. At  late times in the numerical evolution the mode growth saturates and all modes have attained high amplitudes. Therefore, the newly formed non-spinning $l=m=0$ boson star is still highly perturbed and far from a stationary solution. 
We note that the time evolution of azimuthal modes we observe for unstable SBSs is formally identical to what is observed in rotating neutron stars (see, e.g.~Fig.~8 of \cite{Baiotti2007} or Fig.~7 of \cite{Espino:2019xcl}).

As discussed in~\cite{Watts:2005,CERDADURAN2007288} (see also~\cite{Paschalidis:2016vmz}) the low $T/|W|$ dynamical bar-mode instability of differentially rotating neutron stars  develops near the so-called corotation radius. This is the radius where the angular frequency of the unstable mode matches the local angular velocity of the fluid. We proceed next to search for the corotational radius in the case of BSs.

We consider that the azimuthal Fourier modes present in the evolution of $\mathcal{C}_{m}$ have the form
\begin{eqnarray}
\mathcal{C}_{m}\approx e^{(\sigma_{m} + i\, \omega_{m})t}\,,
\end{eqnarray}
where $\sigma_{m}$ is the growth rate of the mode and $\omega_m$ the mode frequency. 
The mode frequencies can be extracted by Fourier-transforming the time evolution of $\mathcal{C}_{m}$.

Figure~\ref{fig7} shows the Fourier transform of $\mathcal{C}_{m}$ for model BS2 with $\lambda=120$. For the Fourier transform we consider only the late-time evolution of the modes (from $t\approx 5000$ to the end of the simulation). The main peaks in the spectrum correspond to the frequencies $\omega_m$ of the unstable modes. For the analysis it is interesting to compute the pattern frequency, $\omega_{\rm p} = \omega_{m}/m$. This frequency corresponds to the rotational frequency of the perturbation pattern. For example, for $m=2$ the time it takes for the bar to make a full rotation would be $2\pi/\omega_{\rm P}$. In principle one could define a different pattern frequency for each of the modes. In practice, however, for instabilities associated with the existence of a corotation radius, the pattern frequency for all modes is very close \citep{CERDADURAN2007288}. To check 
this behaviour we compute the pattern frequency from the main peak in the Fourier transform of the dominant $m=2$ mode as $\omega_{\rm p} = \omega_2/2$, which results in $\omega_{\rm p}=0.0057$  and over-plot the value of $m\omega_{\rm p}$ on top of the Fourier transform. The black dashed vertical lines in Fig.~\ref{fig7} represent $m\omega_{\rm p}$, their values indicated on the top of the figure. We observe that the main peaks in the spectrum of the $m=\lbrace 1,2,3,4\rbrace$ modes approximately follow the relation $\omega_{m} = m\omega_{p}$, indicating that they are harmonics of the $\ell=m=2$ mode. We point out that as the formation of bosonic stars is a very dynamical scenario, the spectrum of the modes appears to be noisy, especially for the lower amplitude modes. It is certainly not as clean as that obtained from a linear perturbation of an equilibrium solution, as shown in \citep{CERDADURAN2007288} for stationary models of neutron stars.

\begin{figure}[t]
\centering
\includegraphics[scale = 0.29]{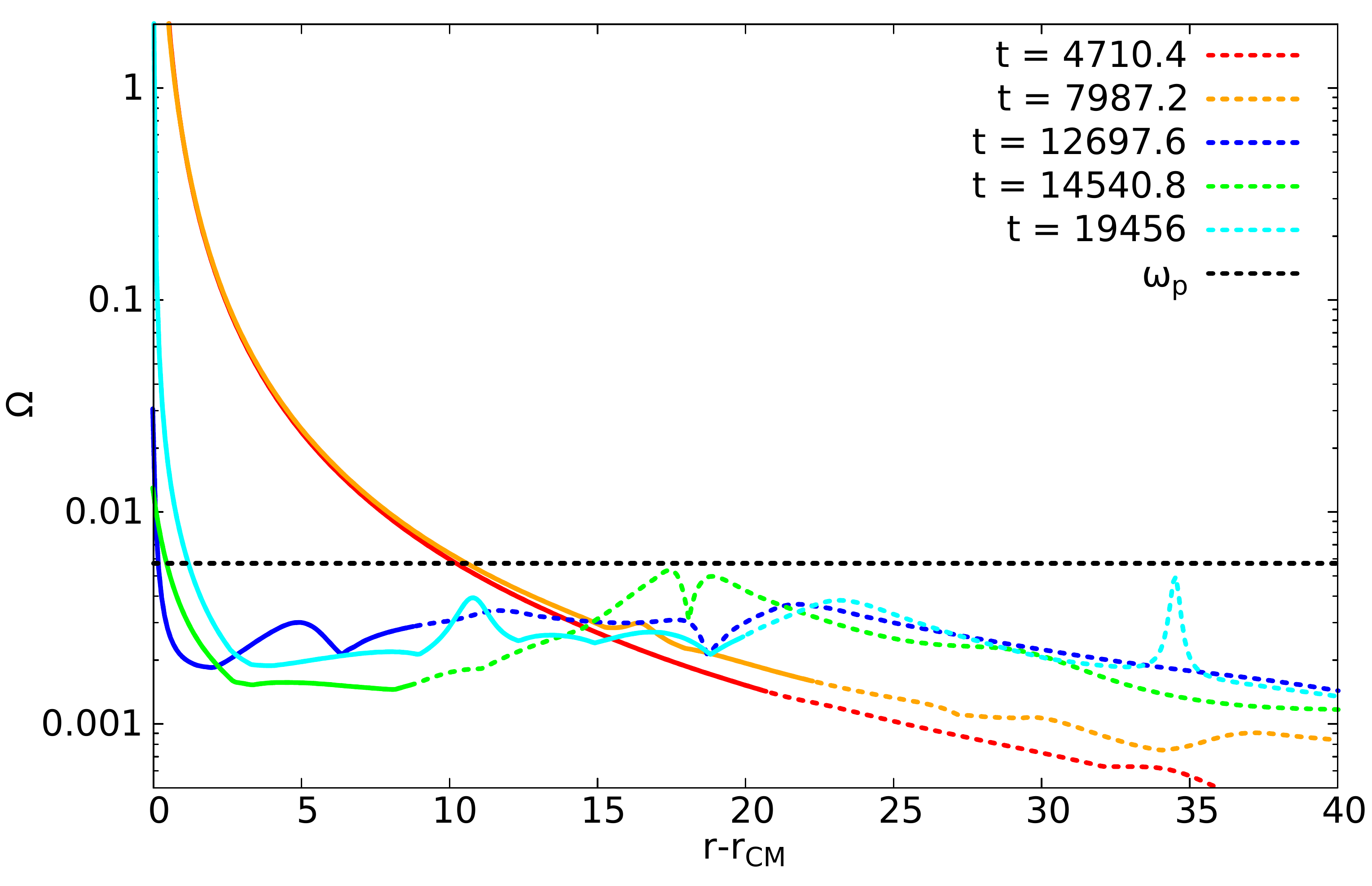} 
\includegraphics[scale = 0.29]{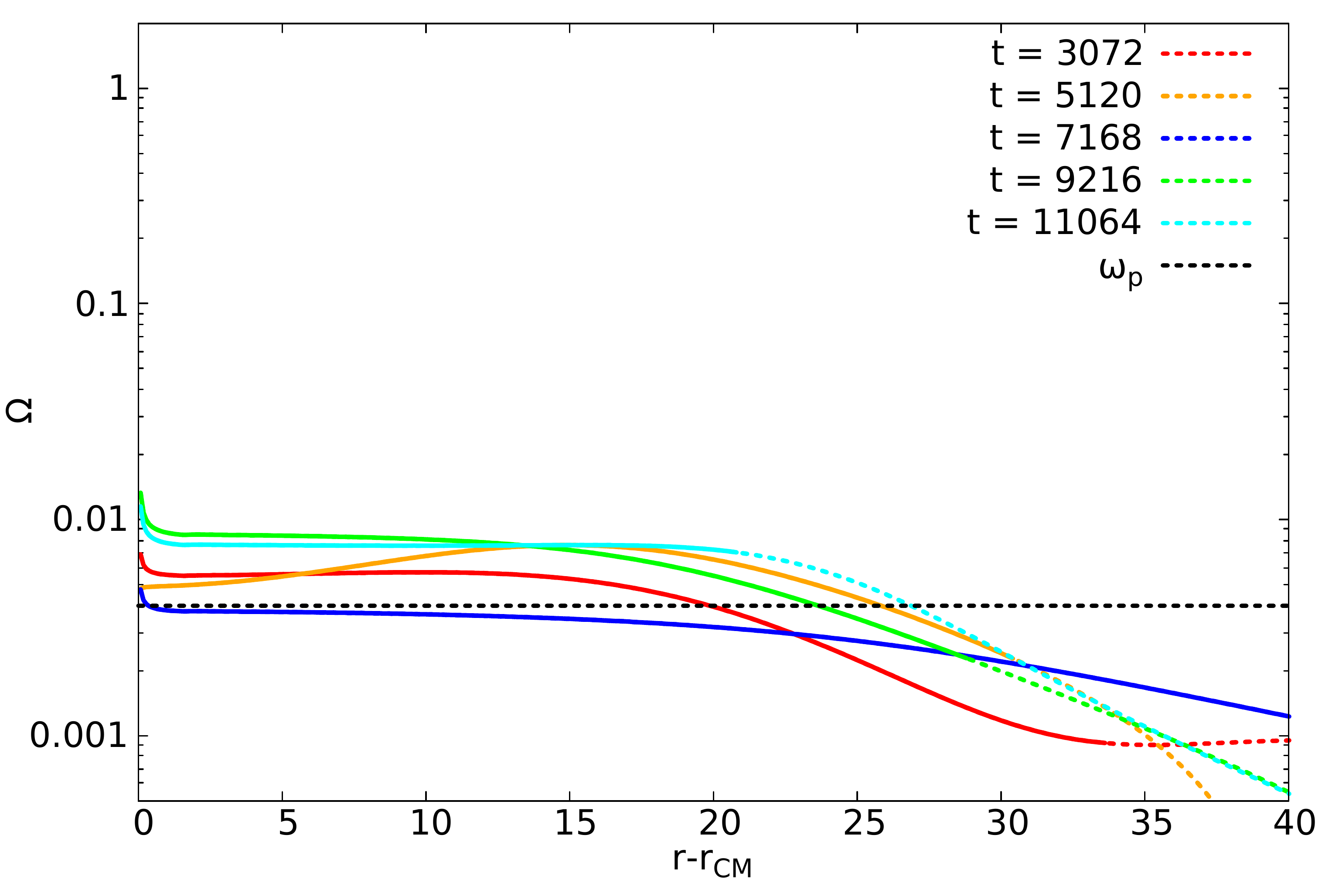} 
\caption{Radial profiles of the angular velocity $\Omega$ for model BS2 with $\lambda=120$
(upper panel) and for an $m=1$ spinning Proca star (lower panel) evaluated at different times. The profiles are shown at the equatorial plane and along the $x$ axis. The horizontal dashed lines indicate the pattern frequency, $\omega_{\rm p}$, computed using the $m=2$ mode frequency in each case. Solid lines represent the region of the star inside $R_{95}$ (interior of the star) and dashed lines outside this radius (exterior). Model BS2 shows a clear corotation radius at $r\approx 10$ at the beginning of the simulation. }
\label{fig8}
\end{figure}

The corotation radius corresponds to the radius at which the pattern frequency is equal to the angular velocity $\Omega$ inside the star, which we define as
\begin{equation}\label{ang_vel}
\Omega = \frac{j^{\phi}}{\rho_{e}}\,,
\end{equation}
by analogy with the definition in relativistic hydrodynamics for a rotating fluid. We warn the reader that this definition of the angular velocity is not a gauge-invariant quantity. The mode pattern speed would be better compared to the angular velocity measured by an observer at infinity, $u^{\varphi}/u^{t}$, rather than to a local gauge-dependent quantity. The reason why we cannot use the latter definition involving a fluid velocity is because we do not have a fluid and, therefore, the only workaround is to compute the angular velocity from the angular momentum of the field.  

The top panel of Fig.~\ref{fig8} shows the radial profile at the equatorial plane of the angular velocity $\Omega$ for model BS2 at different times. Due to the loss of axisymmetry, the radial profile is not the same when evaluated along different directions on the equatorial plane. To obtain clean profiles  we consider time snapshots when the energy density profile is approximately axisymmetric and we evaluate the average from several directions. 
For each profile we estimate $R_{95}$, defined as the radius of a sphere containing $95\%$ of the energy, and highlight this radius in the figure by changing from solid to dashed lines.
As the newly formed object suffers radial oscillations and radiates away energy through the gravitational cooling mechanism, this radius is only a rough estimation. The horizontal dashed line corresponds to the pattern frequency, $\omega_{\rm p}$, estimated above. The red line corresponds to a time when the compact object is already formed but the instability has not yet developed while the orange line shows the profile for a time when the fragmentation process has already started. The  remaining lines correspond to times when the object is already spheroidal. We can observe that for the first two times shown a corotation point exists at radius $r\approx 10$ which is well inside the energy density profile of the star. At later times the angular velocity profile lays entirely below the pattern frequency except at the center. This is an indication that the origin of the observed instabilities is the presence of a corotation point. The instability drives the transport of angular momentum outwards until the corotation point disappears and the instability stops. We point out that the evaluation of the angular velocity $\Omega$ is subject to numerical errors when $r$ approaches $r_{\rm CM}$ due to the fact that $j^{\phi}=j_{\phi}/(r-r_{\rm CM})^{2}$ on the equatorial plane. 

We repeat the same study for an $m=1$ spinning Proca star model that we perturb by hand. In~\cite{sanchis:2019dynamics} we showed that this model does not develop a non-axisymmetric instability, so one would expect not to observe a corotation point. The bottom panel of Fig.~\ref{fig8} shows the radial profile of the angular velocity $\Omega$ for this model extracted at different times, after the formation of the compact object. In this case the measured pattern frequency (horizontal line) is $\omega_{\rm p} =0.004$. For this model the profile of $\Omega$ is much flatter than in the case of the scalar SBS (BS2), and with values close to the pattern frequency. As the density profile radially oscillates, we observe 3 different  phases. When the object is at its maximal extension, the angular velocity $\Omega$ is entirely below the pattern frequency (see $t=7168$, blue line) meaning there is no corotational point. When the object is at its maximal contraction (see $t=11064$, cyan line) $\Omega$ is above the pattern frequency and only crosses it in regions outside $R_{95}$, meaning that there is no corotational point inside the star. For intermediate cases (the other 3 lines in the figure) we can find a corotational point which lies inside the star. This behaviour is an indication of the presence of non-linear oscillations because linear oscillations would have had an amplitude sufficiently small not to modify the background of $\Omega$. The non-linearity is caused by the high amplitude of the oscillations triggered by the collapse of the cloud that leads to the formation of the Proca star in our simulations and is hence unavoidable in our setup. Therefore, we can only conclude that the resulting equilibrium configuration either has no corotational point, and thus is stable to corotational instabilities, or has one but the profile of $\Omega$ is sufficiently shallow not to allow for the grow of instabilities in dynamical timescales.

\subsection{Gravitational waveforms}

The prospects of formation of rapidly-rotating neutron stars following the gravitational collapse of the core of massive stars or through the accretion-induced collapse of a white dwarf, highly motivates the investigation of the GWs produced by non-axisymmetric instabilities (namely, the $f$-mode -- or bar-mode -- and the $r$-mode) that may affect them. Provided that neutron stars do not reach magnetar-like, magnetic-field values (i.e.~for saturation values of the field $B\le 10^{14}$G) the GWs from the $f$-mode (i.e.~bar-mode) instability should be well within the detection capabilities of third-generation interferometers such as the Einstein Telescope, yet they are only marginal for the current LIGO-Virgo detector network~\cite{Glampedakis:2018}. Notwithstanding the simplicity of our setup for the dynamical formation of SBSs, it is worth computing the corresponding gravitational waveforms for such bosonic objects and compare our estimates with those for neutron stars, an exercise we attempt in this section. 

\begin{figure}[]
\centering
\includegraphics[width=3.4in, height = 2.5in]{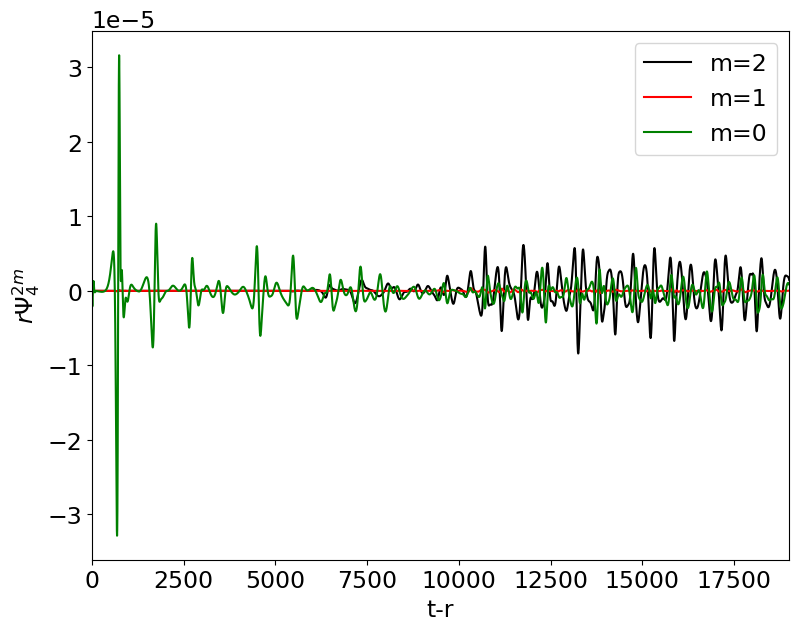} 
\caption{Real part of $r\Psi_{4}^{2,m}$ for $m=1,2$ for model BS2 with $\lambda=120$.}
\label{fig6}
\end{figure}

\subsubsection{Gravitational wave extraction}

The GW emission is computed through the mode decomposition of the Newman-Penrose scalar $\Psi_4$ in spin-weighted spherical harmonics with spin $-2$. The coefficients
 $\Psi_{4}^{l,m}$ for $l=2$ and $m=1,2$ are extracted at radii $r=\lbrace 200,300,500,600,1000 \rbrace$. The GW strain $h = h_+ - i\, h_\times$, where $h_+$ and $h_\times$ are the two polarizations, is related to the second time-derivative of the Newman-Penrose scalar, as $\Psi_4 = -\ddot h$. We evaluate $r\Psi_{4}^{2,m}$ by interpolating with a third-order polynomial fit the values from three different extraction radii, namely $r=\lbrace 300,600,1000 \rbrace$. Figure~\ref{fig6} displays the real part of $r\Psi_{4}^{2,m}$ for $m=0,1,2$, for model BS2 with $\lambda=120$. During the formation process we observe a dominant $m=0$ (axisymmetric) mode in the GW emission. The signal is periodic and it is due to energy emission triggered by the quasi-radial oscillations of the newly formed object. When the non-axisymmetric instability kicks in  we observe, as expected, that the $m=2$ quadrupolar mode becomes the dominant GW emitter  while the $m=1$ mode reaches maximum amplitudes about two orders of magnitude smaller. Correspondingly, the $m=0$ mode is of comparable amplitude or one order of magnitude smaller. 
 
The evolution of the waveform we have just described closely follows the dynamics of this model, displayed in the preceding figures. We observe that the GW emission from non-axisymmetric modes starts around $t\approx 6000$ which, for model BS2 (see e.g.~Fig.~\ref{fig5}), corresponds  to the time when the exponential growth of the $m=2$ mode is about to reach its saturation amplitude. Around that time the object undergoes fragmentation (see second row of Fig.~\ref{fig1}) and starts losing angular momentum (see Fig.~\ref{fig2}). Therefore, we find a direct correspondence between the loss of angular momentum and the emission of GWs.

\subsubsection{Detectability}

\begin{figure*}[t]
\centering
\includegraphics[scale = 0.60]{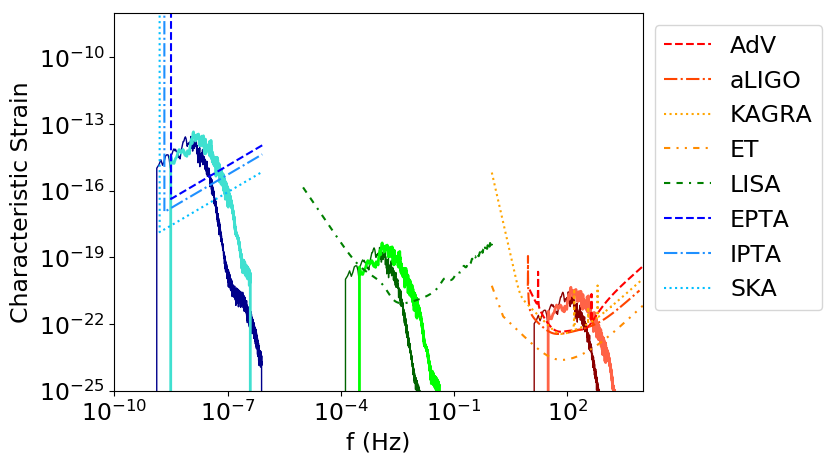} 
\caption{Characteristic GW strain against frequency for model BS2 with $\lambda=120$ (darker colours) and for model PS6 without a perturbation (lighter colours) compared with the sensitivity curves $\sqrt{f S_{n}(f)}$ of a variety of GW detectors. Three different masses are employed, namely $\{5,5\times 10^5, 5\times 10^{10}\}\, M_{\odot}$ for model BS2 and twice those values for model PS6. A source  distance $D=1$~Mpc is assumed for ground-based detectors while for LISA and PTA we assume $D=1$~Gpc.}
\label{fig_sensitivity}
\end{figure*}

For burst-like sources the characteristic GW amplitude is (see e.g.~\cite{Flanagan:1998a})
\begin{equation}
h_{\rm char} (f) = \frac{{1+z}}{\pi D{(z)}}\sqrt{ 2 \frac{dE}{df}[{(1+z)}f]}, \label{eq:hchar}
\end{equation}
where $D$ is the distance to the source, $z$ is the redshift, and $dE/df$ is the energy spectrum of the waves. We use the cosmology calculator described in \cite{Wright:2006} to compute $D(z)$, with values of $H_0=70$~km/s/Mpc and $\Omega_{\rm m}=0.3$ for the Hubble constant and the fraction of energy density of matter, respectively.

For an optimally oriented detector the matched-filtering SNR squared,
averaged over all possible source orientations is \citep{Flanagan:1998a}
\begin{equation}
\rho^2_{\rm optimal} = \int_0^\infty  d (\ln f) \frac{h_{\rm char}(f)^2}{f S_{\rm n}(f)}, \label{eq:SNR}
\end{equation}
where $S_{\rm n}(f)$ is the power spectral density (PSD) of the detector noise.
Therefore, when plotting $h_{\rm char}$ of the signal with $\sqrt{f S_{\rm n} (f)}$
in log-log scale, the area of the former quantity over the latter is directly related to the optimal SNR.
The average SNR square over all possible detector orientations and sky locations is
simply $\langle\rho^2\rangle = \rho^2_{\rm optimal} / 5$.

The energy spectrum can be computed from the local energy flux
\begin{equation}
\frac{1}{r^2}\frac{dE}{d\Omega df} = \frac{\pi f^2}{2} |\tilde h(f)|^2 = \frac{1}{8\pi}\frac{|\tilde \Psi_4|^2}{(2 \pi f)^2}, 
\end{equation}
where we use tilde for the Fourier transform. The energy spectrum can be obtained by integrating in angles
\begin{equation}
\frac{dE}{df} = \int d\Omega \frac{dE}{d\Omega df} = \frac{1}{8 \pi (2 \pi f)^2} \sum_{lm} |r \tilde\Psi_4^{lm}|^2, \label{eq:dedf}
\end{equation}
where we have used the spherical decomposition of $\Psi_4$ and the orthonormality relations of the spin-weighted
spherical harmonics. This expression allows us to compute the energy spectrum directly from the $\Psi_4^{lm}$ extracted in the numerical simulations at the extraction radius $r$.

\begin{figure*}[t]
\centering
\includegraphics[scale = 0.35]{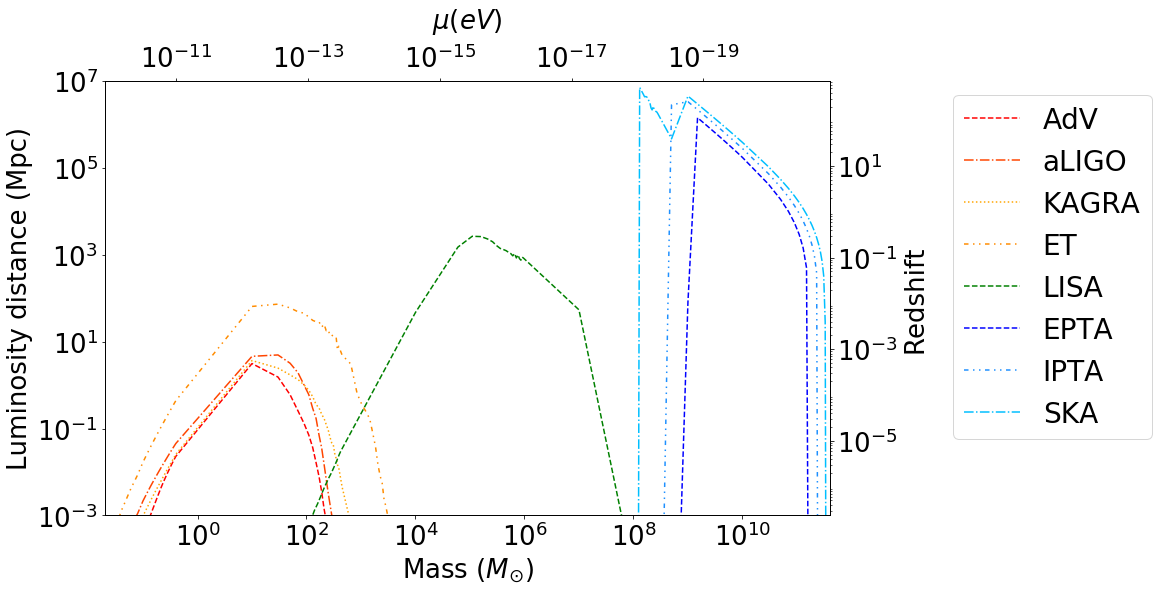} 
\caption{Horizon distances as a function of mass for model BS2 with $\lambda=120$ evaluated for a variety of GW detectors. The top $x$ axis shows the corresponding particle mass $\mu$ and the right $y$ axis the corresponding redshift. We again show the results for ground-based detectors (red colours), LISA (green color), and PTA (blue colors).} 
\label{fig_horizon}
\end{figure*}

Since our system scales with the mass parameter $\mu_{0|1}$, which sets the mass of the system, the typical frequency of the waveform may lie in the frequency band of different GW observatories. We have considered three cases: i) Ground-based laser interferometers, including the ongoing experiments Advanced LIGO (aLIGO) \citep{aLIGO2015}, Advanced Virgo (AdV) \citep{aVirgo2015} and KAGRA \citep{Somiya:2012}, and the future observatory Einstein Telescope (ET) \citep{ET:2010}; ii) the space-based laser interferometer
LISA \citep{LISA:2013}; and iii) pulsar timing arrays (PTA), employing the canonical values used in \cite{Moore:2015:PTA} as a proxy to current observational limits, namely the future International Pulsar Timing Array (IPTA) \citep{IPTA:2013}  for $15$~yr of observation and the Square Kilometre Array (SKA) \citep{SKA:2009} for $20$~yr of observation
(details  can be found in \cite{Moore:2015:sensitivity}). In all cases we use the design sensitivity curves. For PTA we use as sensitivity curves the detection limits for a monochromatic source with a (sky averaged) SNR detection threshold of $\rho_{\rm thr}= 3$.

Figure~\ref{fig_sensitivity} shows the characteristic GW strain for model BS2 with $\lambda=120$ (darker colours) and for model PS6 without perturbation (lighter colours). We consider three different values for the mass of the system, namely $\{5,5\times 10^5, 5\times 10^{10}\} M_\odot$ for model BS2 and twice as much for model PS6. For these ranges of masses the characteristic frequencies of the signals lie within the sensitivity ranges of ground-based detectors, space detectors, and PTA, respectively. For ground-based detectors we show the characteristic strain of a signal from a source at a distance $D=1$~Mpc. For such a distance the signal is only marginally detectable by aLIGO, AdV, and KAGRA but it is within the detection capability of ET. For LISA and PTA we consider a distance to the source of $D=1$~Gpc. Even at such large distance the signal could be detected. We next discuss quantitatively the detectability of these signals for each detector, defining and evaluating the horizon distance.

For the range of masses in the sensitivity range of ground-based and space-based GW observatories, the (scaled) duration of the event is $\sim1$~s and $\sim1$~day, respectively.
The waveform can therefore be regarded as a burst, with limited time duration, and we can compute the SNR using Eq.~\eqref{eq:SNR}.  In these cases we define the horizon as the distance at which the average SNR is $\langle\rho\rangle =\rho_{\rm thr}=8$. For PTA the typical duration of the event is $300$~y, much longer than the duration of the experiment. Here, the waveform can be regarded as a quasi-monochromatic, continuous signal with frequency and characteristic strain corresponding to the peak $h_{\rm char}$ and frequency. In those cases we compute the horizon as the distance at which the peak amplitude is equal to the detection threshold (corresponding to SNR~$\ge 3$) at the peak frequency.

Figure~\ref{fig_horizon} shows the horizon for model BS2 with $\lambda=120$ for a variety of detectors, as a function of the mass of the source. In the right $y$ axis we show the redshift value corresponding to the luminosity distances reported in the left $y$ axis. One must recall that the mass of each SBS model is expressed in dimensionless units and that we can assign a physical mass $M(M_{\odot})$ only after specifying a physical value for the particle mass $\mu_{0}$. The upper $x$ axis indicates the particle mass $\mu_{0}$(eV) corresponding to the values of mass $M (M_{\odot})$ shown in the lower $x$ axis, evaluated as
\begin{eqnarray}
\mu_{0}({\rm eV}) = \frac{\mathcal{M}_{\rm BS2}}{M_{\rm BS2}(M_{\odot})}\frac{M_{\rm Pl}^2}{M_{\odot}}\,,
\end{eqnarray}
 where $M_{\rm BS2}(M_{\odot})$ is the mass in physical units and $\mathcal{M}_{\rm BS2}=0.889$ is the mass of model BS2 in dimensionless units. 
 
A GW signal from a stellar-size SBS with mass in the range $1-100$~$M_{\odot}$ might be detected by current 2nd-generation detectors at distances of a few Mpc. 3rd-generation detectors would increase the range of masses and the horizon to a few 10 Mpc. LISA sources would be in the $10^4-10^6$~$M_\odot$ range and detectable up to a few Gpc, while PTA sources would be in the $10^9-10^{11}$~$M_\odot$ range and detectable up to redshifts of $\sim 100$.
  
The non-detection of this kind of sources by the current GW detectors (aLIGO, AdV, and PTA) allows to set upper limits on the expected rates of such events. A detailed calculation of these rates is out of the scope of this work but we can compute an order-of-magnitude estimate. Given the non-observation, the rate of events per unit volume $R$ cannot be much larger than $1/ (V_{\rm obs} T_{\rm obs})$, where $V_{\rm obs}$ is the observing volume, which can be computed from the horizon estimation, $D_{\rm obs}$, and $T_{\rm obs}$ is the duration of the observation. Using typical values for ground-based detectors and PTA yields rate estimates in two mass ranges:
 %
 %
 \begin{equation}
 R \lesssim 0.2 \left( \frac{D_{\rm obs}}{1\, {\rm Mpc}}\right)^{-3}  \left( \frac{T_{\rm obs}}{1\, {\rm yr}}\right)^{-1} {\rm yr}^{-1} \, {\rm Mpc}^{-3} \,,
\end{equation}
for $M\sim 10 M_\odot$, and
 \begin{equation}
 R \lesssim 2\times10^{-11} \left( \frac{D_{\rm obs}}{1\, {\rm Gpc}}\right)^{-3}  \left( \frac{T_{\rm obs}}{10\, {\rm yr}}\right)^{-1} {\rm yr}^{-1} \, {\rm Mpc}^{-3} \,,
\end{equation}
for $M\sim 10^{10} M_\odot$.
 
Future experiments (LISA, ET, SKA) will put even tighter constraints in the rate of these events and may have implication on the formation rate of bosonic stars or even on their existence. On the other hand, given the relation between the mass of the object and the particle mass $\mu_{0}$ (or $\mu_{1}$), observations of such events would help place tight constraints on the mass of the boson.
 
\begin{figure}[t]
\centering
\includegraphics[scale = 0.42]{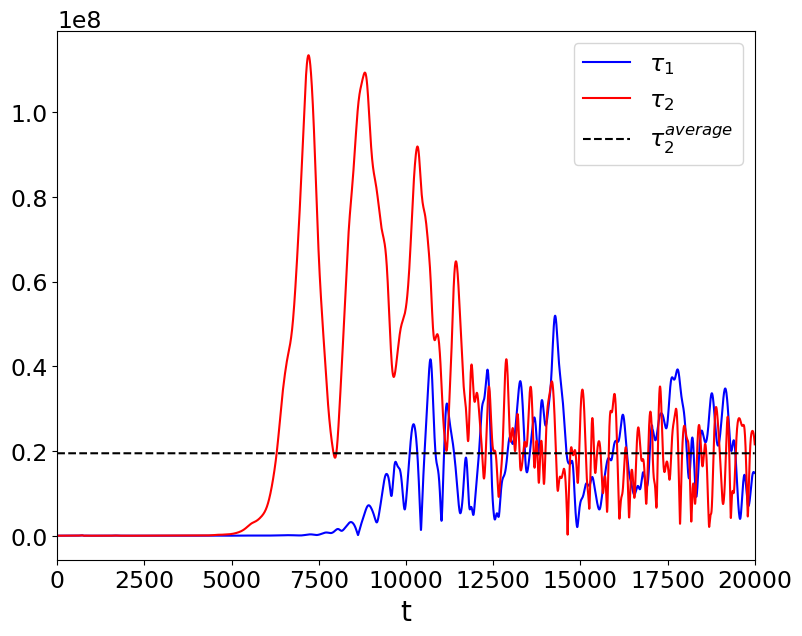} 
\caption{Time evolution of the gravitational wave damping time $\tau_{m}$ for model BS2 with $\lambda=120$. The dashed black horizontal line is the average value of $\tau_2$ evaluated in the time window $t\in[12500,20000]$. }
\label{fig_tau}
\end{figure}

As a final remark we note that by the end of our simulations the bar-like deformation ($m=2$ mode) that was developed during the instability is still present. Even if the condition for the corotational instability is not fulfilled anymore, this deformation may last for a long time emitting GWs. Unlike neutron star matter, bosonic fields do not have efficient dissipation mechanisms such as viscosity to remove the deformation. Therefore, the characteristic damping timescale in which the deformation is erased is set by GW emission. We can estimate the GW damping timescale $\tau_m$ of a non-axisymmetric mode ($m>0$) as the ratio of the energy in each mode  (approximately  $C_m$) to the GW luminosity of the mode $L_m$,  
 \begin{equation}
 \tau_m = \frac{|C_m|}{L_m}\,,
 \label{damping-timescale}
 \end{equation}
where the GW luminosity can be computed integrating Eq.~{\eqref{eq:dedf}} for the relevant values of $m$,  
  \begin{equation}
  L_m 
  = \frac{1}{8\pi} \int_0^\infty df \frac{1}{2\pi f} \sum_l |r \tilde \Psi_4^{lm}|^2\,.
  \end{equation}
Figure \ref{fig_tau} shows the evolution of $\tau_m$ for $m=1,2$ in one of our simulations. After a transient phase associated with the gravitational collapse of the initial cloud and the development of the instability, $\tau_m$ settles to a mean value $\tau_m \approx 2\times 10^7$. This value is about $1000$ times longer than that of the signal connected with the instability. During this time, the emission is expected to be essentially monochromatic  for each of the emitting modes. Therefore,  the post-collapse GW emission is anticipated to be orders of magnitude more energetic than the emission due to the bar-mode instability itself, turning bar-mode-unstable SBSs into hypothetical potentially interesting sources of continuous GWs. For ground-based and space-based interferometers the typical duration of those events would be $\sim 1000$~s and $1000$~days, respectively.
 Using the appropriate detection methods for monochromatic waveforms, whose sensitivity scales with $1/\sqrt{T_{\rm obs}}$, and sufficiently long observation times 
 ($T_{\rm obs}$ of the order of the event duration) the detector horizon for this kind of detectors (Fig.~\ref{fig_horizon}) could be enlarged by a factor $\sim \sqrt{1000}$.
 
We emphasize that the previous estimates should be taken as upper limits as  Eq.~(\ref{damping-timescale}) is overestimating the damping timescale. If it were possible to make a spherical-harmonic decomposition of the background, then $C_m$ would include the energy of all the modes with $l\ge m$. Since GWs are emitted predominantly due to the $l=2$ mode, the numerator of Eq.~(\ref{damping-timescale}) includes the energy of all $ l\ge 2$ while the denominator only includes, essentially, the $l=2$ contribution. The most accurate computation of the damping timescale would involve the extraction of the mode eigenfunctions of the background. While extracting the eigenfunctions from the numerical simulations is possible it would require additional simulations and a complicated analysis which is beyond the scope of the present investigation. 
 
\section{Conclusions}
\label{conclusions}

The interest in studying exotic, horizonless compact objects~\cite{Cardoso2019} as BH mimickers has increased in recent years thanks, in part, to the detection of gravitational waves. Among the simplest, and dynamically more robust, proposals are self-gravitating compact objects made of bosonic particles, either scalar or vector, commonly referred to as boson stars and Proca stars, respectively~\cite{Kaup:1968zz,Ruffini:1969qy,Brito:2015pxa}. In this paper we have studied these systems through three-dimensional numerical-relativity simulations of the Einstein-Klein-Gordon system and of the  Einstein-Proca system, employing complex and massive fields. Using constraint-satisfying initial data representing clouds of scalar/Proca particles with non-zero angular momentum our time evolutions have shown the gravitational collapse of the clouds and the formation of SBSs via gravitational cooling. This paper is a significant  extension of our previous work~\cite{sanchis:2019dynamics} where the transient nature of the newly formed SBS in the scalar case was established. The scalar star is always affected by the growth of a non-axisymmetric instability which triggers the loss of angular momentum and its  migration to a non-spinning boson star. The situation is different for $m=1$ Proca stars, which do not suffer from such instability. In~\cite{sanchis:2019dynamics} we related the different stability properties to the different toroidal/spheroidal topology of the scalar/Proca models. 

The results of the new numerical-relativity simulations reported in the present work have allowed us to draw a more complete picture of the dynamical formation scenario of SBS and of their stability properties in the non-linear regime. Not only have we incorporated additional aspects for the physical description of the system (e.g.~accounting for a quartic self-interaction potential in the scalar case to gauge its effect on the instability or investigating toroidal ($m=2$) Proca stars to confirm our conjecture that they are indeed unstable) but we have also carried out a deeper analysis of the development of the bar-mode instability in SBS and associated GW emission. This analysis has made use of the study of the growth rate of azimuthal density modes in the stars and the search of a corotation point in unstable models. This is an approach commonly employed to study bar-mode unstable neutron stars. Interestingly, we have found that the dynamics of bar-mode unstable SBSs bears a close resemblance with that of their neutron star ``relatives". This parallelism has been discussed to some length in this paper.

Our main results regarding the stability properties of SBSs indicate that: (a) the self-interaction potential can only delay the instability in scalar SBSs but cannot quench it completely; (b)  $m=2$ Proca stars always migrate to the stable $m=1$ spheroidal family; and (c) unstable $m=2$ Proca stars and $m=1$ scalar boson stars exhibit a pattern of frequencies for the azimuthal density modes which crosses the angular velocity profile of the stars in the corotation point. 

An important part of this research has dealt with the analysis of the GWs emitted by SBSs as a result of non-axisymmetric deformations. We have extracted the gravitational waveforms of some representative models and we have investigated their detectability prospects. This has been done by comparing the characteristic strain of the signal with the sensitivity curves of a variety of detectors (current ground-based interferometers Advanced LIGO, Advanced Virgo and KAGRA, the 3rd-generation detector ET, and space missions such as LISA and Pulsar Timing Arrays) and by computing the signal-to-noise ratio for different ranges of masses and for different source distances. Our study has revealed that GWs from a stellar-size SBS in the $1-100$~$M_{\odot}$ mass range might be detected by 2nd-generation detectors up to a few Mpc while 3rd-generation detectors would increase the range of masses and the horizon to a few 10 Mpc. LISA could observe SBS sources in the $10^4-10^6$~$M_\odot$ mass range  up to a few Gpc. For PTA the sources would be in the $10^9-10^{11}$~$M_\odot$ mass range and could be detectable up to redshifts of $\sim 100$. Moreover, by assuming that the characteristic damping timescale of the bar-like deformation in SBSs is only set by GW emission and not by viscosity, unlike what happens for neutron stars where the two effects must be taken into account, we have found that the post-collapse emission could be orders of magnitude more energetic than that of the bar-mode instability itself. As a result, if SBS existed in Nature, the findings reported in this paper would turn them into potentially interesting sources of continuous gravitational waves. The theoretical estimates reported in this work offer the intriguing possibility to probe (or constrain) the existence of bosonic stars and could in turn help place tight constraints on the mass of the constitutive bosonic particle.

\begin{acknowledgments}

We thank Carlos Palenzuela, Dar\'io N\'u\~nez, Juan Carlos Degollado, Sergio Gimeno-Soler, Jens Mahlmann and Christopher Moore for useful discussions and valuable comments, and Nikolaos Stergioulas for a careful reading of the manuscript. This work has been supported by the 
Spanish   Agencia   Estatal   de   Investigaci\'on (PGC2018-095984-B-I00),  by  the  Generalitat  Valenciana  (PROMETEO/2019/071 and  GRISOLIAP/2019/029), by the Funda\c c\~ao para a Ci\^encia e a Tecnologia (FCT) projects PTDC/FIS-OUT/28407/2017, CERN/FIS-PAR/0027/2019, UIDB/04106/2020 and UIDP/04106/2020 (CIDMA) and UID/FIS/00099/2020 (CENTRA), by national funds (OE), through FCT, I.P., in the scope of the framework contract foreseen in the numbers 4, 5 and 6
of the article 23, of the Decree-Law 57/2016, of August 29, changed by Law 57/2017, of July 19. This work has further been supported by  the  European  Union's  Horizon  2020  research  and  innovation (RISE) programme  H2020-MSCA-RISE-2017 Grant No.~FunFiCO-777740. We would like to acknowledge networking support by the COST Action GWverse
CA16104.
MZ acknowledges financial support provided by FCT/Portugal through the IF
programme, grant IF/00729/2015.
PC acknowledges the Ramon y Cajal funding (RYC-2015-19074) supporting his research.
Computations have been performed at the Servei d'Inform\`atica de la Universitat
de Val\`encia, on the ``Baltasar Sete-Sois'' cluster at IST, and at MareNostrum
(via PRACE Tier-0 Grant No. 2016163948).

\end{acknowledgments}



\bibliography{num-rel2}

\end{document}